\documentclass[11pt,a4paper]{article}
\usepackage{jcappub}
\pdfoutput=1
\RequirePackage[T1]{fontenc}
\renewcommand{\thetable}{\arabic{table}}
\usepackage{graphicx}
\usepackage{color}
\usepackage{mathtools} 
\usepackage{amsmath}
\usepackage{caption}
\usepackage{subcaption}
\usepackage{float}
\usepackage{color}
\usepackage{multirow}
\usepackage{array}
\usepackage{tabularx}
\usepackage{calc}
\usepackage[utf8x]{inputenc}
\definecolor{My_red}{cmyk}{0.00,1.00,1.00,0.20}

\usepackage{slashed, latexsym, verbatim, cancel}

\usepackage{graphicx,bm,color}
\usepackage{slashed,verbatim}
\usepackage{gensymb}
\usepackage{footmisc}
\usepackage{hyperref}
\usepackage{tablefootnote}
\renewcommand{\thesection}{\arabic{section}}

\newcommand{\bmat}{\left(\begin{array}}
\newcommand{\emat}{\end{array}\right)}
\newcommand{\beq}{\begin{equation}}
\newcommand{\eeq}{\end{equation}}

\def\bwt{\begin{widetext}}
\def\ewt{\end{widetext}}
\def\be{\begin{equation}}
\def\ee{\end{equation}}
\def\bea{\begin{eqnarray}}
\def\eea{\end{eqnarray}}
\def\bean{\begin{eqnarray*}}
\def\eean{\end{eqnarray*}}
\def\bary{\begin{array}}
\def\eary{\end{array}}
\def\bit{\begin{itemize}}
\def\eit{\end{itemize}}
\def\GeV{\,{\rm GeV}}

\def\su5u1{SU(5) \times U(1)}
\def\fsu5u1{SU(5) \times U(1)'}
\def\so10{SO(10)}
\def\sq20{SO(10) \times SO(10)}

\def\bwt{\begin{widetext}}
\def\ewt{\end{widetext}}
\def\be{\begin{equation}}
\def\ee{\end{equation}}
\def\bea{\begin{eqnarray}}
\def\eea{\end{eqnarray}}
\def\bean{\begin{eqnarray*}}
\def\eean{\end{eqnarray*}}
\def\bary{\begin{array}}
\def\eary{\end{array}}
\def\bit{\begin{itemize}}
\def\eit{\end{itemize}}
\def\GeV{\rm GeV}

\def\su5u1{SU(5) \times U(1)}
\def\fsu5u1{SU(5) \times U(1)'}
\def\so10{SO(10)}
\def\sq20{SO(10) \times SO(10)}

\usepackage{amssymb}

\usepackage{color}
 \usepackage[normalem]{ulem}

\definecolor{darkcyan}{cmyk}{1,0,0,0.4}

\definecolor{darkred}{cmyk}{0,1,1,0.4}

\definecolor{darkgreen}{cmyk}{1,0,1,0.4}

\def\barr{\begin{array}}
\def\earr{\end{array}}
\def\dis{\displaystyle}

\def\gev{\, {\rm GeV}}

\def\lapp{\mathrel{\rlap{\raise.5ex\hbox{$<$}}
                    {\lower.5ex\hbox{$\sim$}}}}
\def\gapp{\mathrel{\rlap{\raise.5ex\hbox{$>$}}
                    {\lower.5ex\hbox{$\sim$}}}}

\begin{document}
\title{Gamma-ray and Synchrotron Radiation from Dark Matter annihilations in Ultra-faint Dwarf Galaxies}

\author[a]{Pooja Bhattacharjee}
\author[b]{Debajyoti Choudhury}
\author[c]{Kasinath Das}
\author[c]{Dilip Kumar Ghosh}
\author[d]{and Pratik Majumdar}

\affiliation[a]{Department of Physics, Bose Institute, EN-80, Sector V, Bidhannagar, Kolkata - 700091, India}

\affiliation[b]{Department of Physics and Astrophysics, University of Delhi, Delhi 110 007, India.}

\affiliation[c]{School of Physical Sciences, Indian Association for the Cultivation of Science, 2A \& 2B Raja S C Mullick Road, Kolkata 700 032, India.}

\affiliation[d]{Astroparticle Physics and Cosmology Division, HBNI,
Saha Institute of Nuclear Physics, 
1/AF Bidhan Nagar, Kolkata 700 064, India.}

\emailAdd{pooja.bhattacharjee06@gmail.com}
\emailAdd{debajyoti.choudhury@gmail.com}
\emailAdd{tpkd@iacs.res.in, kasinath.das91@gmail.com}
\emailAdd{tpdkg@iacs.res.in}
\emailAdd{majumdar.mpratik@gmail.com}

\date{\today}

\abstract{
The very large (100-1000) mass-to-light ratio applicable to the
ultra-faint dwarf galaxies (UFDs) implies a high concentration of dark
matter, thus rendering them ideal theatres for indirect signatures of
dark matter.  In this paper, we consider 14 recently discovered UFDs
and study the electromagnetic radiation emanating from them over a
wide range, from gamma ray down to radio frequencies.  We analyze the
Fermi-LAT data on high energy gamma rays and radio fluxes at the GMRT
and VLA to obtain upper limits on annihilation cross section
$\langle\sigma v\rangle$ in a model independent way. We further
discuss the sensitivity of the Square Kilometer Array radio telescope
in probing the synchrotron radiation from the aforementioned UFDs.  We
also investigate the dependences of the said upper limits on the
uncertainties in the determination of various astrophysical
parameters.}


\maketitle
\section{Introduction}
\label{sec:intro}
The presence of an abundant non-baryonic dark matter (DM) component in
our universe has not only been established through a variety of
independent observations\cite{Peter:2012rz}, but is also expected to
have played a key role in the evolution of structure at a multitude of
scales\cite{Primack:1997av}.  While a particulate nature of the DM is
called for, such an interpretation necessitates physics beyond the
Standard Model (SM) not only in terms of the identity of the said
particle, but also in terms of the dynamics that determines its relic
density. Several different classes of scenarios have been proposed
with the WIMP (Weakly Interacting Massive Particles) paradigm being
one of the most theoretically appealing ones. This is so because not
only are the parameters (masses and couplings) naturally at the
electroweak scale, but the very interaction that decides the relic
density of the DM particle $\chi$ (largely through processes like
$\chi\chi \to {\cal SS}$, where ${\cal S}$ is a generic SM field)
would also render amenable their detection in terrestrial
experiments \cite{Lisanti:2016jxe,Goodman:Witten,Schumann:2019eaa}.
While much effort has gone into theoretical studies that seek to
reproduce a relic density commensurate with what the cosmological data
suggests, unfortunately no confirmatory signals have been seen in any
laboratory experiments. These include not only dedicated {\em Direct
Search} experiments (based on $\chi {\cal S} \to \chi {\cal
S}$)~\cite{Angle:2008we, Aprile:2016swn,
Aguilar-Arevalo:2016ndq,Amole:2017dex,Cui:2017nnn,Akerib:2017kat,Agnes:2018ves},
but also those at collider facilities (exploiting, typically ${\cal
S}_1{\cal S}_2 \to \chi\chi {\cal S}_3$) , whether it be the
LHC~\cite{Kahlhoefer:2017dnp}, or low-energy facilities such as
Belle~\cite{Borodatchenkova:2005ct, Seong:2018gut,Choudhury:2019sxt}
or Da$\Phi$Ne~\cite{Borodatchenkova:2005ct}. Such non-observation
tends to militate against the requirements for obtaining the correct
relic-density, calling into question the entire paradigm. It should be
realised, though, that event rates in such terrestrial experiments are
governed by many assumptions about the nature of the interaction ({\em
e.g.}, spin-dependence) or the identity of ${\cal S}$ for the dominant
process. For example, if ${\cal S}$ were a third generation fermion,
the signal rates would be quite suppressed, without unduly affecting
cosmological evolution. Similarly, if $\chi$ were very heavy, not only
would production at colliders be suppressed, but the consequent
reduction in its number density (so as to maintain the requisite
energy density) would also suppress the direct detection rates.

This, then brings us to the realm of indirect searches, nominally
carried out through satellite-based and ground-based
experiments~\cite{Buckley:2013bha,Gaskins:2016cha,Fermi-LAT:2016uux,Colafrancesco:2015ola,Acharyya:2020sbj}.
The DM particles in open space can annihilate into a pair of SM
particles and impinge on space-based detectors either directly, or at
the end of a cascade of decays. Free from the aforementioned
suppressions that often plague the other avenues, indirect searches
proffer one of the most promising ways to not only detect the DM, but
also to determine its
properties. Signatures~\cite{Fermi-LAT:2016uux,Ahnen:2016qkx,TheFermi-LAT:2017vmf}
include anomalous photon, positron or antiproton production that could
be observed over and above known astrophysical sources.

In this paper, we concentrate on photon signals, which serve to
provide some of the most robust and stringent constraints on DM
annihilation into a variety of final
states~\cite{Bringmann:2012ez,Cirelli:2015gux}.
Unlike charged particles, photons are not deflected by the presence of
interstellar (or intergalactic) magnetic fields and thus we can trace
them back to their origin with a high degree of confidence.
Furthermore, the relatively small attenuation means that spectral
properties of the source remain intact.  As can be easily
appreciated, the simplest and most unmistakable signal would be
monochromatic photon emission resulting from either a two body final
state (where the second particle need not be a photon) or from
internal
bremsstrahlung~\cite{Bergstrom:2004cy,Bergstrom:2004nr,Bergstrom:2005ss,
Bringmann:2007nk,Bringmann:2011ye}. Of course, since the DM is
electrically neutral\footnote{While millicharged DM can be
accommodated in a specific class of models, we shall not consider
such.}, pair-annihilation into a two-body state with a photon can
proceed only as a loop-level process. A second possibility exists,
though, in the form of, say, $\chi \chi \to e^+ e^-$ with the positron
subsequently annihilating with a low-energy (and astrophysically
produced) electron to lead to a nearly monochromatic photon line. In
either case, the effective cross section is small.

Much larger cross sections are obtained for processes that lead to a
gamma-ray continuum (rather than a line
emission)~\cite{Colafrancesco:2005ji, Colafrancesco:2006he}. These
can, roughly, be divided into two classes, namely
\begin{itemize}
\item {\em prompt emissions} from dark matter annihilations
into any SM pair ${\cal S \bar S}$, with photon(s) being either
radiated off or emanating from the decay of a particle that has arisen
either directly from a primary decay product or as a result
of its hadronization (such as $\pi^0 \to \gamma\gamma$). The photon 
energy could range from $E_{\rm max}\, (\lapp M_\chi)$ all the way down.
\item {\em secondary emissions}  from the 
(quasi)stable, charged particles
produced in the primary annihilation event.
Inverse Compton scattering of relativistic particles with the cosmic microwave
background radiation (CMBR) as well as starlight can give photons with energies in the X-ray band, whereas in the presence of astrophysical magnetic fields,
synchrotron radiation (typically in the radio band) result.
\end{itemize}
Various astrophysical systems can be used to look for those
aforementioned signal of DM. For example, the central region of
ordinary Galaxies (like our own Milky Way or M31) are considered
interesting arenas for WIMP searches 
\cite{ANTARES:2019svn, Karwin:2016tsw, Rinchiuso:2017pcx, vitale2009indirect, Abe:2020sbr, Regis:2008ij, McDaniel:2018vam, Paul:2018njd, Vitale:2009hr}.
Signals from these galactic
central (GC) regions may also be used to understand the well known
{\it cusp/core} issue of the innermost part of the DM distribution
profile. However, photon signals of DM annihilation in the GC region
can be contaminated by galactic diffuse emissions as well as
electromagnetic radiations from various other nearby astrophysical
object such as supernova remnants, pulsars etc.  Owing to such
unresolved backgrounds, GCs are not ideal for effecting precision
studies~\cite{Petrovic:2014xra, Gaggero:2017jts, Gaggero:2015nsa,
Cholis:2015dea, Carlson:2015ona}.

N-body simulations indicate the existence of a large number of
DM-dominated sub-halos around typical galaxies~\cite{Kuhlen:2009jv,
Drlica-Wagner:2013bhh}. Some of these sub-halos might be massive
enough to host a dwarf galaxy \cite{Kuhlen:2009jv}, and these appear
as dwarf spheroidal galaxies (dSphs). Constituting the largest
galactic substructures around the Milky Way, these dSphs are expected
harbour the densest DM distributions in the galactic halo, with a
mass-to-light ratio lying in the range (100–1000) $M_{\circ}
/L_{\circ}$, where $M_{\circ}$ and $L_{\circ}$ are the solar mass and
the solar luminosity respectively. Their overwhelming DM content,
minimal Galactic foreground emission, and lack of astrophysical
radiation \cite{Mateo:1998wg, Grcevich:2009gt} render the dSphs
promising targets for the indirect detection of DM.

A decade after the Sloan Digital Sky
Survey (SDSS) revealed a population of ``ultrafaint'' dwarf galaxies
(UFDs) in the northern hemisphere (e.g., \cite{Willman:2004kk,
Zucker:2006he, Belokurov:2006ph}), a new generation of sky surveys has
begun charting the southern hemisphere. In
the past few years, nearly two dozen UFDs have been discovered using
data from Pan-STARRS (\cite{Laevens:2015kla, Laevens:2015una}), the
Dark Energy Survey (\cite{Bechtol:2015cbp, kim:2015abc, Kim:2015ila,
Koposov:2015cua}), and other surveys using the Dark Energy Camera at
Cerro Tololo (\cite{Kim:2015xoa, Kim:2016cgf, Kim:2015ghi,
Martin:2015xla}).

Their small halo mass and negligible baryonic
mass render UFDs extremely valuable laboratories
for exploring the nature of dark matter. The southern UFDs provide new
opportunities to address unanswered old questions about the nature and
origin of Galactic substructure, and, in particular, the galactic
luminosity function.  While the latter continues to be revised, its
faint end sets boundary conditions for galaxy formation within a given
cosmological model (\cite{Koposov:2009ru}). Due to their proximity, high dark-matter content, and the apparent
absence of non-thermal processes, the Milky Way UFDs are excellent targets for the indirect detection of
dark matter (\cite{Evans:2003sc,Bonnivard:2015xpq}).

We consider, here, electromagnetic radiation over a wide range, from
gamma down to radio frequencies, emanating from UFDs.
  Surrounded by
large-scale magnetic fields, the sheer size thereof confines the
$e^\pm $ produced by WIMP annihilation long enough for these to
radiate substantially. Since the dSphs targeted by the space-based
\textit{Fermi}
Large Area Telescope\footnote{\url{http://fermi.gsfc.nasa.gov}}
(Fermi-LAT) have no millisecond pulsars associated with
them \cite{Winter:2016wmy}, the astrophysical gamma-ray background is
essentially negligible.  Consequently, the non-observation of such
high energy gamma-rays or radio-emission from nearby dSphs may be used
to put strong constraints on the dark matter annihilation/decay rates.
To this end, we consider fourteen recently discovered UFDs and analyse
nearly eleven years of gamma-ray data collected by the Fermi-LAT
Collaboration. Amongst them, 13 UFDs are Milky Way satellites and one
(Eridanus II) is from the Local field (LF) (Local Field dwarf
spheroidal galaxies are not bound to Milky Way and M31).  We also
include the classical dSph Draco in our analysis which has been
extensively considered in the literature for indirect detection of
DM \cite{Colafrancesco:2006he, Kar:2019cqo, Cembranos:2019noa}.

The remaining of the paper is organized as follows: in
Section \ref{sec:gamma_flux}, we describe the different components of
the $\gamma$-ray flux and the astrophysical $J$-factor for the
different dSphs. This is followed, in Section \ref{sec:DM_profile}, by
a discussion of different possible dark matter profiles, and the
consequent $J$-factors.  In Section \ref{sec:analysis}, we analyze the
Fermi-LAT data for different dwarf spheroidal galaxies to obtain upper
limits on the annihilation cross sections for different channels. We
also examine the uncertainty accruing from the determination of
astrophysical parameters and the choice of the dark matter profile.
In Section \ref{sec:synchr}, we focus on the synchrotron radiation
from the dwarf spheroidal galaxies, especially in the context of the
existent radio telescopes such as from Giant Metrewave Radio Telescope
(GMRT) and Very Large Array (VLA) telescope and the projected
sensitivity of the Square Kilometer Array (SKA).  This is followed, in
Section \ref{sec:uncertainty}, by an examination of the role of
astrophysical uncertainties in the determination of the
constraints. And, finally we conclude, in
Section \ref{section:conclusion}, along with a discussion of the
relative strengths of various observations and possible future
developments.

\section{$\gamma$-ray flux from pair-annihilation of WIMPs}
\label{sec:gamma_flux}
A given scenario may include multiple new WIMP candidates (${\cal
W}_i$), of which perhaps only one could constitute the DM. Indirect
detection of the said WIMPs depends on two key processes : $(i)$ the
pair annihilation of the WIMPs through processes such as ${\cal W}_i
{\cal W}_j \to \sum_k {\cal S}_k $, or $(ii)$ the decay\footnote{In
the second case, clearly ${\cal W}_i$ is not protected by a discrete
symmetry. It could have arisen, for example, from the
pair-annihilation of a $Z_2$-odd DM, namely $\chi\chi \to {\cal W}_i
{\cal W}_j$. It should be realised though that a decaying DM is also
admissible as long as the decay is very slow, as, for example, happens
in gravity-mediated decays~\cite{Arun:2017zap}.} of a WIMP into SM
particles, namely ${\cal W}_i \to \sum_k {\cal S}_k $. In each case, a
bunch of energetic standard model particles result in the final state,
either directly from the aforementioned processes or as a result of
the subsequent decay cascades of the ${\cal S}_k$.  However, it should
be noted that neither of the two processes are a must in a given
particle dark matter scenario.  Furthermore, they bear relevance
(apart from in the cosmological context) only if they occur at a rate
that can be detected on various experimental facilities based on the
earth or on satellites. This relevance is, of course, determined
jointly by their rates and the detectability (in terms of the final
particle identities and the energies they carry), in such facilities,
earth-bound or satellite-based \cite{Abdallah:2020sas, Hoof:2018hyn,
Alvarez:2020cmw, HAWC:2019jvm, Halder:2019pro, Oakes:2019ywx,
Rinchiuso:2019etv, Rico:2020vlg, Bhattacharjee:2018xem}. These, in
turn, are determined by the details of the underlying particle physics
model, both in terms of the spectrum as well the sizes, or even the
existence, of the couplings.  In addition to these, several
astrophysical input parameters also play a major role in the final
observation (both in the annihilation rate as well as in the
propagation of various standard model particles through the
space). Some of these astrophysical parameters, understandably,
contain large uncertainties rendering any conclusive prediction of the
indirect signature of the dark matter very
challenging\cite{Conrad:2014nna}.

In this analysis, we focus on the annihilation of Majorana dark matter
in an almost model independent manner, with the simplifying assumption
that the pair-annihilation proceeds into a single channel alone, with
a $100\%$ probability. The consequent individual
sensitivities/constraints can, then, be convoluted, in a relatively
straightforward manner, to that applicable for a given realistic
model. The DM annihilation rate per unit volume is given by
$\langle \sigma v\rangle \rho^2_{\rm \chi}/2 M^2_{\rm \chi}$, where
$\rho_{\rm \chi}$ is the density, $\sigma$ the cross section and $v$
the velocity of the second DM particle in the rest frame of the
first\footnote{For a DM annihilation rate of $\langle\sigma
v\rangle \sim 10^{-26}~{\rm cm^3/s}$---one that is consistent with the
correct relic density for $M_{\rm \chi} \sim {\cal O}(10^{2-3})\, {\rm
GeV}$---an observable flux of gamma rays is obtained.}.  The thermal
average $\langle\sigma v\rangle$ is estimated using the knowledge of
particle physics and is model-specific. On the other hand, assuming
that the DM density has a spherical symmetry (a very good
approximation), the radial dependence of $\rho_{\rm \chi}$ is modelled
based on astrophysical observations, as we shall discuss later.  For a
specific energy $ E (\equiv E_\gamma)$, the differential $\gamma$-ray
flux $\phi_{\rm{WIMP}} (E, \Delta \Omega)$ (in units of photons
cm$^{-2}$s$^{-1}$GeV$^{-1}$) lying within a solid angle
$\Delta \Omega$ and having arisen from the annihilations of WIMPs of
mass $M_{\rm \chi}$, can be expressed~\cite{Abdo:2010ex} as a product
of two terms, one each accounting for the particle-physics and
astrophysics factors, namely
\begin{equation}
\phi_{\rm{WIMP}}(E, \Delta \Omega)~ = ~ \Phi^{pp}(E) \times J(\Delta \Omega) \ .
\label{eqn:dm_flux}
\end{equation}
The particle physics factor can be written as \cite{Abdo:2010ex}:
\begin{equation}
\Phi^{pp}(E)~ = ~\frac{\langle\sigma v\rangle}{8 \pi ~M^{2}_{\rm{\chi}}} \sum_{f} 
\frac{dN_{\gamma,f}}{dE}B_{f} \ ,
\label{eqn:dm_pp}
\end{equation}
where $dN_{\gamma, f}/dE$ is the differential photon spectrum (per
annihilation event) for a given final state '$f$', and $B_{f}$ is the
corresponding branching fraction. Several numerical packages like
Pythia \cite{Sjostrand:2007gs}, DarkSUSY (\cite{Gondolo:2004sc}),
DMFit \cite{Jeltema:2008hf} etc. are designed to estimate differential
photons yields from each annihilation channel. While the selection of
standard model final states, through which annihilation would occur
(e.g. $b\bar{b}$, $\tau^{+}\tau^{-}$, $\mu^{+}\mu^{-}$ etc.), is
theoretically motivated, as stated above, we remain agnostic and
consider only a single channel dominance.
Thus, unless otherwise mentioned, in the rest of the analysis, 
only a single final state $(b \bar b, \tau^+\tau^-, 
\mu^+\mu^-)$ will be considered to have been produced ({\em i.e.}, with $100\%$ probability) 
in a DM annihilation process, and consequent limits obtained on the
cross sections.

\subsection{Astrophysical Factor (J)}

Since the total flux is proportional to the factor $J$, it behaves us
to examine it closely. As we are primarily concerned with pair
annihilations, it should depend on $\rho^{2}_{\rm \chi}$.  While the
galactic center, where $\rho_{\rm \chi}$ is the largest, is associated
with the highest flux, it is also associated with an intense
astrophysical background. In contrast, the UFDs present features that
make them ideal sources. The typical values of the
$J$-factor \cite{Funk:2013gxa} for the GC are $J\approx
10^{22}-10^{23} {\rm GeV}^2 {\rm cm}^{-5} $, while $J\approx
10^{16}-10^{19} {\rm GeV}^2 {\rm cm}^{-5} $ for UFDs and $ J\approx
10^{15}-10^{19} {\rm GeV}^2 {\rm cm}^{-5} $ for galaxy clusters.

While the aforementioned quadratic dependence is indicative, a true
measure of the effective $J$ involves the line-of-sight (l.o.s)
integration of the same, namely~\cite{Abdo:2010ex}
\begin{equation}
J (\Delta \Omega) = \int \int \rho^{2}_{\rm \chi} (r(\lambda)) d\lambda ~ d\Omega 
= 2 \pi \int_{\theta_{\rm{min}}}^{\theta_{\rm{max}}} d\theta  \,
     \rm{sin} \theta \int_{\lambda_{\rm{min}}}^{\lambda_{\rm{max}}} d\lambda \; \rho^{2}_{\rm \chi} \left(r(\lambda)\right) \ ,
\label{eqn:Jfactor_analytical}
\end{equation}
where $\lambda$ is the l.o.s
distance and $\theta$ is the angle between the l.o.s and the center of
the UFD. The galactocentric distance $r(\lambda)$ is obtained in terms
of the UFD's distance $d$ from the Sun through
\begin{equation}
r(\lambda) = \sqrt{\lambda^{2} + d^{2} - 2~ \lambda ~d~ \rm{cos \theta}} \ .
\label{eqn:r_lambda}
\end{equation}

The DM density profile in UFDs remain a topic of debate.  Two broad
classes are popularly used to fit the observational data, namely those
with a cusp-like profile\cite{Navarro:1996gj} or a cored
profile\cite{Burkert:1995yz, Salucci:2011ee, Gunn:1972sv}
respectively.  While the lack of sufficient kinematical observational
prevents the selection of a particular profile type, N-body
cosmological simulations, with the most recent results, favors the
cuspy Navarro-Frenk-White (NFW) form~\cite{Navarro:1996gj}, specially
for dSphs and UFD galaxies. This is parameterized
as~\cite{Navarro:1996gj}
\begin{equation}\label{eqn:density_NFW}
\rho_{\chi} (r)=\frac{\rho_{s}r_{s}^{3}}{r(r_{s} + r)^{2}} \ ,
\end{equation}
where $\rho_{s}$ and $r_{s}$ are the characteristic density and scale
radius respectively. These, as well as another one, namely, $r_h$ (that we would
find useful when we discuss synchrotron radiation in Sec.\ref{sec:synchr})
can be obtained using $d$, $\theta_{max}^0$, $\sigma_{l.o.s}$ and $r_{1/2}$. 
Here $d$, $r_{1/2}$ and $\sigma_{l.o.s}$ denote the heliocentric distance, the half-light radius and the velocity dispersion of each UFD galaxy. 
$\theta_{max}^o$ is the angle made by the outer most star of the UFD. While
the details can be found in ref.\cite{Evans:2016xwx}, we discuss the relations
briefly.

To begin with, consider the mass $M_{1/2}$ contained within the
half-light radius ($r_{1/2}$) of the UFD and approximately
expressed in terms of $r_{1/2}$ and the UFD velocity dispersion $\sigma_{l.o.s}$ as
\begin{equation}\label{eqn:mhalf}
 M_{1/2} = M(r_{1/2}) \approx \frac{2.5}{G_N} \sigma_{l.o.s}^2 r_{1/2}.
\end{equation}
where $G_N$ is Newton's constant. The NFW parameter $r_s$ can
be approximated in terms of $r_{1/2}$, {\em
viz.}
\begin{equation}\label{eqn:rs}
 r_s = 5 \, r_{1/2} \ ,
\end{equation}
whereas $\rho_s$ is given by
\begin{equation}\label{eqn:rhos}
 \rho_s = \frac{M_{1/2}}{4\pi r_s^3}
 \left[\log\left(\frac{r_s + r_{1/2}}{r_s}\right)
       -\, \frac{r_{1/2}}{r_s + r_{1/2}}\right]^{-1} \ .
\end{equation}
The distance to the outermost star of the UFD from the center of UFD is,
of course,
\begin{equation}\label{eqn:rmax}
 r_{\rm max} = d\, \, \sin \theta^\circ_{\rm max}.
\end{equation}
The definition of the diffusion radius ($r_h$) is somewhat ambiguous, and 
depends on the spatial extent of both the gas and the magnetic field,
quantities that are poorly understood for dSphs. One could draw
inspiration from the Milky Way where the height of the diffusion
cylinder is a few times the size of the stellar disc
width. Given that this is the typical extent of
the interstellar gas for galaxies, we assume
\cite{Jeltema:2008ax, Colafrancesco:2006he} that
\begin{equation}\label{eqn:rh}
 r_h \approx 2 \, r_{\rm max} = 2\, d\, \, \sin \theta^\circ_{\rm max}.
\end{equation}
We have checked that varying $r_h$ by a factor 0.5--2 does
not alter the results to a great extent.

Using eqns.(\ref{eqn:rhos}--\ref{eqn:rh}) with the values of the
parameters $d$, $\sigma_{l.o.s}$, and $r_{1/2}$ given in
Table~\ref{table:astro_fundamental_param_dwarfs} of the
Appendix, we calculate the parameters $\rho_s$, $r_s$ and $r_h$ and
list them in Table~\ref{table:astro_param_dwarfs}. The parameters in
Table~\ref{table:astro_param_dwarfs} correspond to the central values
of the parameters in Table \ref{table:astro_fundamental_param_dwarfs}
of the appendix.

\begin{table}[!ht]
\begin{center}
 \begin{tabular}{c c c c c}
  dSphs        & d(Kpc) & $r_h(Kpc)$ & $\rho_s (GeV/cm^3)$ & $r_s (Kpc)$ \\ \hline
  Aquarius II  & 107.9& 0.42  & 2.27  & 0.615  \\
  Carina II    & 37.4 & 0.3 & 1.78 & 0.38 \\
  Draco II     & 20 & 0.07 & 71.73  & 0.06 \\
  Eridanus II  & 366 & 0.792 & 1.454 & 0.88 \\ 
  Grus I       & 120.2 & 0.39 & 6.7 & 0.26 \\
  Horologium I & 79 & 0.188 & 30.55 & 0.16 \\
  Hydra II      & 151 & 0.448 & < 8.24 & 0.335 \\
  Leo V        & 173 & 0.465 & 23.83 & 0.15 \\
  Pegasus III  & 215 & 0.228 & 40.73  & 0.185  \\
  Pisces II    & 183 & 0.438 & 8.93 & 0.24 \\
  Reticulum II & 30 & 0.251 & 10.08  & 0.16 \\
  Tucana II    & 57.5 & 0.452 & 3.6  & 0.575 \\
  Tucana III    &  25 & 0.174 & < 2.29 &  0.215 \\  
  Triangulum II & 30 & 0.157 & < 46.1 &  0.14  \\ \hline \hline
  Draco & 80 & 2.5   & 1.4    & 1.0 \\  \hline
 \end{tabular}
\end{center} 

\caption{\em The values of the astrophysical parameters for the 13 newly discovered UFDs \cite{Pace:2018tin} and the classical dSph Draco.
The parameters for the 13 UFDs in this table correspond to the central values of the parameters in Table \ref{table:astro_fundamental_param_dwarfs}.
The parameters for Draco has been taken from \cite{McDaniel:2017ppt}.}
\label{table:astro_param_dwarfs}
 \end{table}

Given the measurements of $\rho_s$ and $r_s$, it is a straightforward
task to determine $J$. In Table~\ref{table:table-1}, we present this
for a set of UFDs, adopting, in each case, the standard choices of
$\theta_{\rm min} = 0$ and $\theta_{\rm max}=
0.5^\circ$. 
While the point spread function (PSF) of the
Fermi-LAT varies with the energy of the incident photons, for the
energy range of our interest in the Fermi-LAT analysis, the average PSF
value of the LAT detector roughly equals
$0.5^\circ$ \cite{Hoof:2018hyn, Fermi-LAT:2016uux, Ackermann:2015zua,
Ackermann:2013yva}.

The calculation of $J$ for a dSph is beset with several uncertainties.
The relative paucity of identifiable stars in such galaxies renders
spectroscopic studies difficult. The small sample size, for one,
results in significant statistical errors. Moreover, the lack of
sufficiently large stellar kinematic data from the spectroscopic
observations results in large systematic uncertainties in the
determination of the velocity dispersions and, therefore, the mass~\cite{Bonnivard:2015xpq, Pace:2018tin}.
To handle this, we use an algorithm developed by some of us~\cite{Bhattacharjee:2019jce}, using Monte Carlo methods and distributions of the variables $d$,
$\sigma_{\rm l.o.s}$ and $r_{1/2}$. The asymmetrical error bars (including systematics) are modelled by skew normal distributions with 
unequal values for the standard deviation on each side of the
mean. The resultant uncertainties in the $J$-factor are displayed in Table 
\ref{table:table-1}.

For comparison, we also list the corresponding
values derived by Pace {\em et al.}~\cite{Pace:2018tin}, using an
empirical scaling relation motivated by the analytical work of Evans
{\em et al.}~\cite{Evans:2016xwx}. For the NFW profile, the empirical
relation reads
\begin{equation}\label{eqn:jfactor_pace}
\frac{J(0.5^{\circ})}{\GeV^{2}{\rm cm}^{-5}} \approx 10^{17.72}
\left(\frac{\sigma_{\rm l.o.s}}{5 \, {\rm km\,s}^{-1}}\right)^{4}
\left(\frac{d}{100 \, {\rm kpc}}\right)^{-2}\left(\frac{r_{1/2}}{100 \, {\rm pc}}\right)^{-1} \ ,
\end{equation}
It is worth pointing out the remarkably good agreement between the
exact numerical result and the empirical formula. Indeed, at the level
of accuracy of our results, the two are virtually
indistinguishable. Nonetheless, we use the exact results (as
derived using \ref{eqn:Jfactor_analytical} along with the data listed
in the Appendices) for our subsequent numerical calculations. It
should be borne in mind, though, that a different choice of the
density profile would lead to a substantially different value for $J$,
a point that we would return to at a later section. It should also be
noticed that, for certain UFDs, only upper bounds on the $J$-factor
exists. This can be traced to the insufficiency of the kinematic data,
which leads to only an upper bound on the velocity dispersion and,
hence, on the $J$-factor \cite{Bhattacharjee:2018xem}.

\begin{table}[!t]
\centering
\begin{tabular}{|p{2.5cm}|c|c|c|c|}
\hline \hline
Galaxy & \multicolumn{4}{c|}{$\log_{10}(J/{\rm GeV}^2\, {\rm cm}^{-5})$}
\\
\cline{2-5}
& Pace {\em et al}~\cite{Pace:2018tin} & \multicolumn{3}{c|}{Direct Integration}\\
\cline{3-5}
& (NFW) & NFW & Burkert & ISO \\
\hline \hline
Aquarius II & $18.27^{+0.66}_{-0.58}$ & $18.11^{+0.68}_{-0.63}$   & $18.53^{+0.72}_{-0.66}$ & $18.01^{+0.73}_{-0.66}$ \\     
\hline \hline                                                                                                                
Carina II & $18.25^{+0.55}_{-0.54}$ & $18.16^{+0.55}_{-0.53}$     & $18.45^{+0.60}_{-0.56}$ & $18.05^{+0.58}_{-0.54}$ \\     
\hline \hline                                                                                                                
Draco II & $18.93^{+1.39}_{-1.70}$ & $19.07^{+1.33}_{-1.69}$      & $19.54^{+1.35}_{-1.70}$ & $18.90^{+1.34}_{-1.70}$ \\     
\hline \hline                                                                                                                
Eridanus II & $17.28^{+0.34}_{-0.31}$ & $17.14^{+0.35}_{-0.30}$   & $17.68^{+0.35}_{-0.31}$ & $17.06^{+0.35}_{-0.31}$ \\     
\hline \hline                                                                                                                
Grus I & $16.88^{+1.51}_{-1.66}$ & $16.94^{+1.57}_{-1.74}$        & $17.48^{+1.60}_{-1.75}$ & $16.76^{+1.54}_{-1.67}$ \\     
\hline \hline                                                                                                                
Horologium I & $19.27^{+0.77}_{-0.71}$ & $19.01^{+0.83}_{-0.73}$  & $19.37^{+0.85}_{-0.75}$ & $18.73^{+0.85}_{-0.75}$ \\     
\hline \hline                                                                                                                
Hydra II & $<~17.77$ & $<~17.92$                            &  $<~18.46$   &                    $<~17.84$ \\                                                
\hline \hline                                                                                                                
Leo V & $17.65^{+0.91}_{-1.03}$ & $17.91^{+1.03}_{-1.06}$         & $18.51^{+1.02}_{-1.08}$ & $17.84^{+1.01}_{-1.07}$ \\     
\hline \hline                                                                                                                
Pegasus III & $18.30^{+0.89}_{-0.97}$ & $18.46^{+0.94}_{-1.05}$   & $19.06^{+1.02}_{-1.07}$ & $18.39^{+1.03}_{-1.05}$ \\     
\hline \hline                                                                                                                
Pisces II & $17.30^{+1.00}_{-1.09}$ & $17.53^{+1.02}_{-1.09}$    & $18.10^{+1.04}_{-1.09}$ & $17.45^{+1.03}_{-1.09}$ \\     
\hline \hline                                                                                                                
Reticulum II & $18.96^{+0.38}_{-0.37}$ & $18.76^{+0.43}_{-0.42}$ & $19.21^{+0.43}_{-0.42}$ & $18.66^{+0.43}_{-0.42}$ \\     
\hline \hline                                                                                                                
Triangulum II & $<~19.73$ & $<~19.74$                      &$<~20.18$ & $<~19.64$ \\                                                
\hline \hline                                                                                                                
Tucana II & $19.02^{+0.58}_{-0.53}$ & $18.93^{+0.62}_{-0.58}$   & $19.22^{+0.64}_{-0.61}$ & $18.83^{+0.66}_{-0.62}$ \\     
\hline \hline                                                                                                                
Tucana III & $<~17.71$ & $<~17.87$                        & $<~18.20$   & $<~17.76$ \\                                                
\hline \hline                                                                                                                
Draco & $18.83^{+0.12}_{-0.12}$ & $18.85^{+0.12}_{-0.12}$       & $19.08^{+0.13}_{-0.13}$ & $18.75^{+0.13}_{-0.13}$ \\     
\hline \hline
\end{tabular}

\caption{\em The $J$-factors for the various UFDs as obtained by directly integrating
eqn.~\ref{eqn:Jfactor_analytical} for three density profiles and for $\theta_{max}=0.5^{\circ}$.
Also shown, for comparison,
are the values obtained by Pace {\em et al.}~\cite{Pace:2018tin}, for the NFW profile, using 
an approximate scaling relation.} 
\label{table:table-1}
\end{table}


\subsection{Dependence of $J$ on the density profiles}
\label{sec:DM_profile}

While the NFW is a traditional benchmark choice for the DM profile
motivated by the $N$-body simulations, its cusp-like nature at the
center of the galaxy is quite different from the alternate cored
profiles. While we would use the NFW for most of our numerical
results, it is an imperative to ensure that the cuspy nature does not
lead to extreme answers, To this end, we examine two examples of the
second category, namely the pseudo-isothermal(ISO)~\cite{Gunn:1972sv}
profile and the one originally proposed by
Burkert~\cite{Burkert:1995yz, Salucci:2011ee}. These are given by 
\begin{equation}
\rho_{\rm ISO}(r)=\frac{\rho_{c} \, r_c^2}{r^{2}+ r_{c}^{2}}  \ , \qquad
\rho_{\rm Burkert}(r)=\frac{\rho_{B}r_{B}^{3}}{(r_{B}+r)(r_{B}^{2} + r^{2})} 
\end{equation}
where $r_B$ and $r_{B}$ represent the respective core radii, whereas
$\rho_{c}$ and $\rho_B$ are the corresponding densities at the very
center.  While $ \rho_{\rm Burkert}(r) $ resembles an isothermal profile
in the inner regions $(r \ll r_B)$, for large $r$ it falls off much
faster ($r^{-3}$ versus $r^{-2}$).  As in the case for the NFW
profile, the parameters are to be determined from observational data
related to a given galaxy.
While the density profiles are undoubtedly different, having been
deduced from the same data, they ought to generate virtually identical
galactic rotation curves. Demanding this,  ref.\cite{Boyarsky:2009rb}
deduced approximate relations between the parameters
for the three profiles, {\em viz.},
$r_s \simeq 6.1 r_c \simeq 1.6 r_B$ and $\rho_s \simeq
0.11 \rho_c \simeq 0.37 \rho_B$. With such choices, the
consequent rotation curves were found to be virtually
indistinguishable for $r \lapp 2 r_s \simeq 12 r_c$, {\em i.e.},
almost the entire observed range.

While $N$-body simulations~\cite{Navarro:2008kc, Wagner:2020opz} tend
to favour a cuspy profile (such as the NFW) over the smooth (and
relatively flat at the center) profiles such as the Burkert or the
ISO, insufficient kinematic data (such as those on rotational curves)
prevent us from strongly favoring any one profile. Indeed, some
observations \cite{de_Blok_2001}
actually favour cored halos over cuspy ones. Fortunately
for us, the $J$-factor does not have too strong a dependence on the
choice. Indeed, as Table~\ref{table:table-1} demonstrates, the NFW
profile leads to values that are not too far from the average of those
obtained from the three profiles. In other words, the theoretically
favoured choice is also a good representative.





\section{Analysis of $\gamma$-ray fluxes from ultra-faint dwarf galaxies}
\label{sec:analysis}
Several features of gamma rays make them an ideal medium for indirect
DM detection. Unlike the charged particles, these do not get deflected
by the presence of strong magnetic fields and thus we can trace back
the origin of the emission. In addition, gamma rays suffer very little
attenuation and hence the spectral properties of the source remain
intact.  In studying $\gamma$-rays originating from pair annihilations
of DM into SM particles within UFDs, we consider the data from the
Large Area Telescope (LAT) on board the Fermi observatory.  The
Fermi-LAT is a $\gamma$-ray pair conversion space-based detector that
scans the whole sky every 3 hours from a low-Earth orbit of 565 km
altitude at a 25.6-degree inclination with an eccentricity
$<$0.01. Launched on June 11, 2008, by the Delta II Heavy launch
vehicle from Cape Canaveral, the principal objective of the Fermi-LAT
was to conduct a long term high sensitivity $\gamma$-ray observations
of celestial sources in the energy range from $\approx$ 20 MeV to $>$
500 GeV \cite{Atwood:2009ez} and we analyze over ten years 
(2008-09-01 to 2019-02-04) worth of this sky survey data.  While several
dwarf spheroidals have been observed over the past few years, no
conclusive signal has been found from any of them.  Given that such
negative results too can provide important information about the dark
matter\cite {Ackermann:2011wa,GeringerSameth:2011iw,Ackermann:2013yva,
Ackermann:2015zua, Fermi-LAT:2016uux}, we study a set of 14 UFDs which
have been recently discovered in several surveys (see Appendix A for a
list). Using the latest version of Fermi ScienceTools (v1.2.1) for our
analysis, we process the data with an improved PASS 8 instrument
response function (IRF), in particular the source class IRF,
$\rm{P8R3\_SOURCE\_V2}$. Furthermore, the tool \textit{`gtmktime'} has
been used to extract the ``good time interval'' data from the whole
data set. Extracting the LAT data within a $15^{\circ}$ radius of
interest (ROI) around each source, we consider only a limited range
for the reconstructed energy $E$, {\em viz.} $E\in [0.1, 300]$~GeV, so as
to reduce possible uncertainties at low energies on the one hand and
background contamination at high energies on the other. To remove the
possible contamination from the earth albedo, we apply a zenith-angle
cut at $90^{\circ}$ as recommended by the Fermi-LAT analysis team.
With the Earth’s limb lying at a zenith angle of $113^{\circ}$, the
application of a zenith cut at $90^{\circ}$ eliminates a large
fraction of the background atmospheric gamma-rays.

The binned likelihood analysis for the extracted data set was performed
with the `gtlike' tool \cite{Cash:1979vz, Mattox:1996zz}. To this end,
we first generate a source model file with the inclusion of all the
sources from the 4FGL catalog \cite{Fermi-LAT:2019yla} within a
20$^{\circ}$ ROI from the position of the `source of interest'. Note
that we extend the spatial coverage up to $20^{\circ}$ ROI to
account for possible overlapping between the point spread functions of
nearby sources. In addition, to
eliminate possible background effects resulting from galactic and
extra galactic diffuse emission, we add the Galactic diffuse
model ($\rm{gll\_iem\_v07.fits}$) and the isotropic extra galactic
diffuse model ($\rm{iso\_P8R3\_SOURCE\_V2\_v1.txt}$) to the source
model. The spectral parameters of all the 4FGL sources
\cite{Fermi-LAT:2019yla} within ROI, as well as the normalization
parameters of two diffuse models have been left free in the fitting
procedure. The rest of the background sources within the
$20^{\circ}~\times~20^{\circ}$ ROI have been kept fixed at their
values given in the 4FGL catalog \cite{Fermi-LAT:2019yla}.
Table~\ref{table:fermi_lat_parameters}, in the Appendix, lists all the
parameters used at different stages of the analysis of the data.

\subsection{Constraints on DM annihilation with eleven years of Fermi-LAT data}\label{section:gamaray_sigmav_constraint}

To search for $\gamma$-ray emissions coincident with our targets,
we first model our sources with a power-law spectral model
($dN/dE \propto E^{-\Gamma}$) with spectral index $\Gamma$ =
2 \cite{Bhattacharjee:2018xem, Ackermann:2013yva, Ackermann:2015zua,
Fermi-LAT:2016uux}. As a statistical discriminator, we use the ratio
of the maximum likelihoods for two hypotheses, namely, $TS =
-2\ln\Big(L_{\rm {(max, 0)}}/L_{\rm {(max, 1)}}\Big)$, where $L_{\rm
{(max, 1)}}$ and $L_{\rm {(max, 0)}}$ respectively denote the maximum
likelihood in the presence of a signal and the null hypothesis.
In other words, a value $TS > 25$ would correspond to a ``$5 \sigma$
discovery''.

\begin{figure}[h!]
\begin{minipage}[c]{0.3\textwidth}
{\small
\begin{center}
    \begin{tabular}{|l|r|r|}
          \hline
  UFD & $TS_{\rm peak}(b)$ & $TS_{\rm peak}(\tau)$ \\[-0.3ex]
  \hline
  Aquarius II  & 2.88 & 2.94 \\[-0.3ex]
  Carina II    & 1.24 & 1.81 \\[-0.3ex]
  Draco II     & 1.37 & 1.88 \\[-0.3ex]
  Eridanus II  & 0.81 & 1.23 \\[-0.3ex] 
  Grus I       & 1.59 & 1.65 \\[-0.3ex]
  Horologium I & 4.21 & 4.71 \\[-0.3ex]
  Hydra II     & 2.21 & 2.31 \\[-0.3ex]
  Leo V        & 0.88 & 0.92 \\[-0.3ex]
  Pegasus III  & 1.91 & 2.13 \\[-0.3ex]
  Pisces II    & 1.22 & 1.96\\[-0.3ex]
  Reticulum II & 4.85 & 4.95 \\[-0.3ex]
  Tucana II    & 11.87 & 12.47 \\[-0.3ex]
  Tucana III    &  4.36 & 4.53 \\[-0.3ex]  
  Triangulum II & 1.19 & 1.25 \\[-0.3ex] 
  \hline
  \end{tabular}
\end{center} 
}

\end{minipage}
\hfill
\begin{minipage}{0.5\textwidth}
\includegraphics[width=\textwidth,clip,angle=0]{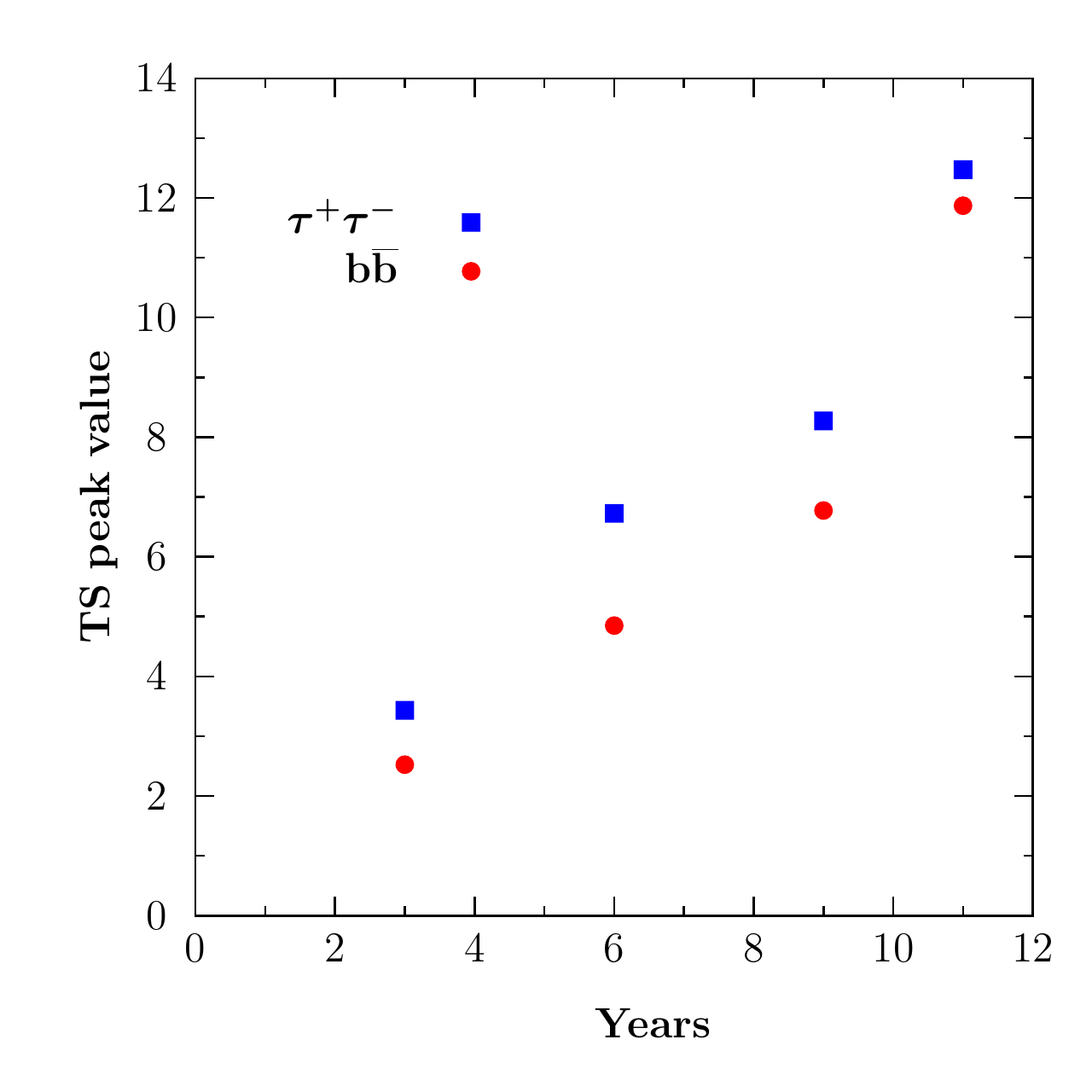}
\end{minipage}
\caption{\em The TS peak value for two annihilation final states as
evinced by the Fermi-LAT data (left panel). The change in the TS value for 
Tucana II, as a function of telescope operation time, for excess $\gamma$-ray 
emission from the direction of Tucana II (right panel).}
\label{figure:ts_tucana}
\end{figure}

Unfortunately, as Fig.\ref{figure:ts_tucana}$(a)$ attests, except for
Tucana II, the other target UFDs have not yet shown any faint emission
from their location (i.e. TS $\preceq$ 5). And while an intriguing
hint of a faint gamma-ray signal has been
reported~\cite{Bhattacharjee:2018xem}, from the direction of
Tucana II, the excess is still too small to claim a real discovery.
Quite hearteningly, though, the significance of the putative excess
has only grown with more data, as is evinced by
Fig.\ref{figure:ts_tucana}$(b)$, thereby raising hope that a signal
may yet be established.

Meanwhile, in the absence of a discernible signal from an UFD can be
used to impose 95$\%$ C.L. upper limits on the $\gamma$-ray flux from
the site. In this, we use the Bayesian
approach~\cite{Helene:1990yi}---already implemented in the {\tt
pyLikelihood} module of \texttt{ScienceTools}---as this is more
sensitive~\cite{Rolke:2004mj, Barbieri:1982eh} than the profile
likelihood method for low statistics.

\begin{figure}[h]
  \centering
  \vskip  -0.5in
  \includegraphics[width=0.49\textwidth,height=0.44\textwidth]{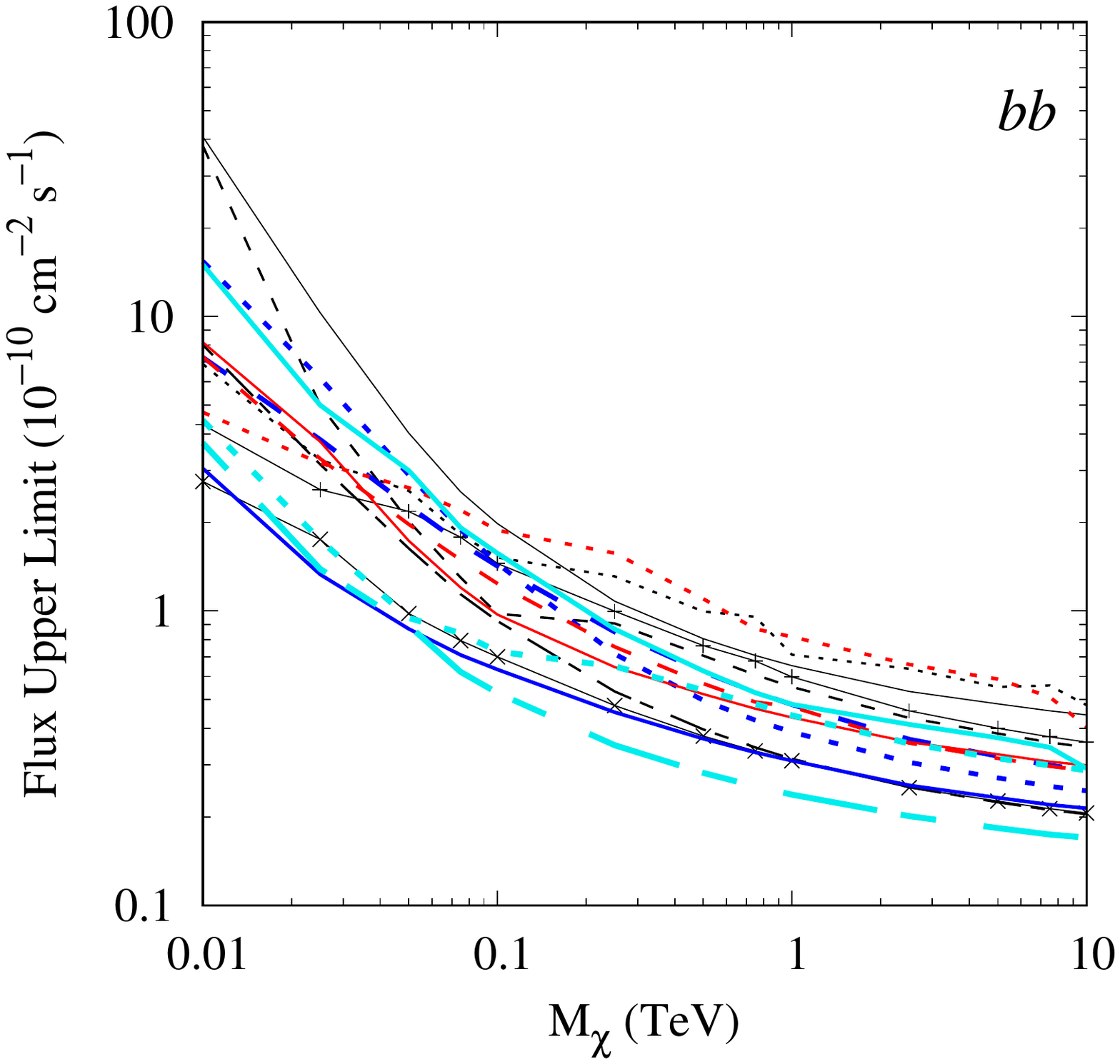}
  \includegraphics[width=0.49\textwidth,height=0.44\textwidth]{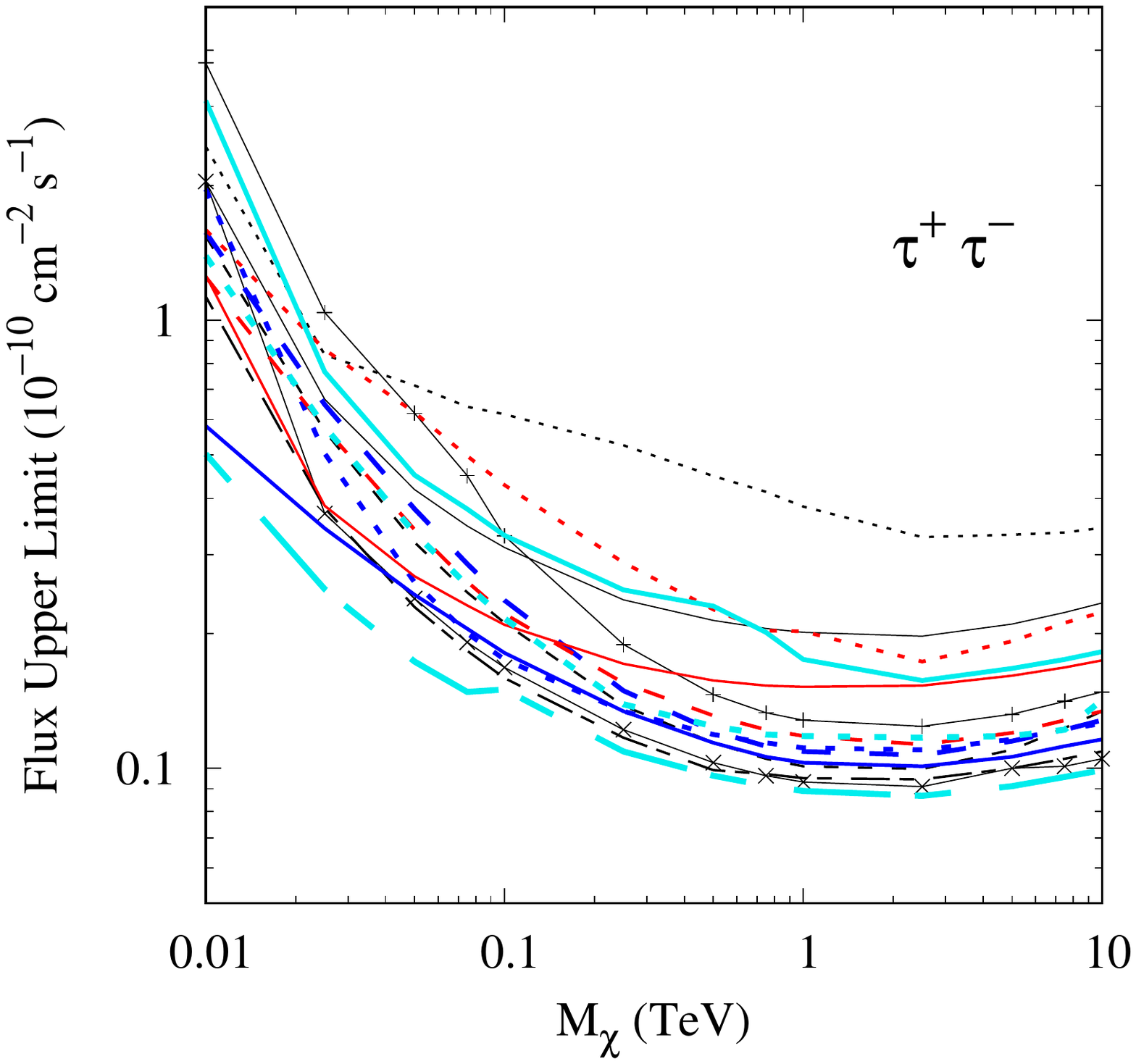} \\
  \vskip -1.4in
  \includegraphics[width=0.49\textwidth,height=0.44\textwidth]{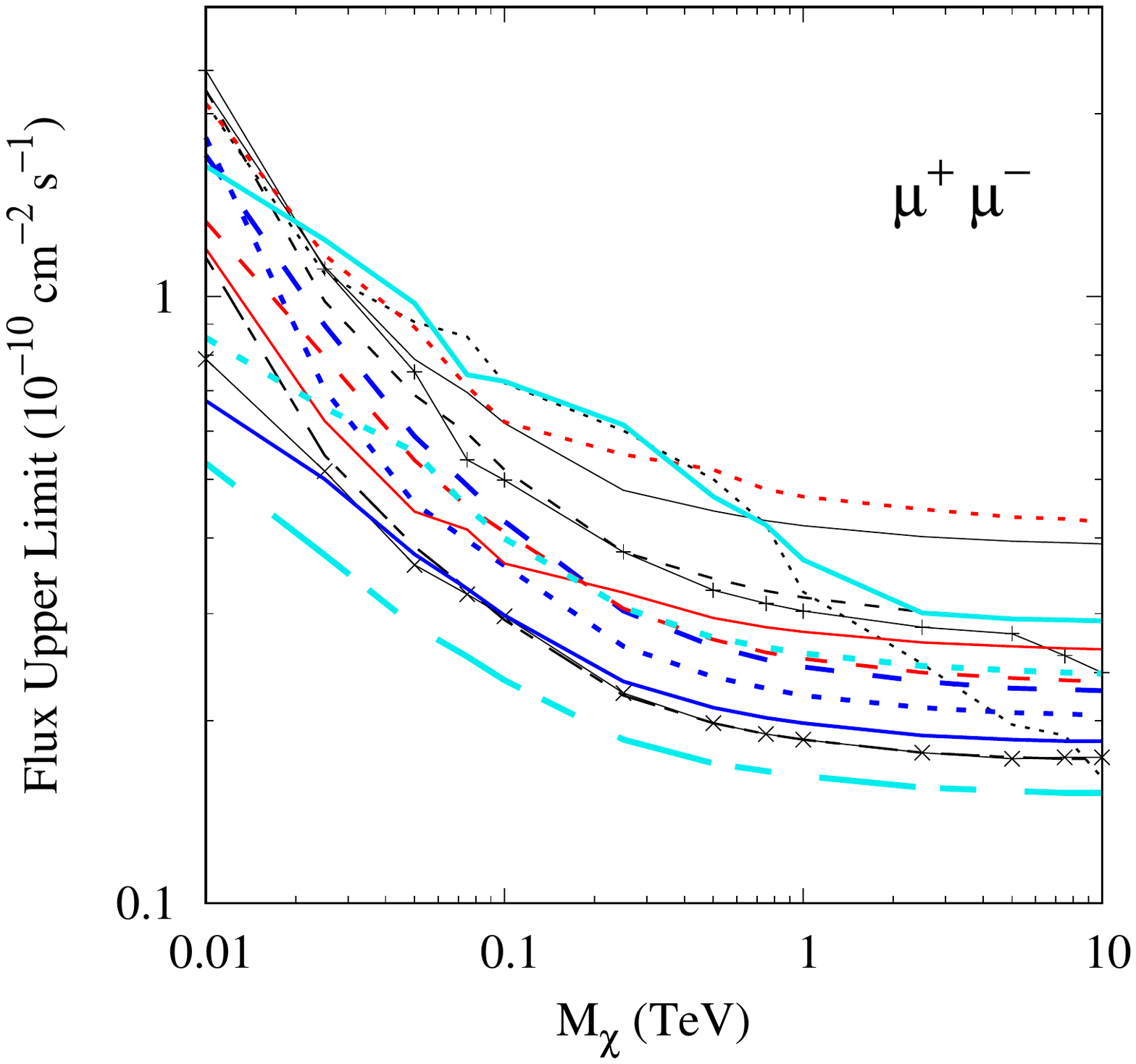}
  \includegraphics[width=0.49\textwidth,height=0.44\textwidth]{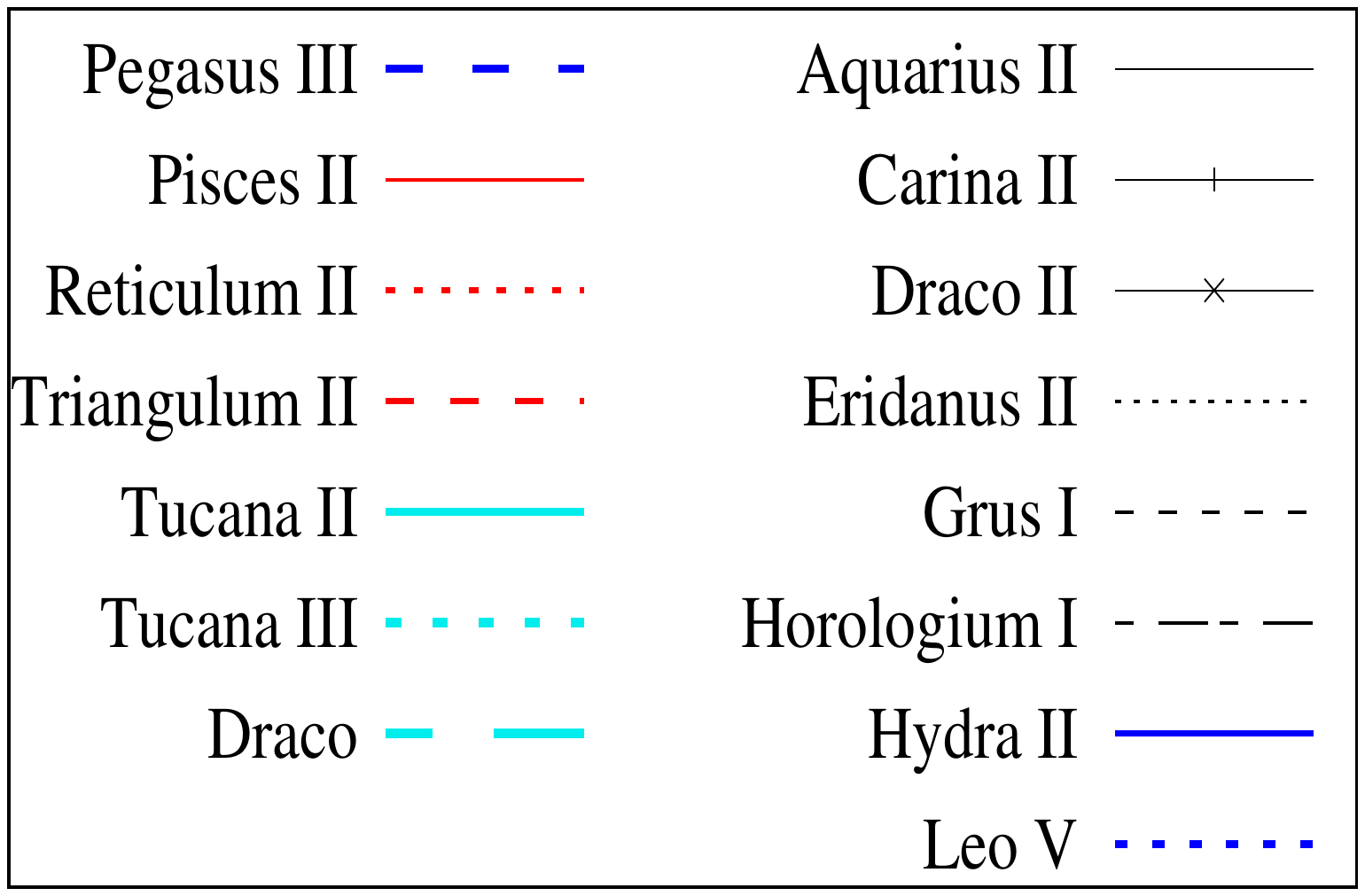}
  \vskip -0.8in
  \caption{\em $95\%$ C.L. upper limits on the $\gamma$-ray fluxes
  (from DM pair-annihilations in UFDs) as a function of
  $\rm{M_{\chi}}$. In deriving these, the indicated channel is assumed
  to be an exclusive one.}  \label{figure:fermi_flux}
\end{figure}

These limits are displayed in Fig.~\ref{figure:fermi_flux}.  Naively,
one might expect that such upper limits on the flux should be
independent of the mass of the DM or the channel it annihilates into.
We need to appreciate though that the $\gamma$-ray spectrum is
strongly dependent on each of these factors. Various factors
contribute to this. These include the charge and mass of the radiating
particle as well as whether cascade decay play a major role.
And since the detector acceptances and efficiencies (not to speak of
the galactic backgrounds) have strong energy-dependence, it stands to
reason that the limits thus derived would also do so.  This is
reflected by Fig.~\ref{figure:fermi_flux}.  It is interesting to note
that, for the most of range of $M_\chi$ considered here, the strongest
limits are obtained for the classical dwarf Draco.  However, much of
this is due to astrophysical factors, and as we shall soon see, this
does not imply that the best bounds on the annihilation rate would be
associated with Draco.

\begin{figure}[h]
  \centering
  \includegraphics[width=0.49\textwidth,height=0.44\textwidth]{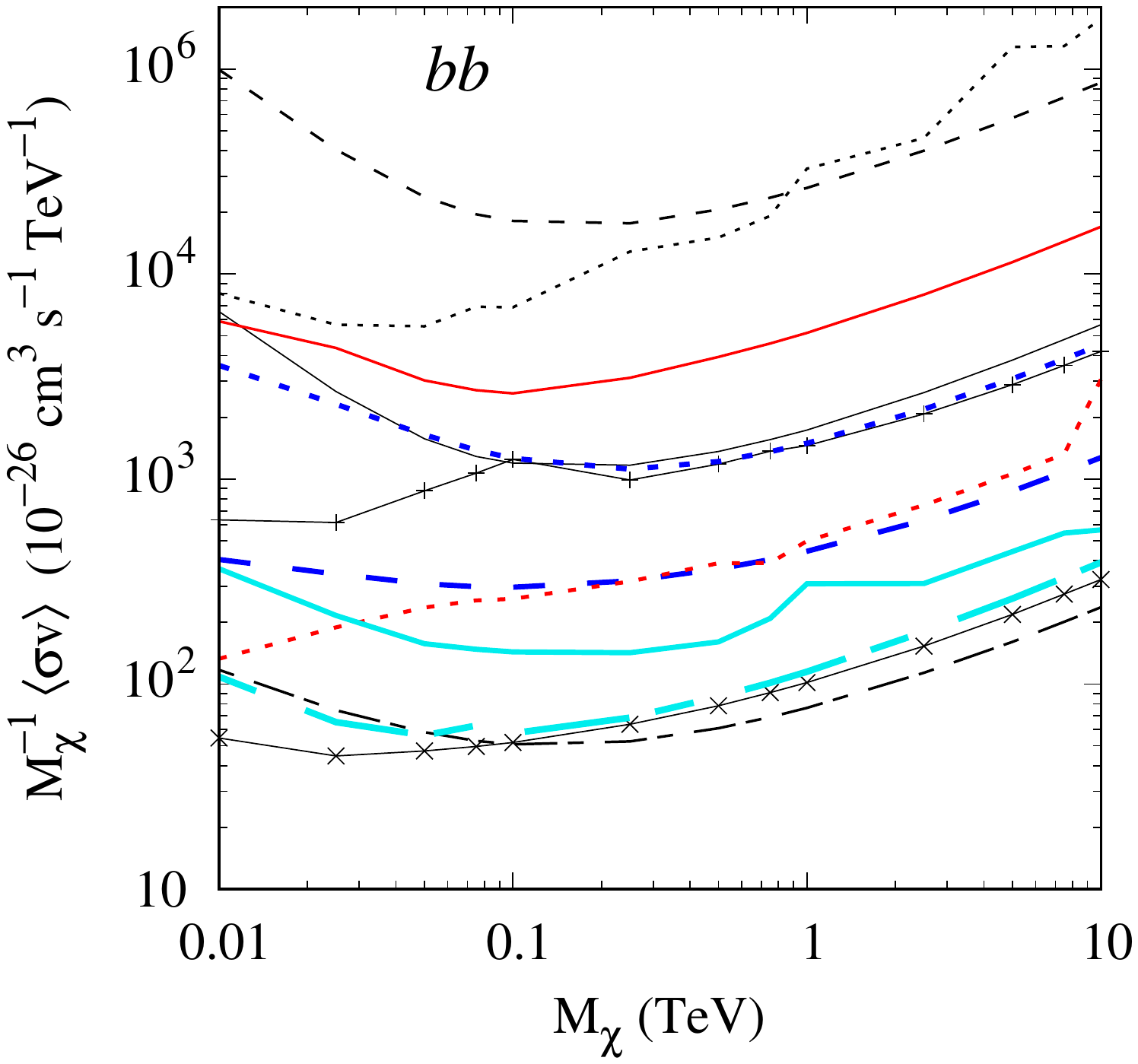}
  \includegraphics[width=0.49\textwidth,height=0.44\textwidth]{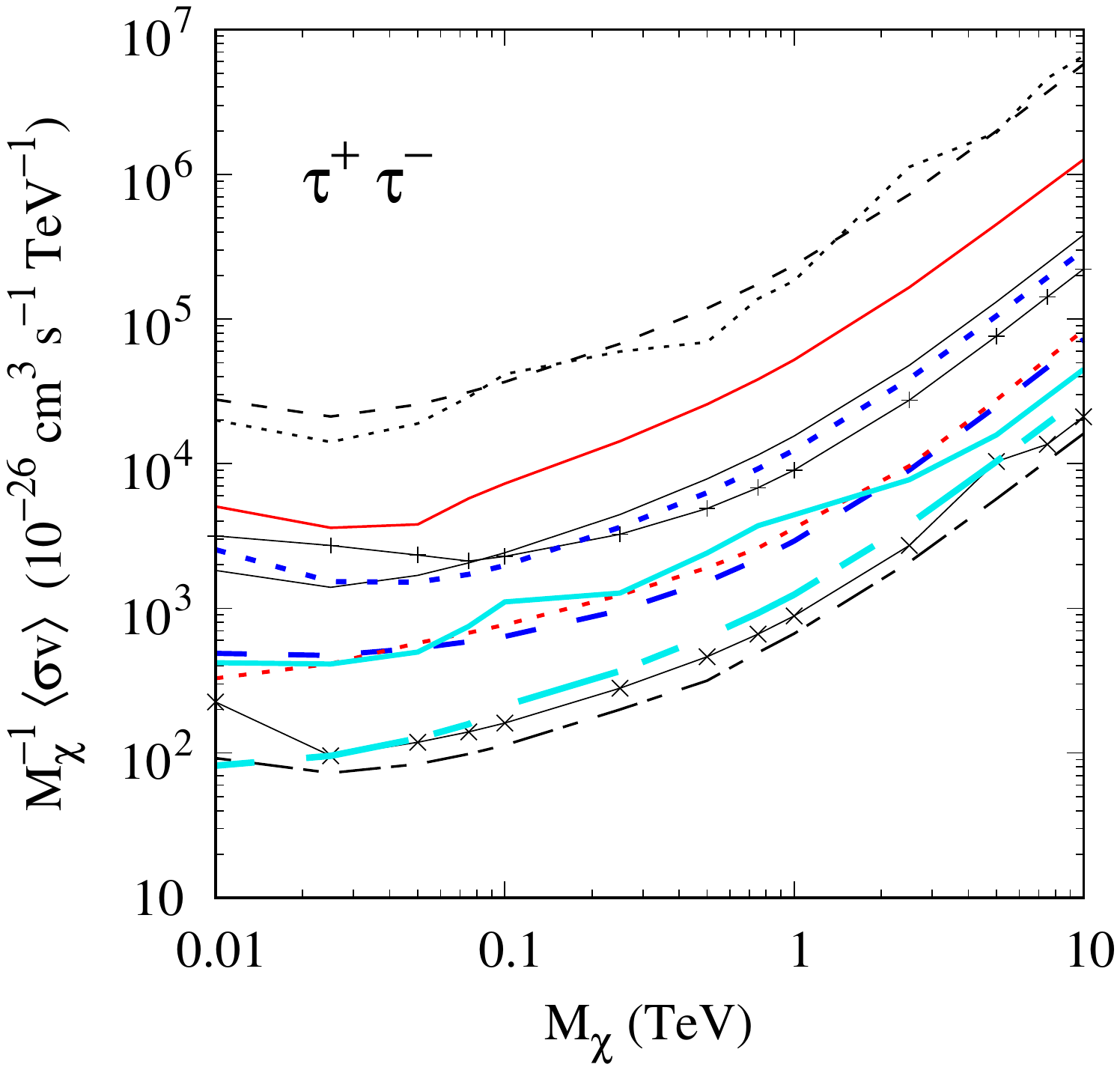} \\
  \vskip -1.1in
  \includegraphics[width=0.49\textwidth,height=0.44\textwidth]{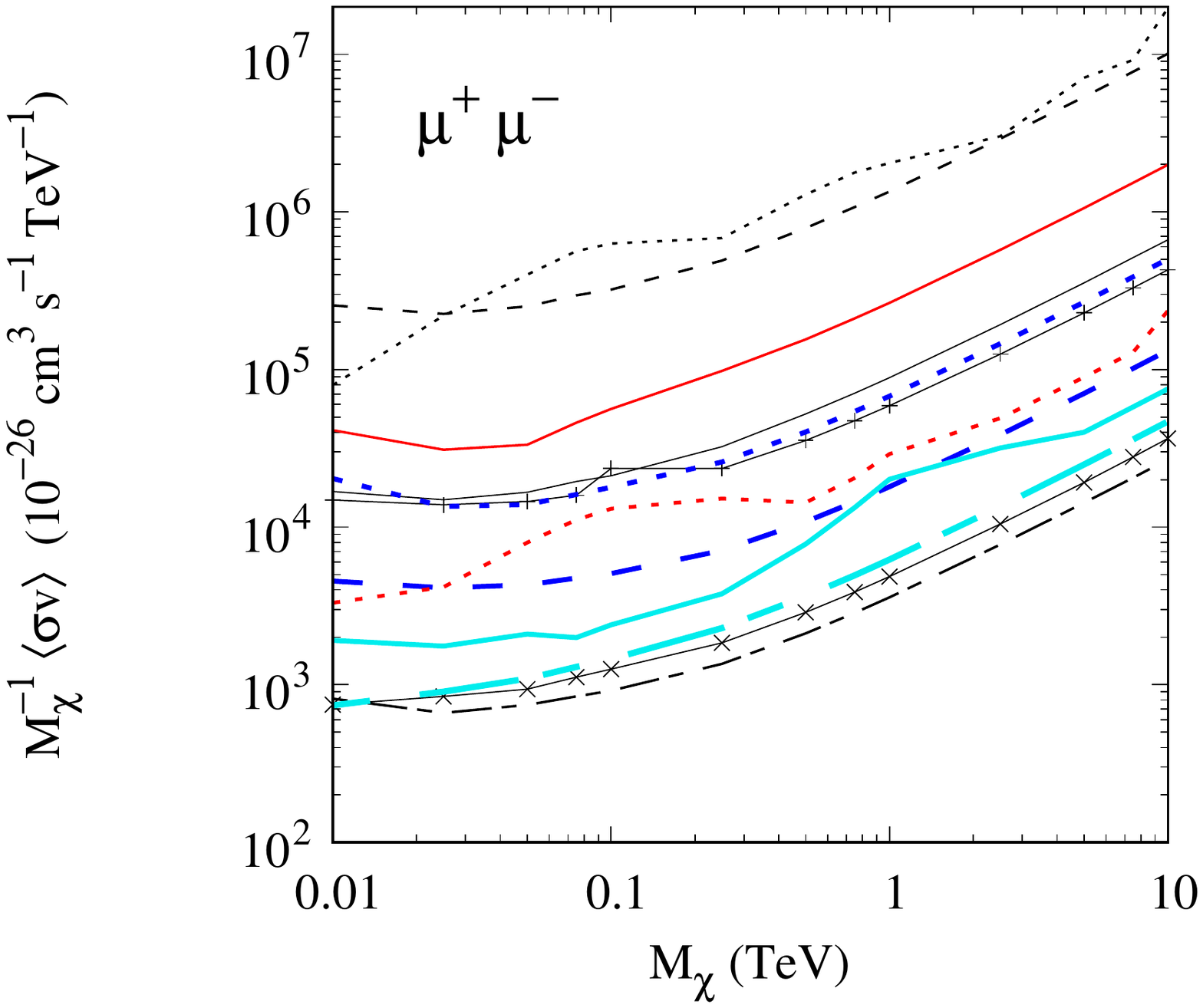}
  \includegraphics[width=0.49\textwidth,height=0.44\textwidth]{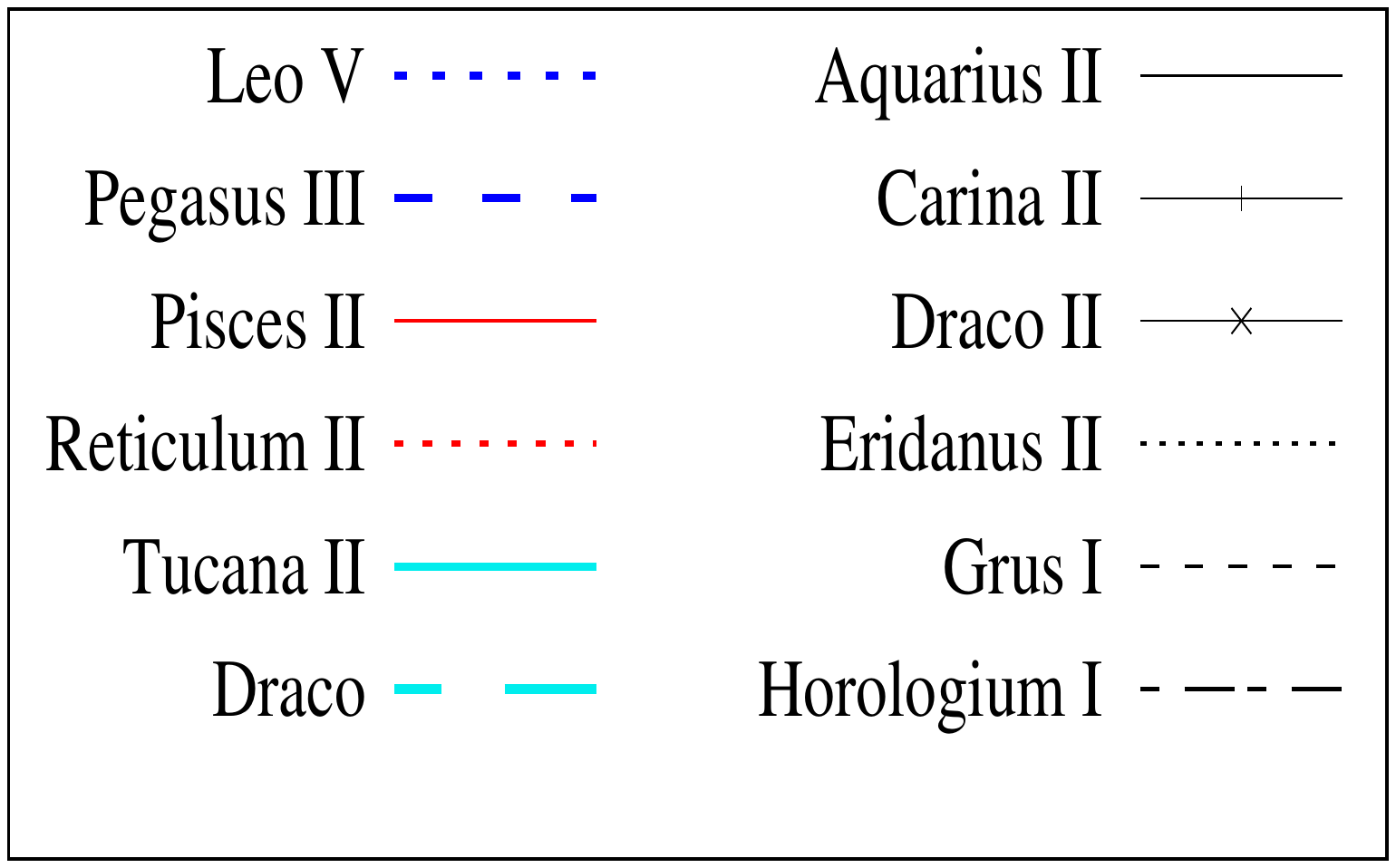}
  \vskip -0.6in
  \label{fig:cross_legends}
 \caption{\em 95\% C.L. upper limit on the thermally
 averaged WIMP pair-annihilation $\langle \sigma v \rangle$ as derived
 from the upper limits on the gamma-ray fluxes from individual UFDs.
 In each case, the annihilation channel is assumed to be
 exclusive. (No limits result from Hydra II, Triangulum II and Tucana III
 as there exist only upper limits on the corresponding $J$-factors.).}
   \label{figure:fermi_cross}
\end{figure}

Such upper limits on the $\gamma$-ray fluxes from DM
annihilation can be translated to constraints in the two dimensional
plane of the WIMP mass $(M_\chi)$ 
and the thermally averaged pair-annihilation
cross section $(\langle \sigma v\rangle )$. This exercise, though,
depends on the final states resulting from the annihilation processes
and, hence, on the details of the model. However, as indicated
at the outset, we adopt an agnostic standpoint and consider three
{\em exclusive} channels, namely $b \bar b$, $\tau^+\tau^-$ and $\mu^+\mu^-$. 
To estimate the limits on
$\langle \sigma v \rangle $, we fit the
$\gamma$-ray spectrum arising from the DM-dominated UFDs with an
MC-simulated DM self-annihilation spectrum, {\tt DMFitFunction}\footnote{\url{https://fermi.gsfc.nasa.gov/ssc/data/analysis/scitools/source_models.html}}
\cite{{Jeltema:2008hf}}. The {\tt DMFit} package is based on the particular set
of MC simulations of hadronization and/or decay of the annihilation
products as used by the DarkSUSY \cite{Gondolo:2004sc} team. 

The consequent limits, for individual UFDs, are displayed in
Fig.~\ref{figure:fermi_cross} for each of the three annihilation
channels. The limits thus obtained are strongly dependent on various
astrophysical parameters that enter in the calculation through the
$J$-factor and the dark matter density profile (see eqns.(1) \&
(2)). These parameters vary substantially between different UFDs and,
as a result, Horologium I, due to its $J$-factor being the largest,
provides the strongest limits on $\langle \sigma v \rangle $ for all
three DM annihilation channels.
However, one should keep in mind that among the newly found UFDs, 
Horologium I is plagued with large uncertainties in its $J$-value
compared with the classical dwarf Draco. Thus the obtained limit from
it may not be as robust as the one from Draco. In
section~\ref{section:uncertainties_horo_tuc}, we provide a detailed
discussion on the $J$-factor uncertainties for Horologium I and
Tucana II.

\subsection{Comparison between our obtained result from the Fermi-LAT analysis with the estimated limits provided by Planck}

Several disparate cosmological observables can probe the DM sector.
In the early epoch of the universe, WIMP annihilation could inject
electromagnetic radiation as well as energetic particles into the
plasma thereby altering the fine-grained evolution. Of the myriad such
effects, two particular ones, namely those pertaining to Big Bang
Nucleosynthesis (BBN) and the Cosmic Microwave Background (CMB) are
important in our context (with others like primordial black hole
formation or HI line strengths {\em etc.} playing only subservient
roles. The injection of high energy particles (hadrons) and radiation
during or after BBN can modify light nuclei abundances, thus coming
into conflict with the data. The limits, understandably, depend on
both the mass of the DM and the annihilation channel.  For
$M_\chi \sim 1\, {\rm TeV}$, typical limits range over $\langle \sigma
v \rangle \sim 10^{-23}-10^{-21} ~{\rm
cm^3/sec}$~\cite{Hisano:2011dc}, depending on the details of the
annihilation mechanism and the consequent effect on the particular
light element being synthesised. Evidently, the BBN limit is a few
orders of magnitude higher than the thermal cross-section and, thus,
do not pose any threat to such heavy WIMPs.

\begin{figure}[h!]
\begin{center}
 {\includegraphics[width=.5\linewidth,height=3in]{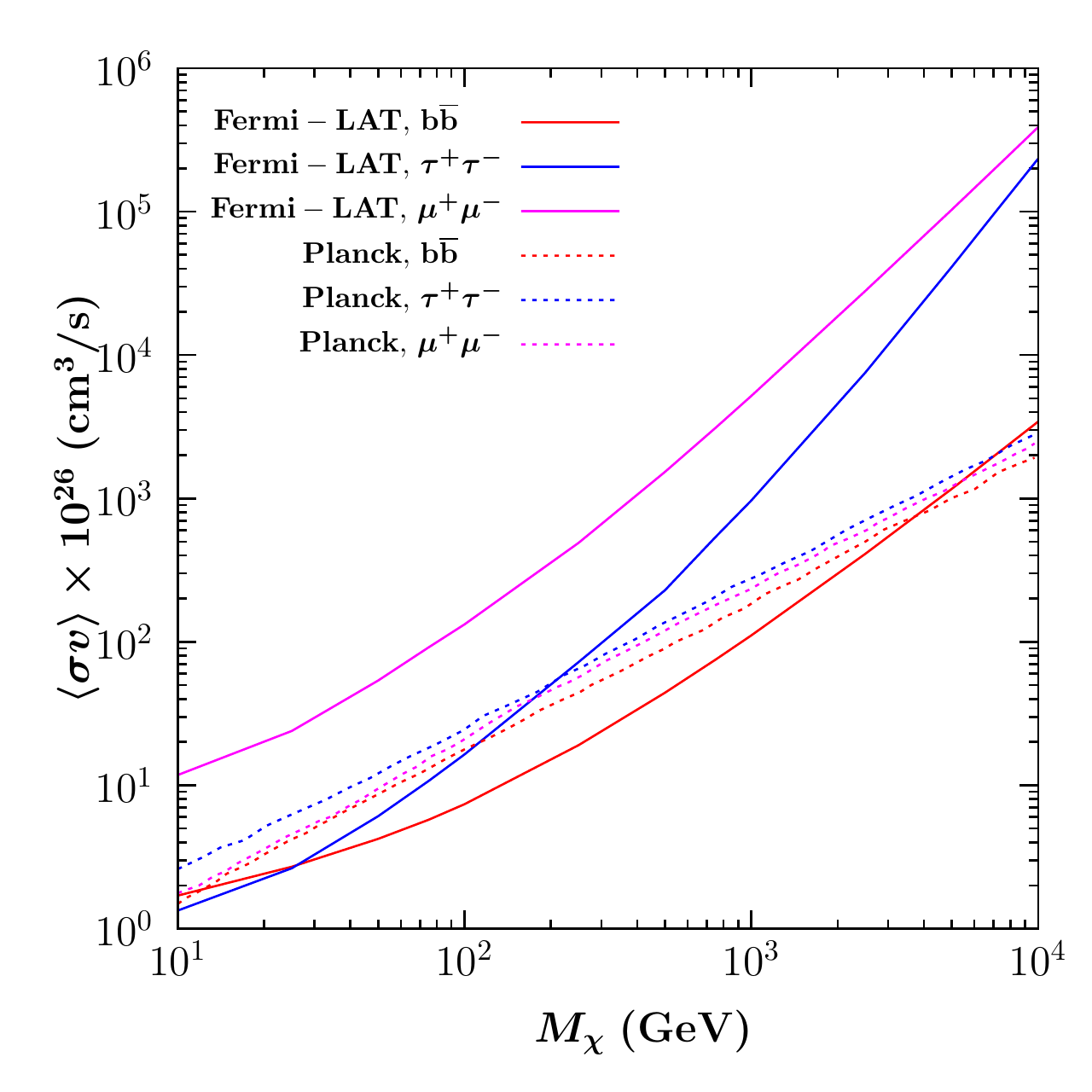}}
\caption{\em Comparison between the limits obtained for Horologium I from Fermi-LAT with the limits provided by Planck. }
\label{figure:planck_comparison}
\end{center}
\end{figure}

On the other hand, the injection of such charged particles could
significantly increase the residual ionization fraction which, in
turn, modifies the last scattering surface as well the observed CMB
anisotropy\cite{Gruzinov:1998un, Lewis:1999bs}.  The
WMAP \cite{Bennett_2013} and, more recently, the
Planck~\cite{Aghanim:2018eyx} satellite experiments are highly
sensitive to any such modifications in the CMB, leading to strong and
model-independent limits on the aforementioned
energy-injections~\cite{Aghanim:2018eyx}.  These limits can, further,
be translated into constraints on WIMP annihilation cross sections.
As the Horologium I provided the strongest limits on $\gamma$-ray
fluxes as seen by Fermi-LAT, in Fig.~\ref{figure:planck_comparison},
we compare the same with the constraints from Planck data.  It is
clear that, for the $b{\bar b}$ channel, the Fermi-LAT limits are
stronger than the Planck limits over the entire range $M_\chi \in [10,
10^4]\gev$, with the CMB constraints like to dominate only for
$M_\chi \gapp 20\,{\rm TeV}$.  For the $\mu^{+}\mu^{-}$ channel, the
situation is quite the opposite, with the Planck measurement providing
the most stringent limits. And, finally, the situation for the
$\tau^{+}\tau^{-}$ channel is halfway in between with the Fermi-LAT
constraints being stronger for $M_\chi \lapp 500$~GeV and the Planck
limits being dominant for larger masses. This puts into perspective
the fact that a comprehensive probe of the DM parameter space is best
achieved by a combination of search strategies, an issue we return to later.

\section{Synchrotron radiation from ultra-faint dwarf galaxies}
\label{sec:synchr}
A charged particle propagating through the interstellar medium would
lose energy owing to a variety of electromagnetic processes such as
inverse Compton radiation, synchrotron radiation, Coulomb losses and
bremsstrahlung. While this applies to any charged particle, the
radiation would be substantial only if the particle is sufficiently
long-lived, or in other words if it is one of $e^\pm$ or $p, \bar
p$. The contributions from the last two are much smaller, both on
account of their larger masses as well as the small probability for a
quark (from the hard process) fragmenting\footnote{Similarly,
the probability for a charged particle to traverse unmolested from the
UFD to the earth-- or satellite-bound detectors is rather
low. Consequently, the corresponding bounds from antiproton or
positron detection are much weaker.}  into a $p, \bar p$. Given this,
we develop the subsequent arguments for $e^\pm$ alone. The other
species can be treated analogously.

A complete treatment of the DM-initiated synchrotron radiation must
consider the diffusion of the secondary particles, and the
aforementioned energy loss contributions
therefrom. The formalism, developed in
refs.\cite{Colafrancesco:2005ji, Colafrancesco:2006he,
McDaniel:2017ppt}, can be summarised by the transport equation for
$n_e(\mathbf{r},E)$, the number density of $e^\pm$ of a given energy
$E$ at the position $\mathbf{r}$ with respect to the center of the
UFD), {\em viz.},
\begin{align}\label{eqn:diffusion}
\frac{\partial}{\partial t} \frac{dn_e(\mathbf{r},E)}{dE}
= \nabla . \Big( D(E,\mathbf{r}) \nabla \frac{dn_e(\mathbf{r},E)}{dE}\Big)
     + \frac{\partial}{\partial E} \Big( b(E,\mathbf{r}) \frac{dn_e(\mathbf{r},E)}{dE}\Big)
     + Q_e (E,\mathbf{r}).
\end{align}
Here, $ D(E,\mathbf{r})$ is the space-dependent diffusion coefficient,
$b(E,\mathbf{r})$ encapsulates the energy loss term and the source
term $Q_e$ is given by
\begin{align}
 Q_e (E,\mathbf{r}) = \frac{\rho^2_\chi(\mathbf{r}) \langle \sigma v\rangle}{2 m_\chi^2}    \frac{dN_{e}}{dE},
\end{align}
where $N_e$ is the number of $e^{\pm}$ produced with a given energy $E$
per DM annihilation.

The energy loss term consists of several independent contributions from the processes listed earlier and is given by \cite{Colafrancesco:2005ji,McDaniel:2017ppt}
\begin{equation}
\barr{rcl}
b(E) & = & \dis  b_{\rm IC}(E) + b_{\rm syn}(E) + b_{\rm Coul}(E) + b_{\rm brem}(E)  \\[1ex]
     & = & \dis b_{\rm IC}^0 E^2 + b_{\rm syn}^0 B^2 E^2 
         + b_{\rm Coul}^0 n ( 1 + \log(\gamma /n)/75) + b_{\rm brem}^0 
n (\log(\gamma / n) + 0.36).
\earr
\end{equation}
where the dependence on $\mathbf{r}$ has been suppressed.  Here, $B$
is the magnetic field in $\mu G$, $n$ is the number density of thermal
electrons in cm$^{-3}$ and $\gamma = E/m_e$ is the time-dilation
factor. The various energy loss parameters $b^0_{\rm IC}, b^0_{\rm
syn}, b^0_{\rm Coul} $ and $b^0_{\rm brem}$ have approximate values of
$ 0.25,0.0254,6.13$ and $1.51 $ respectively in units of $ 10^{-16}$ GeV
s$^{-1}$ \cite{Colafrancesco:2005ji}.  A position-dependence in
$D(E, \mathbf{r})$ would be expected to be occasioned only by
deviations from a homogeneous mass distribution. Given the low
light-to-mass ratio for the UFDs, this is not expected to be a major
concern.  In the absence of a detailed knowledge of the structure of
the UFDs, we neglect such dependencies, limiting ourselves to
$D(E, \mathbf{r}) = D(E)$ with~\cite{Spekkens:2013ik,
McDaniel:2017ppt}
\begin{equation}\label{eqn:diffusion_coefficient}
 D(E) = D_0 \left(\frac{E}{1 \GeV}\right)^{\gamma_D} \ .
\end{equation}

Assuming spherical symmetry, a uniform magnetic field and a uniform
number density of thermal electrons, the stationary state solution of
the diffusion equation is given by
\begin{align}\label{eqn:solutionndifusion}
  \frac{dn_e}{dE}(r,E) = \frac{1}{b(E)} \int_{E}^{M_\chi} dE^\prime \,
            G\Big(r, v(E)-v(E^\prime)\Big) Q_e(E^\prime,r),
\end{align}
where the  Green's function is given by 
\begin{equation}
\begin{array}{rcl}
G(r, \Delta v) & = & \displaystyle
\frac{1}{\sqrt{4\pi \Delta v}} \sum_{k=-\infty}^{\infty} (-1)^k 
                     \int_{0}^{r_h} dr^\prime \frac{r^\prime}{r_k}
                     \left(\frac{\rho_\chi(r^\prime)}{\rho_\chi(r)}\right)^2
                     \\[2ex]
& & \displaystyle \hspace*{10em}
\left[ \exp\left(-\, \frac{(r^\prime -r_k)^2}{4 \Delta v}\right)
                     - \exp\left(-\, \frac{(r^\prime + r_k)^2}{4 \Delta v}\right)
                     \right],
\end{array}
\label{greens_func}
\end{equation}
with
\beq
v(E) \equiv \int_E^{M_\chi} d\tilde{E} \frac{D(\tilde{E})}{b(\tilde{E})} \ ,
\qquad
r_k \equiv (-1)^k r + 2k r_h \ .
\eeq
Here $r_h$ defines the diffusion zone of the UFD, namely the
radius at which the free escape boundary condition $dn_e(r_h,E)/dE =
0$ may be imposed.  Typically, $r_h$ is approximately twice the radius
of the last stellar component of the galaxy ({\em i.e.}, twice the
distance of the outermost star from center).

The synchrotron power spectrum or the total power radiated per unit
frequency at $\nu$ by an electron of energy $E$ present
in a magnetic field $B$, {\em viz.}, $P_{\rm synch}(\nu,E,B)$ is defined as: 
\begin{equation}
 P_{\rm synch}(\nu, E, B) = \pi \sqrt{3} r_0 m_e c \nu_0 \, \int_0^\pi \, d\theta \, \sin^2\theta \, F\big(\frac{x}{\sin\theta }\big),
\end{equation}
where $\theta$ is the pitch angle, $r_0 = e^2/(m_e c^2)$ is the
classical electron radius and $\nu_0 = eB/(2\pi m_e c)$ is the
non-relativistic gyro-frequency. While
\begin{equation}
 F(y) = y \,  \int_y^\infty d\zeta \, K_{5/3}(\zeta) \simeq 1.25 \, y^{1/3}\,
      e^{-y} \, (648 + y^2)^{1/12} \ ,
\end{equation}
the quantity $x$ is given by
\begin{equation}
 x = \frac{2 \, \nu\, m_e^2 \, (1+z)}{3 \, \nu_0\, E^2} 
\end{equation}
with $z$ being the redshift of the source. For the UFDs under consideration,
$z \approx 0 $. 

We can, now, estimate the total energy radiated or the local
emissivity (i.e. the amount of energy radiated at a given
$r$, per unit volume per unit time) at a given frequency
$\nu$ in the form of synchrotron radiation in terms of $P_{\rm synch}$
and $dn_e/dE$ {\em viz.},
\begin{equation}\label{eqn:emissivity}
j_{\rm synch}(\nu, r)
 = \int_{m_e}^{M_{\chi}} dE \left(\frac{dn_{e^+}}{dE}
 + \frac{dn_{e^-}}{dE}\right) P_{\rm synch}(\nu,E,B) = 
 2 \int_{m_e}^{M_{\chi}} dE \, \frac{dn_{e^-}}{dE}\,
 P_{\rm synch}(\nu,E,B) \ .
\end{equation}
The integrated synchrotron flux density spectrum is, now, given by 
 \begin{equation}\label{eqn:syn_flux}
S_{\rm synch}(\nu) = \frac{1}{4\pi d^2}\int d^3 r \, \,  j_{\rm synch}(\nu,r),  
 \end{equation}
where $d$ is the distance to the UFD and
the integration is over the whole diffusion volume.

It is worth remembering that, unlike in the case for gamma rays, the
synchrotron flux is not related to the $J$-factor, especially on
account of the dependence on the diffusion and energy loss processes.
Rather, the magnetic field $B$ inside the UFD as well as the
diffusion coefficient (parameterized by $D_0$ and $\gamma_D$) are the
important attributes.

The magnetic field in the interior of a dSph is neither well
understood nor well-measured.  Rather, ranges for it are inferred
using a variety of theoretical
arguments \cite{Regis:2014koa}. Typically, star-forming
low-mass galaxies such as dwarf irregulars carry a
magnetic field of ${\cal O}(1-10) \mu$G, and this gives an order of
magnitude estimate. This could be further refined by examining the
observed correlations between the magnetic field and the star
formation rate in Local group galaxies \cite{Chyzy:2011}. An
independent source of the magnetic fields could be the Milky Way
itself; large outflows of the galactic field have been
observed \cite{Carretti:2013}, and these could well magnetize the
dSphs. To circumvent the uncertainties associated with such modelling
exercises, it has been suggested to use local equipartition between
the magnetic field and the charged particles in the plasma (as evinced
from cosmic rays). Each such mechanism indicate different preferred
ranges, typically within an order of magnitude of 1 $\mu$G, often at
slight variance with each other \cite{Regis:2014koa}.  A further
complication is introduced by the fact that the magnetic field can
have a nontrivial profile .  If the magnetic field owes its origin to
the stars of the dSph, then it stands to reason that the field
configuration would be maximum close to the center and decaying
outwards. Certain studies have assumed an exponential profile having spherical symmetry and of the form $B = B_0 \, e^{-r/r_*}$, where $r_*$ would be of the order of the half-light radius of the dSph~\cite{Regis:2014koa, Regis:2014tga,
McDaniel:2017ppt, Regis:2017oet, Cembranos:2019noa}.
On the other
hand, if the dSph is primarily magnetized by the Milky Way magnetic
field, then the magnetic field would be expected to be almost constant
over the dSPh, owing to the smallness of the latter's size in
comparison to its distance from Milky Way. Given these uncertainties,
we assume a uniform profile of strength
$B=1 \, \mu$G~ \cite{Colafrancesco:2006he, Spekkens:2013ik} for most
of our indicative calculations and discuss the variation of our result
on the size and profile of the magnetic field strength in
Sec. \ref{sec:uncertainty}.

The diffusion coefficient, for which we assume the simplified form of
eqn.\ref{eqn:diffusion_coefficient}, has large uncertainties.  For
galaxy clusters, a value for the coefficient $D_0$ as large as
$10^{28}$--$10^{30}\, {\rm cm}^2/{\rm s}$ has been argued
for~\cite{Natarajan:2015hma,Jeltema:2008ax}.  Constraints on the Milky
Way diffusion parameters can be inferred from
data~\cite{Webber:1992dks,Baltz:1998xv,Maurin:2001sj} and
typically range between $10^{27}$--$10^{29}\, {\rm cm}^2/{\rm s}$.
Similarly, the parameter $\gamma_D$ is expected to lie in the range
$0\leq \gamma_D \leq 1$ \cite{Jeltema:2008ax}.
To be specific, we choose values close to the geometric means of the
individual ranges, namely 
 $D_0 = 3 \times 10^{28}\, {\rm cm}^2/{\rm s}$ and $\gamma_D= 0.3$
\cite{McDaniel:2017ppt}, postponing the discussion
of the dependence on the choices to Sec.\ref{sec:uncertainty}. 

For a given DM particle, the synchrotron flux, understandably, depends
on the states to which it pair-annihilates and their subsequent
cascades. As in the preceding sections, we consider three annihilation
channel states, {\em i.e.}, $b\bar{b}$, $\tau^+ \tau^-$ and
$\mu^+ \mu^-$, and for the sake of simplicity, continue to assume that
a single channel dominates overwhelmingly.  While we use the RX-DMFIT
code \cite{McDaniel:2017ppt} to calculate the synchrotron
flux, it is useful to undertake a simplified discussion to understand
the different aspects of the dynamics. For this exercise, we use a
typical value for the velocity averaged DM annihilation cross section, namely,
$\langle \sigma v \rangle = 10^{-26} \, {\rm cm}^3/{\rm
s}$, postponing a derivation of constraints on the same until
later. For all the UFDs we have used a thermal electron density
$n \approx 10^{-6}$
cm$^{-3}$ \cite{Colafrancesco:2006he,McDaniel:2017ppt},
and the NFW density profile for the DM distribution within.

\begin{figure}[!h]
  \begin{subfigure}[b]{0.49\textwidth}
     \centering
  \includegraphics[width=0.8\textwidth]{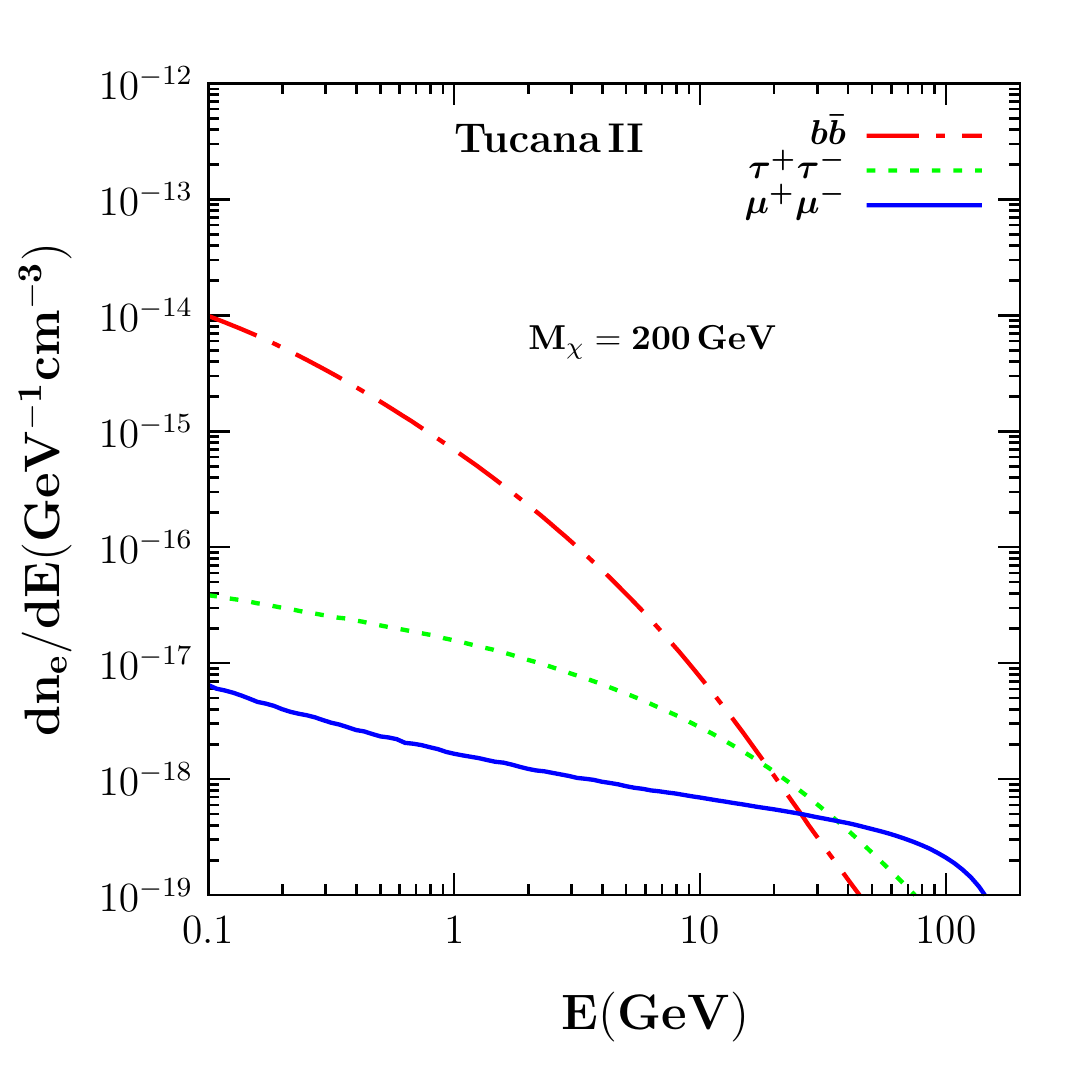}
    \caption{}
    \label{fig:dnde_200GeV}
  \end{subfigure}
\hfill
  \begin{subfigure}[b]{0.49\textwidth}
     \centering
  \includegraphics[width=0.8\textwidth]{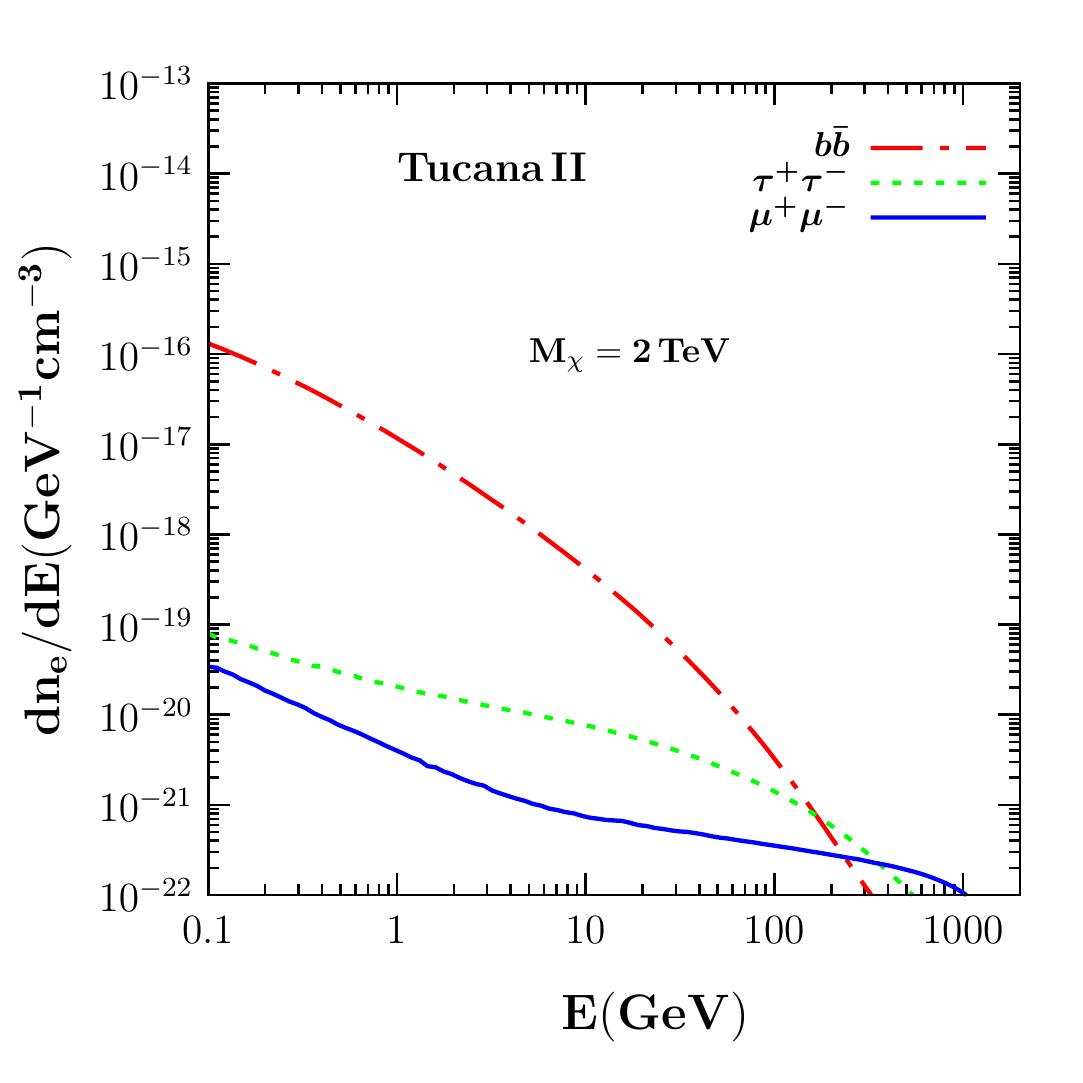}
    \caption{}
    \label{fig:dnde_2TeV}
  \end{subfigure}
\caption{\em Equilibrium electron number density spectrum at a radial distance $r=0.1$~kpc 
for Tucana II for three different exclusive annihilation channels, to
  $b\bar{b}$ (red), to $\tau^+ \tau^-$ (green) and to $\mu^+ \mu^-$
  (blue), using $B \, = \, 1\, \mu$G, $D_0 =
  3 \times 10^{28}$ cm$^2$/s, $\gamma_D = 0.3$ and $\langle \sigma
  v\rangle \, = 10^{-26}$ cm$^3$/s and the NFW density
  profile. The two panels correspond to different values for the DM mass.
  } \label{figure:dnde}
\end{figure}

 Fig.\ref{figure:dnde} shows the stationary electron distribution
spectrum for Tucana II at a radial distance 0.1 kpc for two different
DM mass values, 200 GeV and 2 TeV. With the cascades from a $b$-decay
being capable of producing more $e^\pm$ than a $\tau$ or a $\mu$ can
(the latter, only one), it is understandable that the integrated
spectrum is much larger for the $b\bar b$ channel than it is for the
others. The existence of the cascades also explains the relative
softness of the three spectra in Fig.\ref{figure:dnde}.

\begin{figure}[!h]
  \begin{subfigure}[b]{0.49\textwidth}
     \centering
 \includegraphics[width=\linewidth,height=3in]{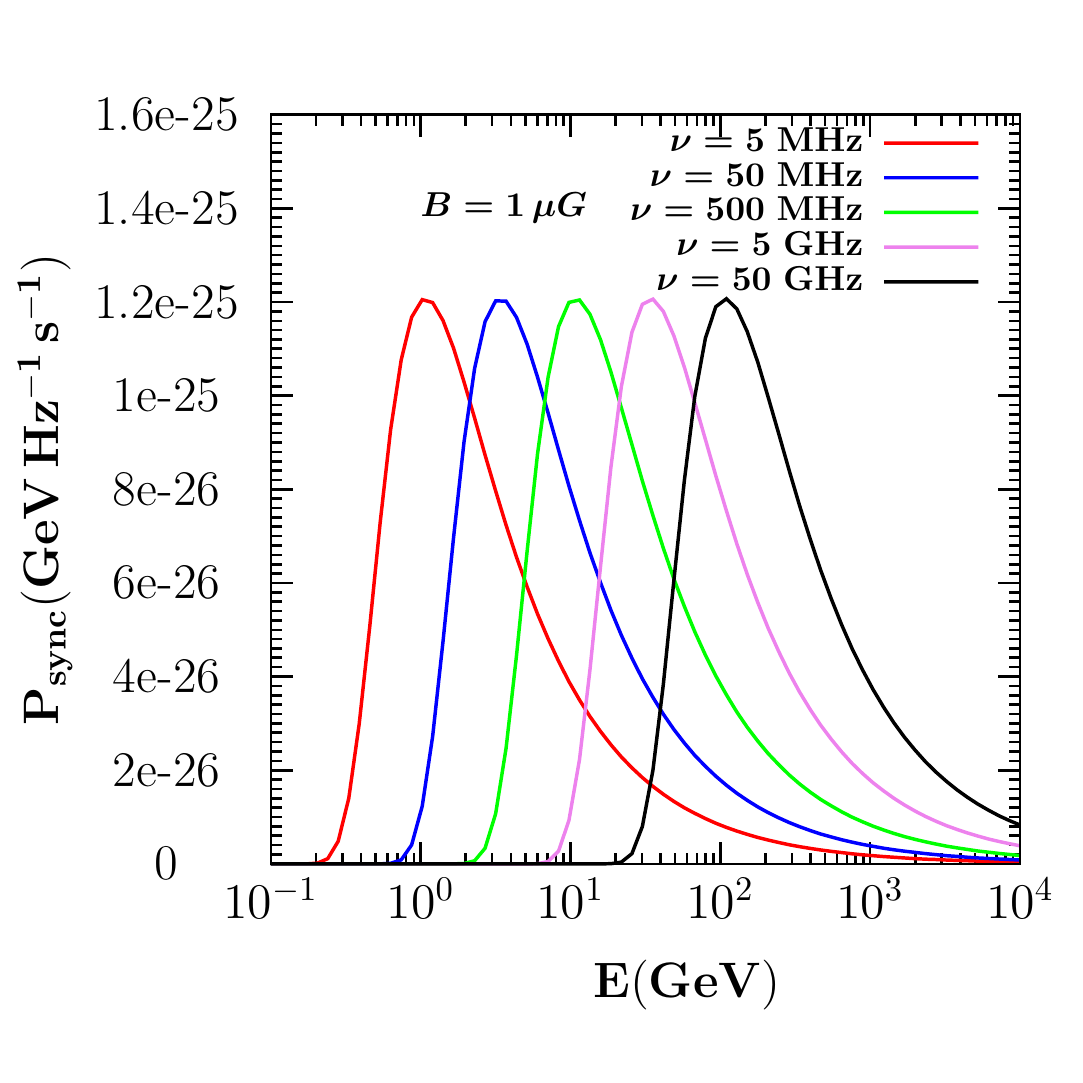}
    \caption{}
    \label{fig:synpower}   
  \end{subfigure}
  \begin{subfigure}[b]{0.49\textwidth}
\centering
\includegraphics[width=0.95\linewidth,height=2.8in]{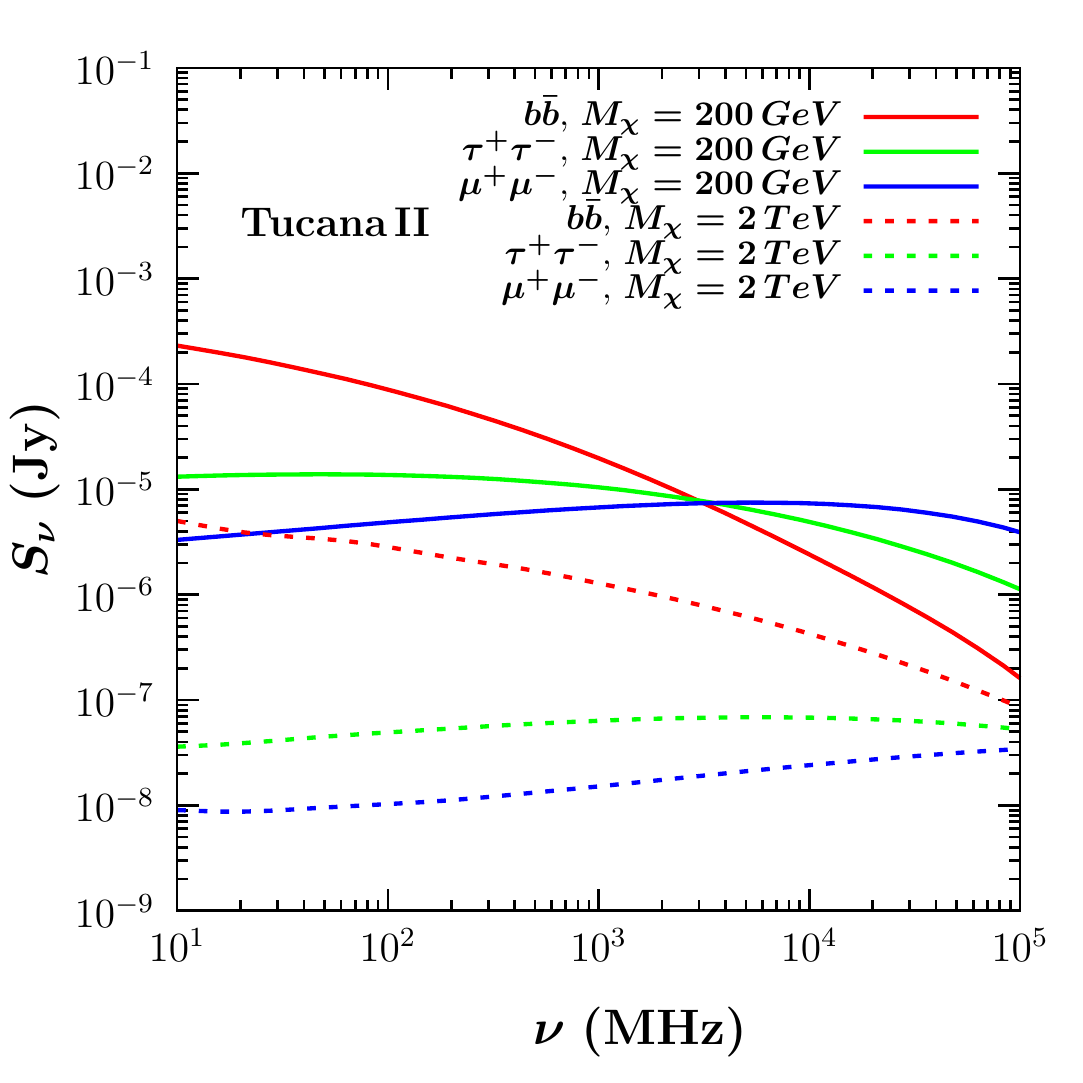}
\caption{}
\label{fig:bbbartautau_flux_comp}
\end{subfigure}
\caption{\em {\em (a)} Synchrotron power spectra at different frequencies
for an arbitrary galaxy with a magnetic field 1 $\mu$G.  {\em (b)}
  Synchrotron fluxes for three exclusive final states for DM
  annihilation, i.e., $b\bar{b}$ (red), $\tau^+ \tau^-$ (green) and
  $\mu^+ \mu^-$ (blue).  The solid (dashed) lines correspond to DM
  masses of 200 GeV and 2 TeV respectively.  We have used $B \, = \,
  1\, \mu$G, $D_0 = 3 \times 10^{28}$ cm$^2$/s, $\gamma_D = 0.3$,
  $\langle \sigma v\rangle \, = 10^{-26}$ cm$^3$/s and the NFW density
  profile for DM distribution inside the UFD.} \end{figure}

For a given frequency, the energy corresponding to the peak of the
synchrotron power spectrum contributes significantly to $j_{\rm
synch}(\nu,r)$ in eqn.\ref{eqn:emissivity}. The power spectrum $P_{\rm
synch}(\nu,E,B)$ for $B= 1$ $\mu$G and for different frequencies in
the range 5 MHz--50 GHz are shown in
Fig.\ref{fig:synpower}. Understandably, for higher frequencies, the
synchrotron power peaks at a higher value of energy. Clearly, for a
given frequency, the channel resulting in a larger number of $e^\pm$
with energies closer to the peak of the synchrotron power spectrum
will result in a larger synchrotron flux. Hence, for higher
frequencies, the synchrotron flux from a leptonic channel will
dominate over that from a hadronic channel. This feature can be
observed in Fig.\ref{fig:bbbartautau_flux_comp} where, for a given
DM mass ({\em e.g.}, 200 GeV), the $\tau^+ \tau^-$ channel dominates
over the $b \bar{b}$ channel for higher frequencies; by the same
token, for lower frequencies, the $b\bar{b}$ channel dominates over
the $\tau^+ \tau^-$ channel. Since the electrons originating from the
pair-annihilation of a DM of mass $M_\chi$ can have a maximum energy
$M_\chi$, the spectrum corresponding to a heavier DM would be harder
(as shown by Fig.\ref{figure:dnde}).  Consequently, for larger
frequencies, the synchrotron power spectrum peaks at a higher value of
the electron energy. This is reflected in
Fig.\ref{fig:bbbartautau_flux_comp} where, for a larger $M_\chi$,
the crossover from $b\bar b$ dominance to $\tau^+ \tau^-$ dominance
occurs at progressively higher frequencies.

\subsection{Results pertaining to the Ultra Faint Dwarf Galaxies}
\label{sec:synchro_ufd}

We begin by considering existing 
radio-frequency observatories, in particular the data from the
following two telescopes:
\begin{itemize}
\item the Giant Metrewave Radio Telescope (GMRT) \cite{Intema:2016jhx} with its sky-survey, covering the expanse over $-53^{\circ}$ to $+90^{\circ}$, with 
a particularly useful set corresponding to $\nu =0.1475~{\rm GHz}$, and
\item the NVSS survey by the Very Large Array 
(VLA) telescope \cite{condon1998}, ranging over $-40^{\circ}$ to $+90^{\circ}$,
and at $\nu = 1.4~{\rm GHz}$.
\end{itemize}
The non-detection of radio emission from any of the UFDs can be
translated to 95$\%$ C.L. upper limits on the fluxes \footnote{Since
all radio interferometric maps are made per unit beam where the beam
is convolved with the respective point spread functions (PSF), we can
directly obtain the flux density by using the final image.} as listed
in Table \ref{table:radio_flux_upper_limits}. Note here that the
limited sky coverage implies that no statement can be made for certain
UFDs, such as Tucana II.

\begin{table}[!h]
\centering
\begin{tabular}{|p{2.5cm}|p{4.2cm}|p{4cm}|}
\hline \hline
Galaxy &  GMRT ($\nu = 147.5$ MHz) & VLA ($\nu = 1.4$ GHz)  \\
\hline \hline
Aquarius II & $6.8 $ & $0.86$ \\
\hline \hline
Draco II & $9 $ & $1.1$ \\
\hline \hline
Eridanus II & $7.8 $ & No Data \\
\hline \hline
Grus I & $4.1$ & No Data \\
\hline \hline
Hydra II & $8.8 $ & $1.1 $ \\
\hline \hline
Leo V & $6 $ & $0.98$ \\
\hline \hline
Pegasus III & $10$ & $0.96$ \\
\hline \hline
Pisces II & $3.5$ & $0.88$ \\
\hline \hline
Triangulum II & $6 $ & $1$ \\
\hline \hline
Draco & $7.2$ & $9.2$ \\
\hline \hline
\end{tabular}
\caption{\em 95 $\%$ C.L. upper limits (in units of mJy)
on the radio flux density from the UFDs obtained from GMRT and VLA. 
For Carina II, Horologium I, Reticulum II and Tucana II\&III, neither
observatory provides any data.}
\label{table:radio_flux_upper_limits}
\end{table}

\begin{figure}[h!]
 \includegraphics[width=.3\linewidth]{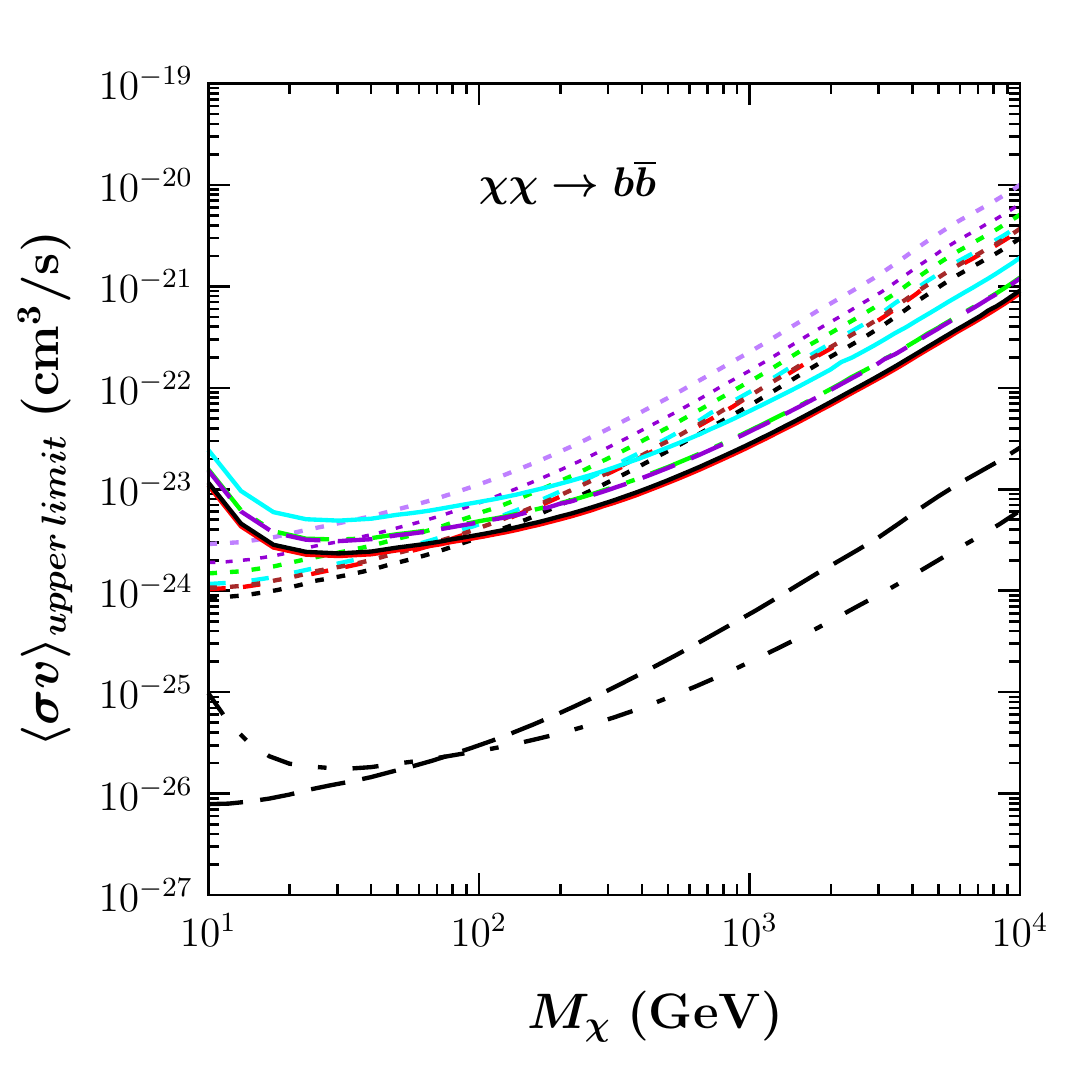}
\hskip 10pt
 \includegraphics[width=.3\linewidth]{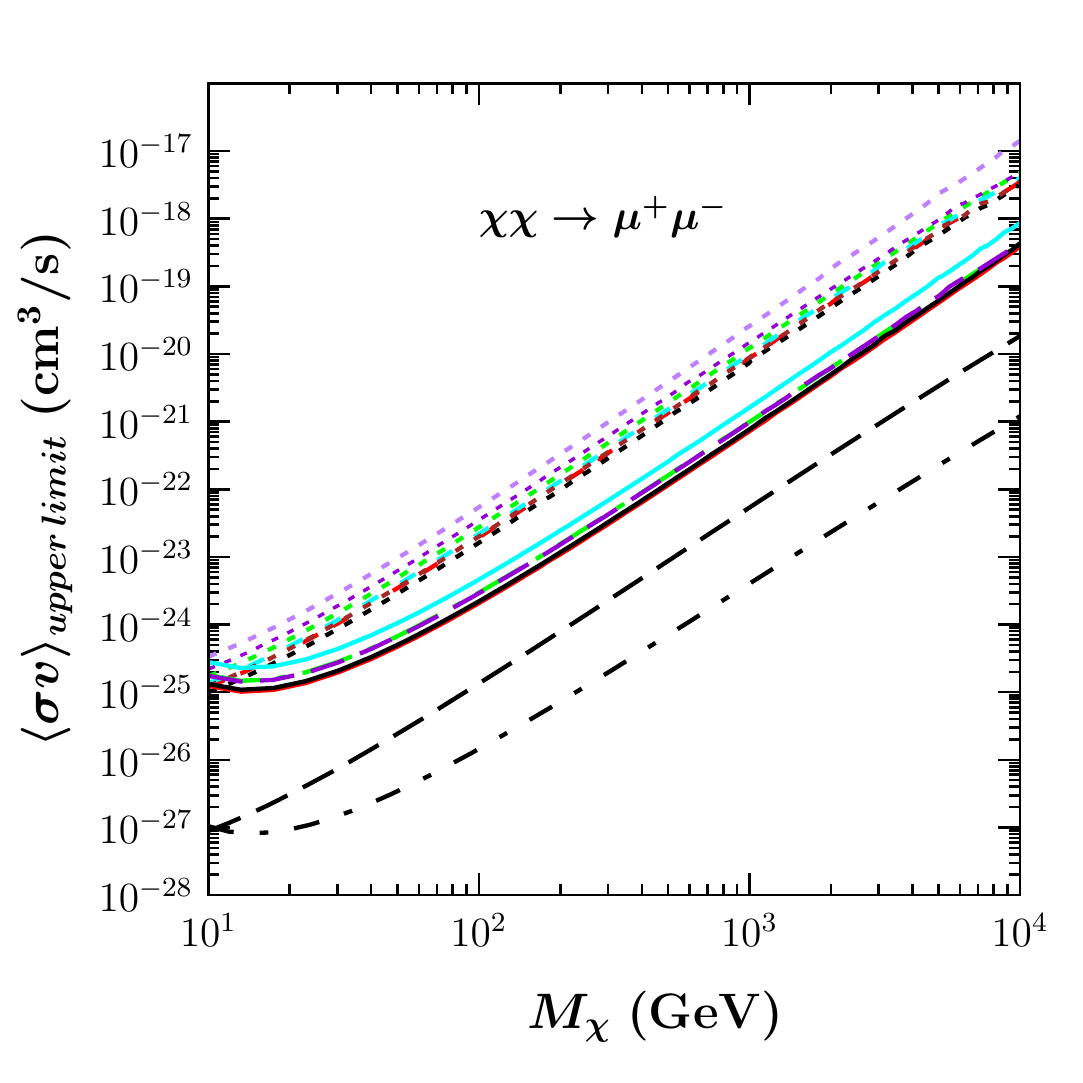}
\hskip 10pt
 \includegraphics[width=.3\linewidth]{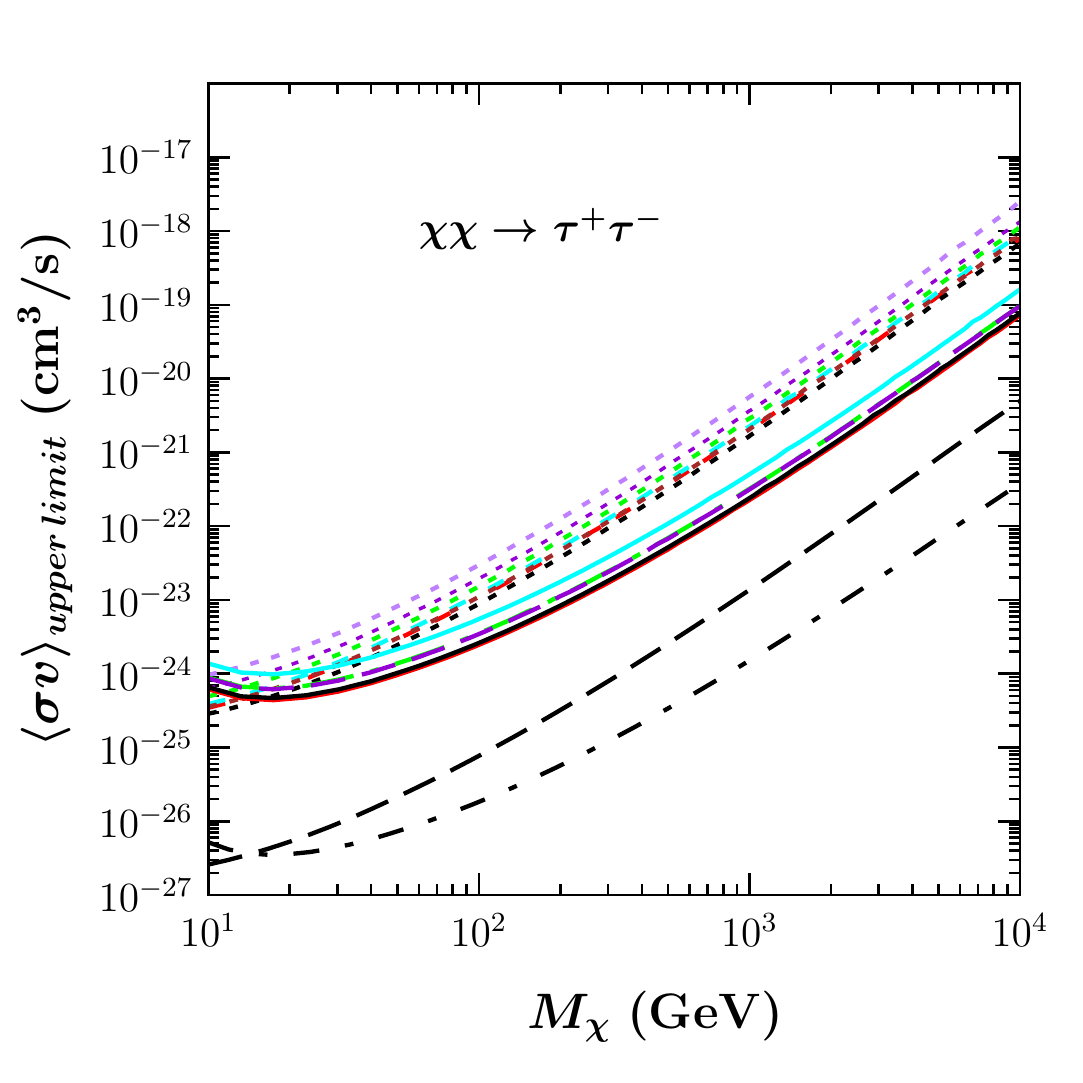}

 \hskip 10pt
 \includegraphics[width=0.9\linewidth]{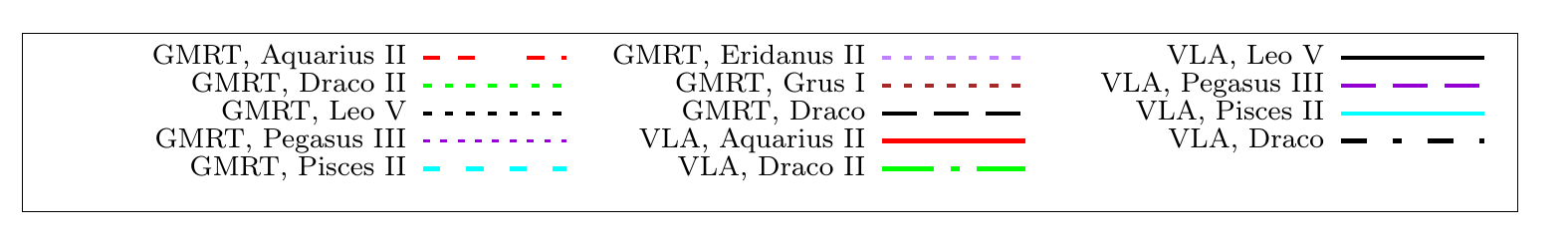}
\caption{\em 95 $\%$ C.L. upper limits on $\langle\sigma v\rangle$ for different values of DM mass obtained using data from GMRT and VLA. The left, center and right
panels are for $\chi \chi \rightarrow b \bar{b}\ , \ \tau^+ \tau^-\
, \ \mu^+\mu^- $ respectively. We have used $B \, = \, 1\, \mu$G, $D_0
= 3 \times 10^{28}$ cm$^2$/s, $\gamma_D = 0.3$ and the NFW density
profile for the DM distribution inside the UFDs, with the values of
the parameters $d$, $r_h$, $\rho_s$ and $r_s$ taken from
Table \ref{table:astro_param_dwarfs}.}
\label{figure:sigmav_Exclusion_curve}
\end{figure}

These limits can, then, be translated to upper limits on $\langle
\sigma v \rangle$ for different annihilation channels, as shown in 
Figure \ref{figure:sigmav_Exclusion_curve}. As with the case of the
gamma-ray observations, no such limit can be derived for Hydra II and
Triangulum II as current observations only admit upper limits on their
NFW density parameter $\rho_s$ (see Table \ref{table:astro_param_dwarfs}).

The comparison between the GMRT and the VLA results should be made
carefully. For a given UFD, the limits obtained from VLA are,
typically, stronger compared to those obtained from the GMRT. This is
particularly true for larger $M_\chi$. This is primarily because at
lower frequencies in which GMRT operates, the rms noise is higher owing to 
radio interference and atmospheric turbulence. 
Additionally, at higher operating frequency the galactic backgrounds
are smaller. On the other hand, at low $M_\chi$, the situation reverses
somewhat. This, again, is a consequence of the dependence of the
synchrotron spectrum on $M_\chi$ and the relative efficiencies of the two
telescopes.

\subsection{Future projections}

\begin{figure}[h!]
 \includegraphics[width=.3\linewidth]{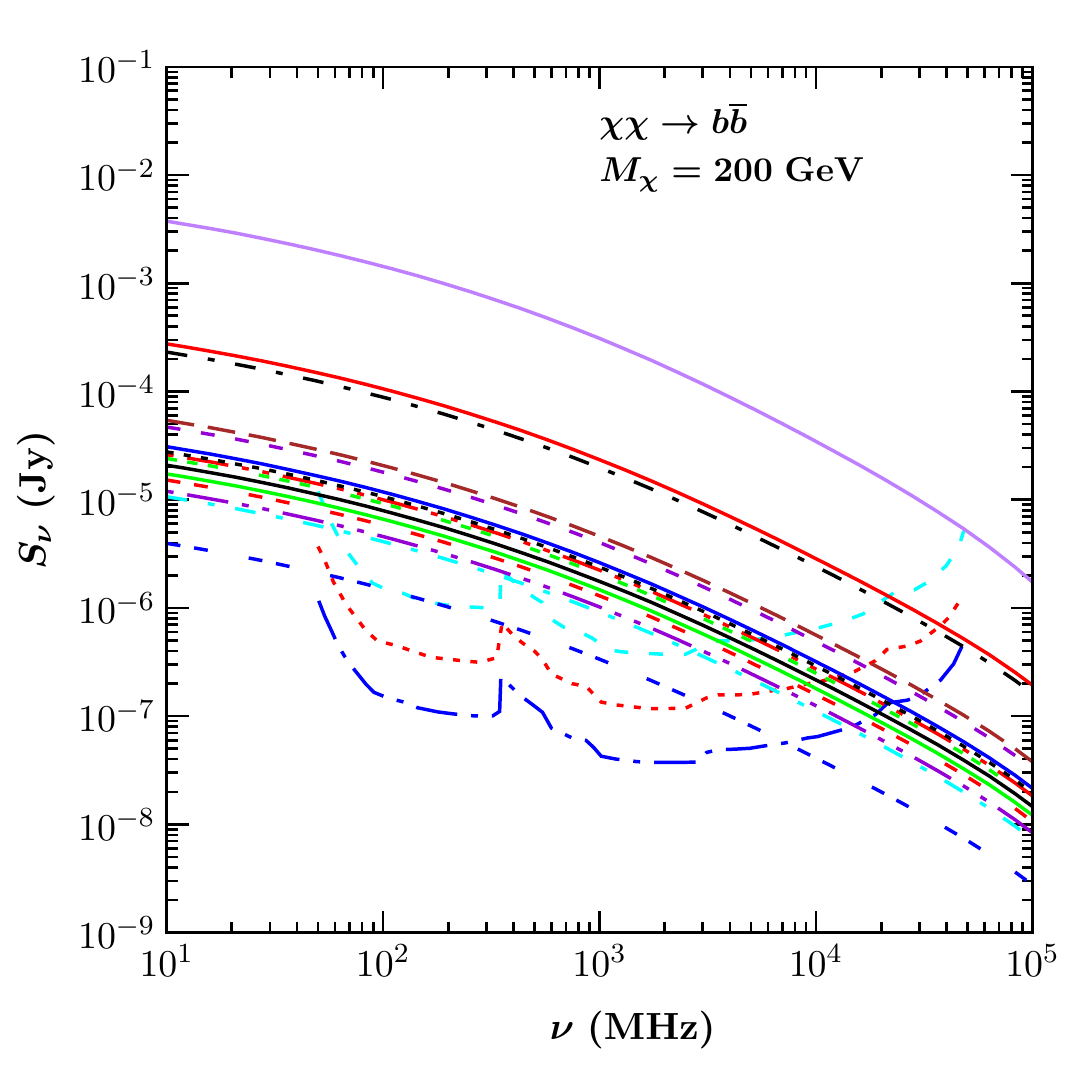}
\hskip 10pt
 \includegraphics[width=.3\linewidth]{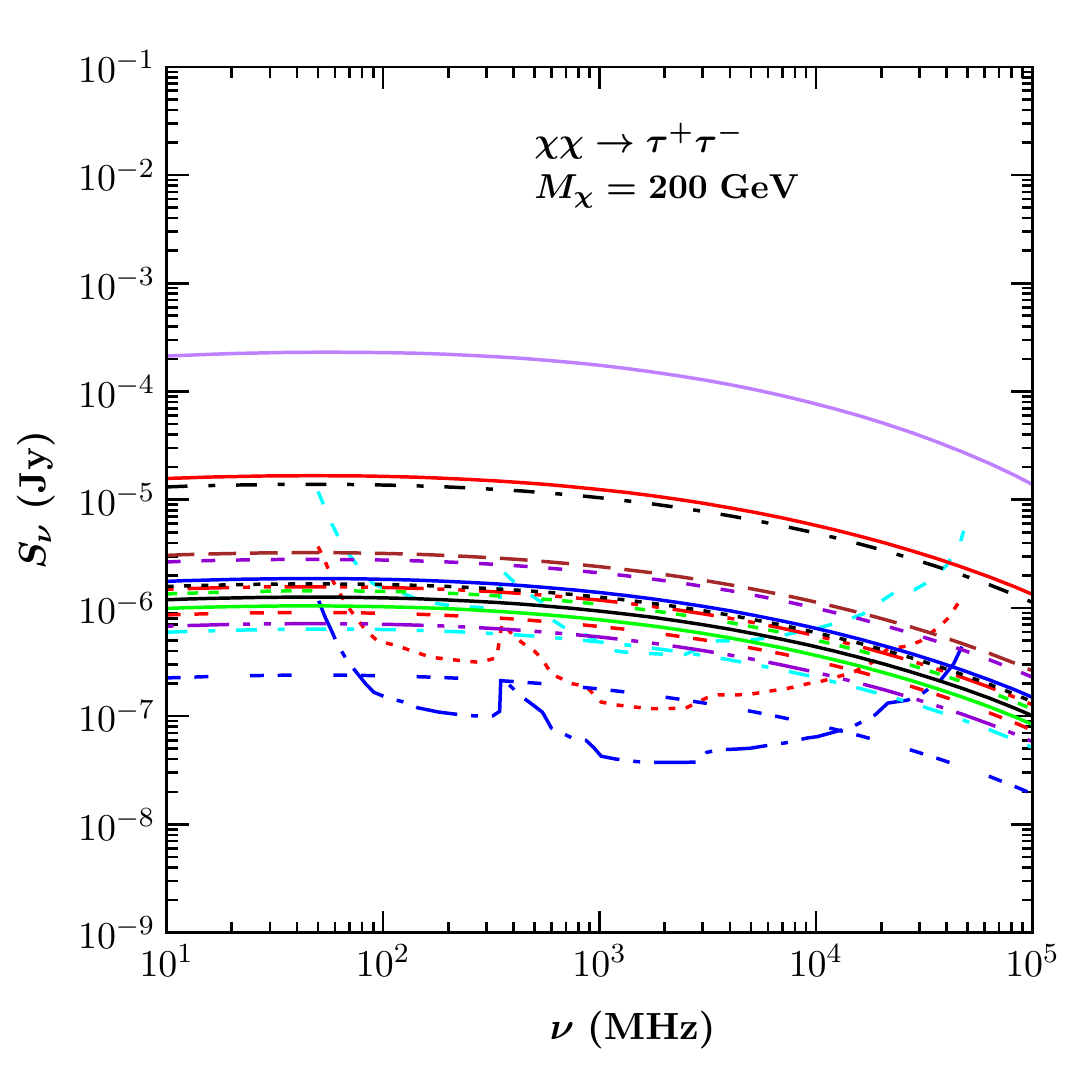}
\hskip 10pt
 \includegraphics[width=.3\linewidth]{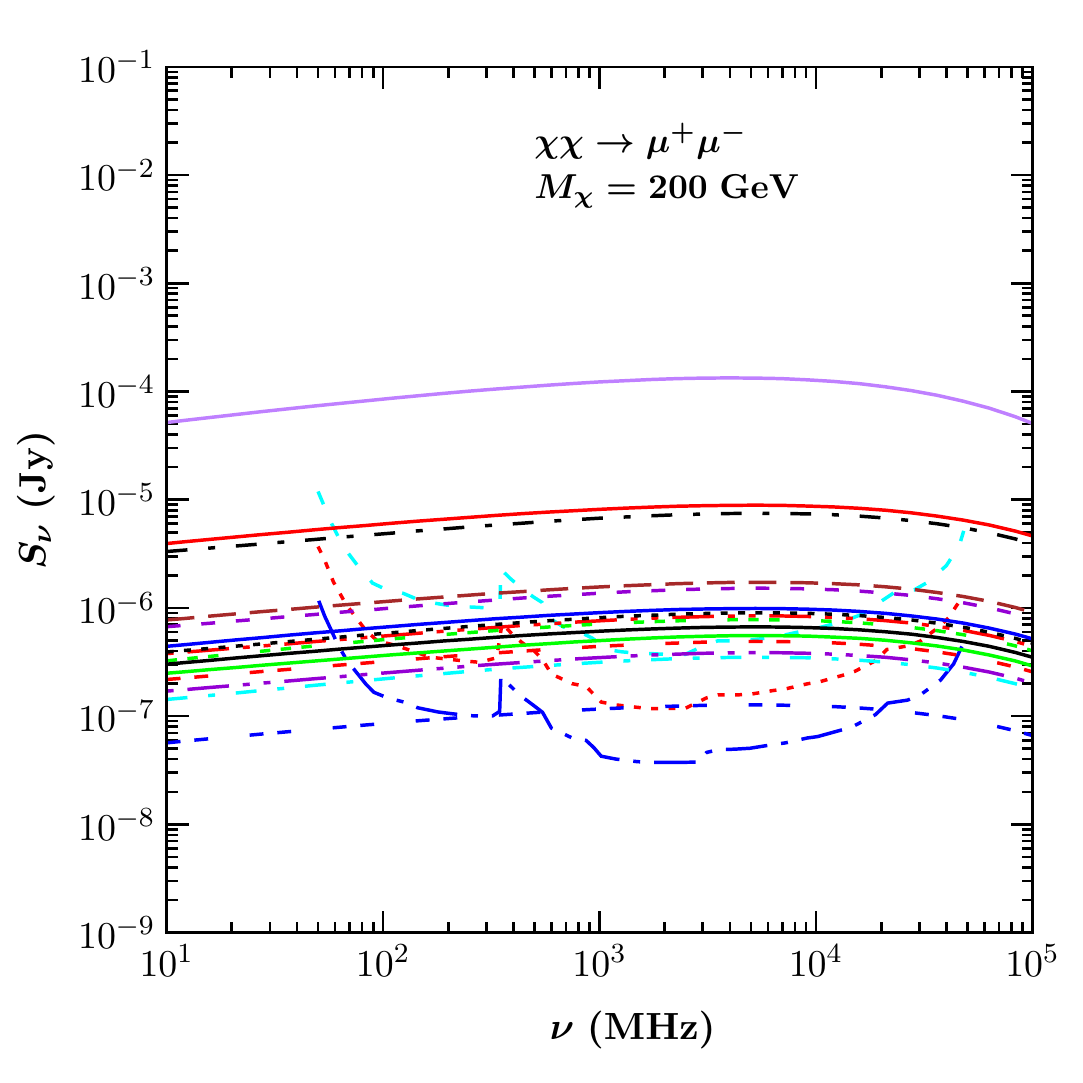}

 \includegraphics[width=.3\linewidth]{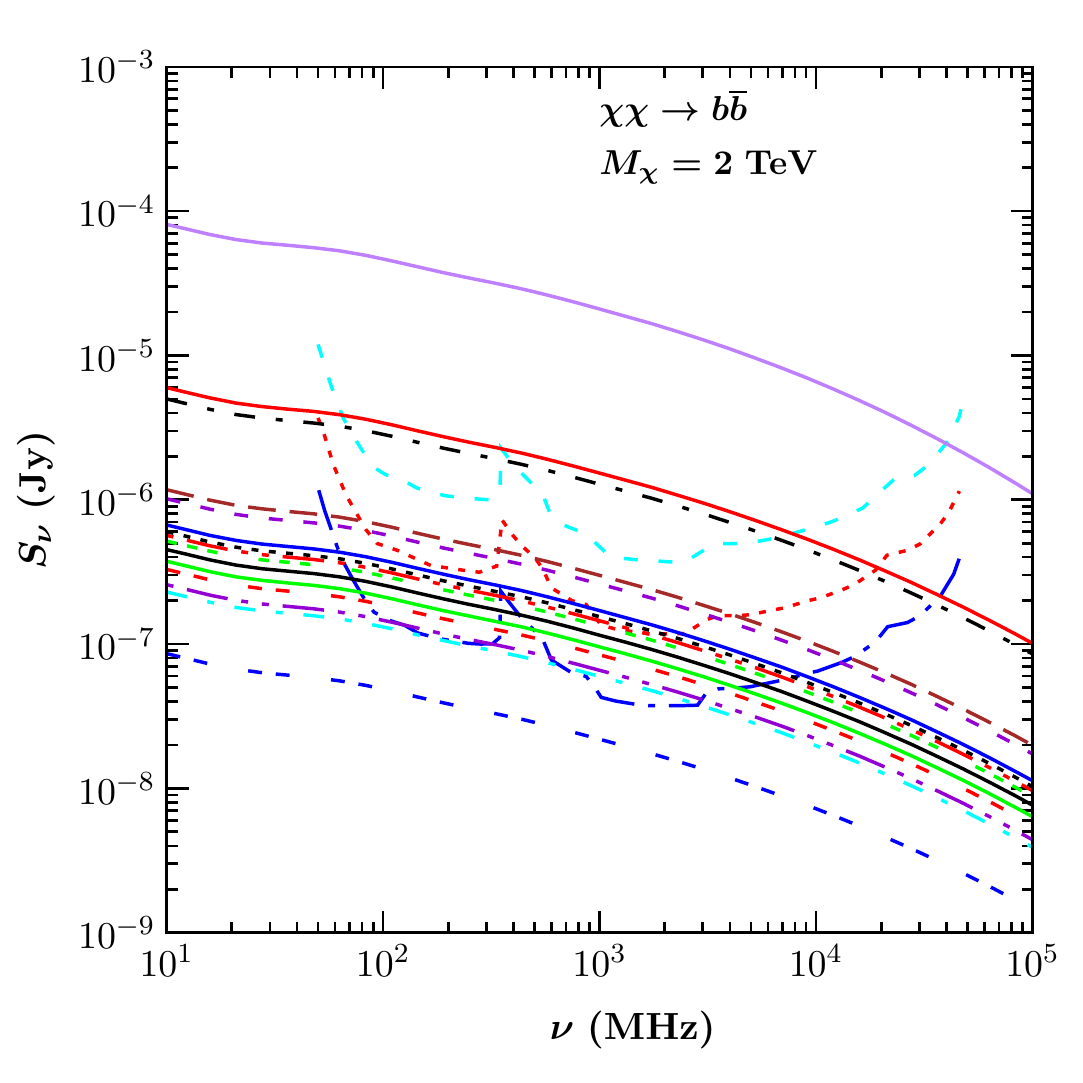}
\hskip 10pt
 \includegraphics[width=.3\linewidth]{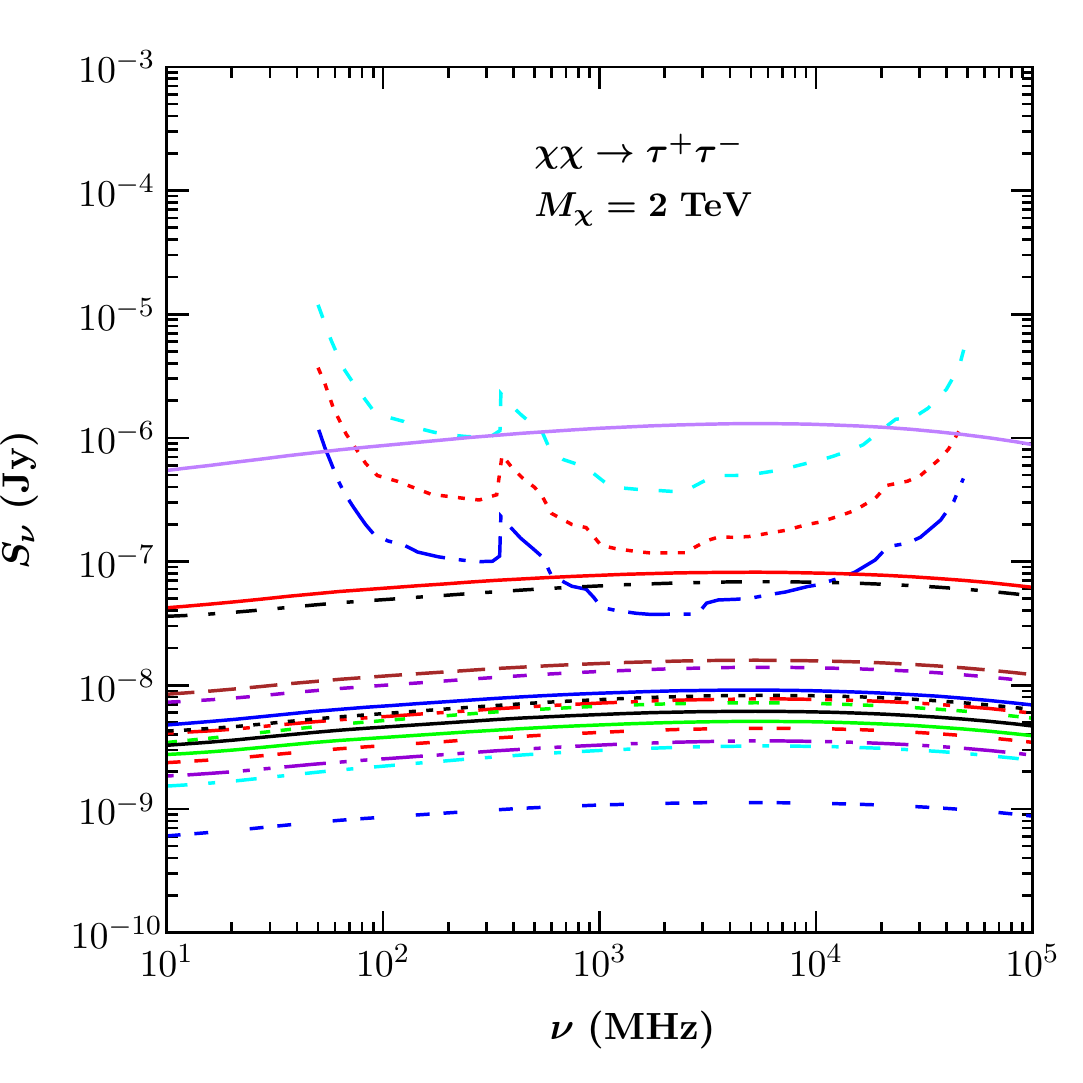}
\hskip 10pt
 \includegraphics[width=.3\linewidth]{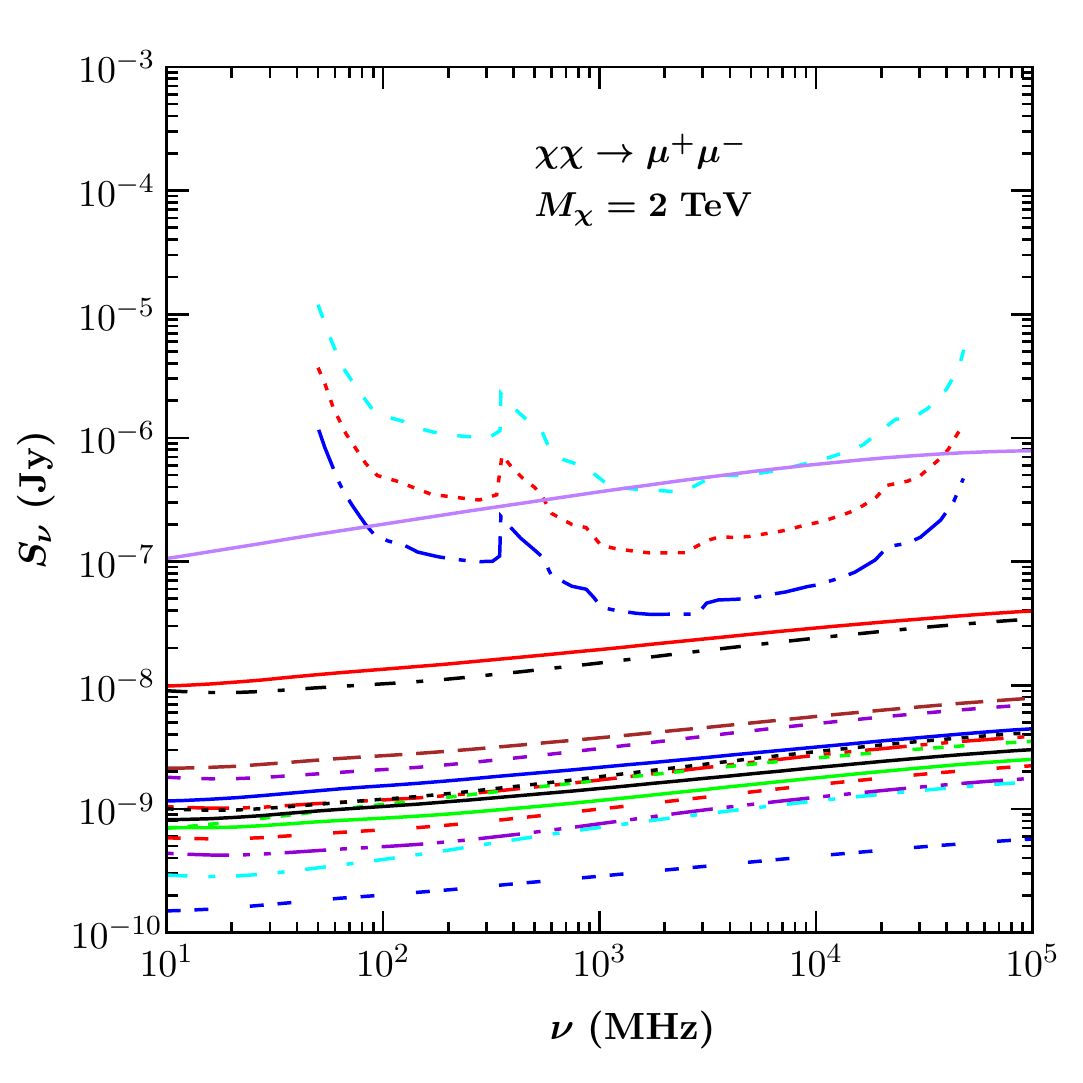}
 \hskip 10pt
 \includegraphics[width=0.9\linewidth]{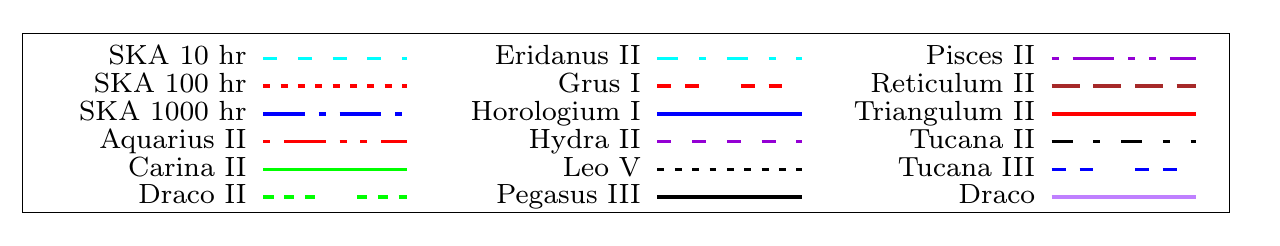}
\caption{\em Synchrotron fluxes from the new galaxies including Draco. The left, center and right
panels are for $\chi \chi \rightarrow b \bar{b}\ , \ \tau^+ \tau^-\ , \ \mu^+\mu^- $ respectively, while the top (bottom) rows correspond to $M_\chi = 200 \GeV (2 \, {\rm TeV})$. 
For each plot $B \, = \, 1\, \mu$G, $D_0 = 3 \times 10^{28}$ cm$^2$/s, $\gamma_D = 0.3$ and $\langle \sigma v\rangle \, = 10^{-26}$ cm$^3$/s have been used.
NFW density profile has been used for DM distribution inside the UFDs. 
For each UFD the values of the parameters $d$, $r_h$, $\rho_s$ and $r_s$  have been used from Table \ref{table:astro_param_dwarfs}~. 
Since upper limits on $\rho_s$ are available for Hydra II, Tucana III and Triangulum II (see Table \ref{table:astro_param_dwarfs} ), 
the corresponding curves for these three UFDs show the maximum possible amount of synchrotron flux .
}
\label{figure:synflux_newgalaxies}
\end{figure}

The Square Kilometer Array (SKA) is the largest radio telescope ever
planned and the search for particulate DM is one of the primary goals
of the project \cite{Bull:2018lat, Colafrancesco:2015ola}.  We exploit
the wide radio frequency range (50 MHz---50 GHz) of the SKA to examine
its sensitivity to synchrotron emissions resulting from DM
annihilations in the UFDs. To facilitate a qualitative comparison, we
present, in Fig.\ref{figure:synflux_newgalaxies}, the resultant
synchrotron fluxes for each of three exclusive annihilation channels
($b\bar{b}$, $\tau^+ \tau^-$ and $\mu^+ \mu^-$). To this end, we have
used a reference value of $\langle \sigma v \rangle $ $=$ $10^{-26}$
$cm^3/s$, consistent with the upper limits obtained earlier.  Once
again, of the UFDs listed in Table \ref{table:astro_param_dwarfs}, we
do not consider Hydra II, Tucana III and Triangulum II as only upper
limits on $\rho_s$ are available for these. It should be noted that
the wide angular coverage of the SKA implies that, unlike in the cases
the GMRT or the VLA, none of the other UFDs need to be discounted.
Also shown in Fig.\ref{figure:synflux_newgalaxies} are the predicted
SKA sensitivities~\cite{braun2017ska,braun2019anticipated}
corresponding to three different exposure times, {\em viz.}  10, 100
and 1000 hours.

As we go to higher DM mass, two effects become more important. For
one, the relic number density would fall, leading to a smaller rate of
annihilation (since it scales as the square of the number density) for
the same annihilation cross section. In addition, very
energetic primaries leads to a harder synchrotron spectrum, thereby
going beyond the sensitivity range of the SKA (just as very low
frequencies are not well-suited for the SKA).  Consequently, the
feasibility of detecting radio signals at the SKA would decrease. As
Fig.\ref{figure:synflux_newgalaxies} demonstrates, the synchrotron
radiation from a 200 GeV DM pair-annihilating to any one of the three
channels $b \bar{b}$, $\mu^+ \mu^-$ and $\tau^+ \tau^-$ from each of
the 12 UFDs would be detectable with just 100 hours of SKA data (for
an individual UFD, however, it could be much less). For $M_\chi =
2$~TeV, though, even with 1000 hours of data, detection is guaranteed
only for the $b\bar b$ channel.  Interestingly, it is only for the
classical dwarf galaxy Draco, that detection is assured for each of
the three channels, and that too with only $\sim 10$ hours of SKA
data. This is in stark contrast to the case of $\gamma$-ray signal at
Fermi-LAT where several of the UFDs proffer better prospects than
Draco.

\section{Astrophysical Uncertainties and the Constraints}
\label{sec:uncertainty}
The constraints derived thus far have all been based on `central'
values of the various astrophysical parameters. In view of the rather
restrictive nature of these constraints and even more so, the apparent
sensitivity of future observatories such as the SKA, it behaves us to
examine how crucially the said constraints depend on the exact values
of the parameters. This assumes particular significance in view of the
fact that, for several of the astrophysical measurements, the
uncertainties are significantly large.

\subsection{Uncertainties in the Gamma-Ray bounds}
\label{section:uncertainties_horo_tuc}

\begin{figure}[h!]
\begin{center}
 \includegraphics[width=0.5\textwidth,clip,angle=0]{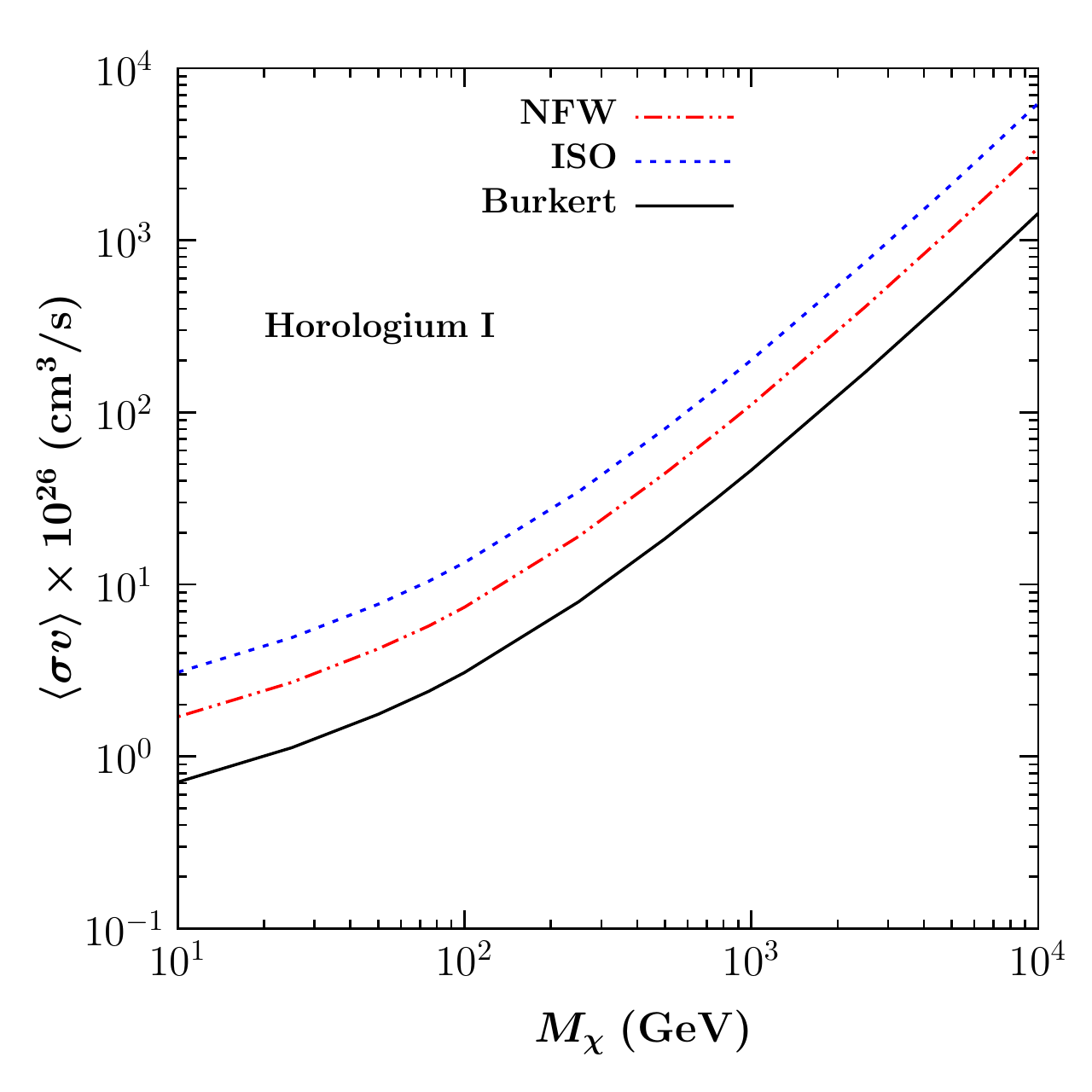}
\end{center}
\caption{\em Comparison between the $\langle \sigma v \rangle $ upper limits for three density profiles for $100\%$ $b\overline{b}$ final state.}
  \label{figure:profile_comparison}
 \end{figure}

A prime source of errors is, of course, the DM distribution. With very
few visible stars in the UFDs, the kinematic data is
sparse\cite{Funk:2013gxa} and this has proved an obstacle in
understanding the DM distribution in UFDs. Indeed, while $N$-body
simulations seem to argue for the NFW distribution, the data has also
been interpreted to favour isothermal
distributions~\cite{de_Blok_2001}. Since Horologium I had led to the
strongest constraints, we choose to illustrate the role of the density
profile for this particular UFD. Using the central values for the
$J$-factor from Table~\ref{table:table-1}, we depict, in
Fig.~\ref{figure:profile_comparison}, the consequent upper limits on
$\langle \sigma v \rangle $ for all three density profiles. For
brevity's sake we only present the results for $b\bar b$ channel
alone, with the understanding that the relative change in the
constraints are entirely analogous for the two other channels.  It is
worthwhile to note that the Burkert profile would have led to stronger
constraints than we have, while the isothermal-like profiles would,
typically, lead to weaker constraints. This feature repeats for most
other UFDs as well.

This dependence can be recast in terms of the uncertainties in the
$J$-factor. Indeed, in the context of the $\gamma$-ray signal,
ambiguities in the density profile can be subsumed, to a great extent,
in uncertainties in the $J$-factor. Indeed, while the $J$-factor
corresponding to the NFW profile gives the median value, a look at
Table~\ref{table:table-1} convinces us that the central values
associated with the two other profiles do fall within the $1\sigma$
band of the former. 
With the $\gamma$-ray flux being proportional to the $J$-factor (see
eqn.\ref{eqn:dm_flux}), it is understandable that a
large error in the latter would translate to correspondingly large
errors in the upper limit on $\langle \sigma v\rangle$.  As the
  dependences are straightforward, we do not discuss these any
  further.


\subsection{Uncertainties in the synchrotron fluxes}
\label{sec:synchrotron_uncertainty}

\begin{figure}[h]
  \begin{subfigure}[b]{0.32\textwidth}
     \centering
  \includegraphics[width=1\textwidth]{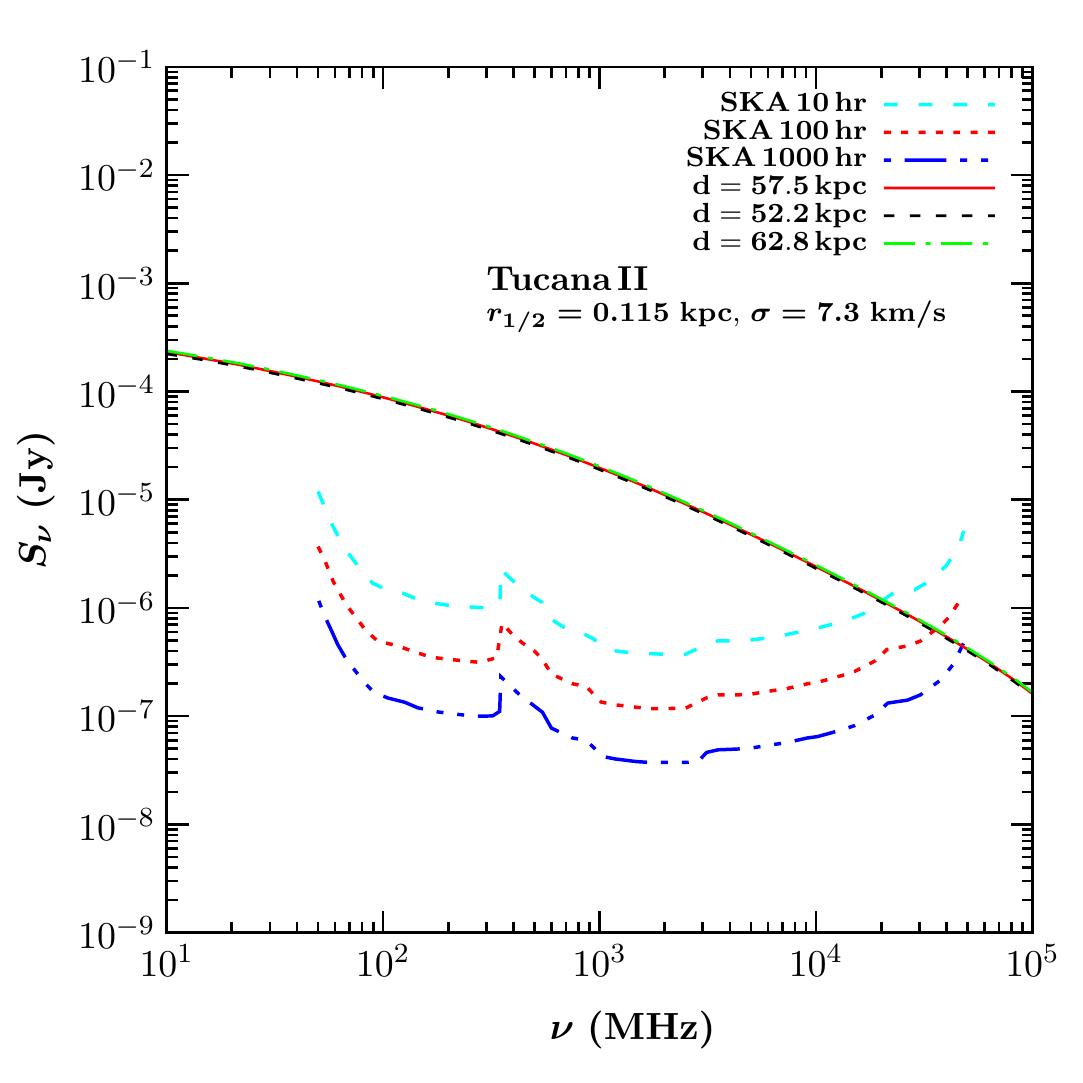}
    \caption{}
    \label{fig:d_uncertainty}
  \end{subfigure}
\hfill
  \begin{subfigure}[b]{0.32\textwidth}
     \centering
  \includegraphics[width=1\textwidth]{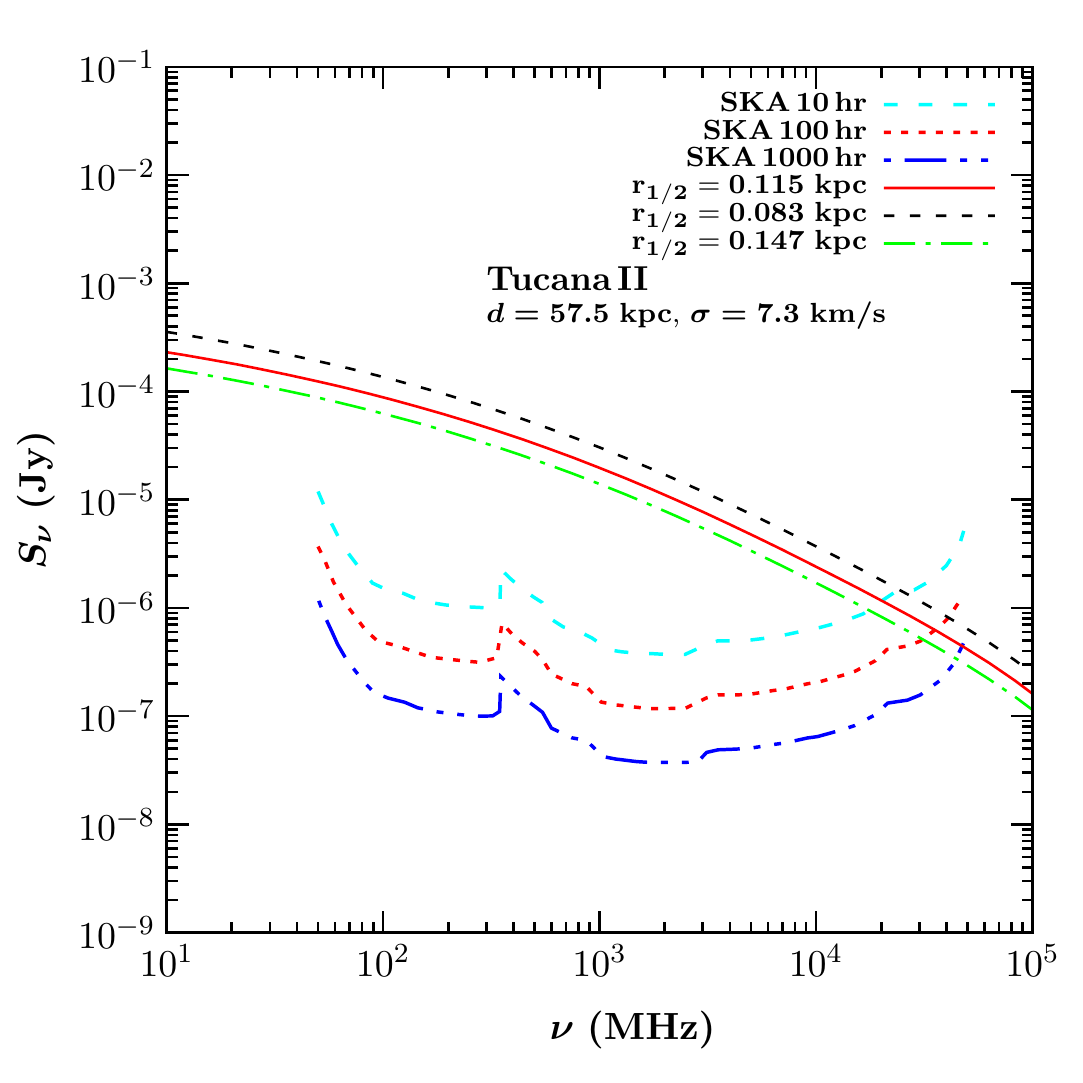}
    \caption{}
    \label{fig:rhalf_uncertainty}
  \end{subfigure}
    \begin{subfigure}[b]{0.32\textwidth}
     \centering
  \includegraphics[width=1\textwidth]{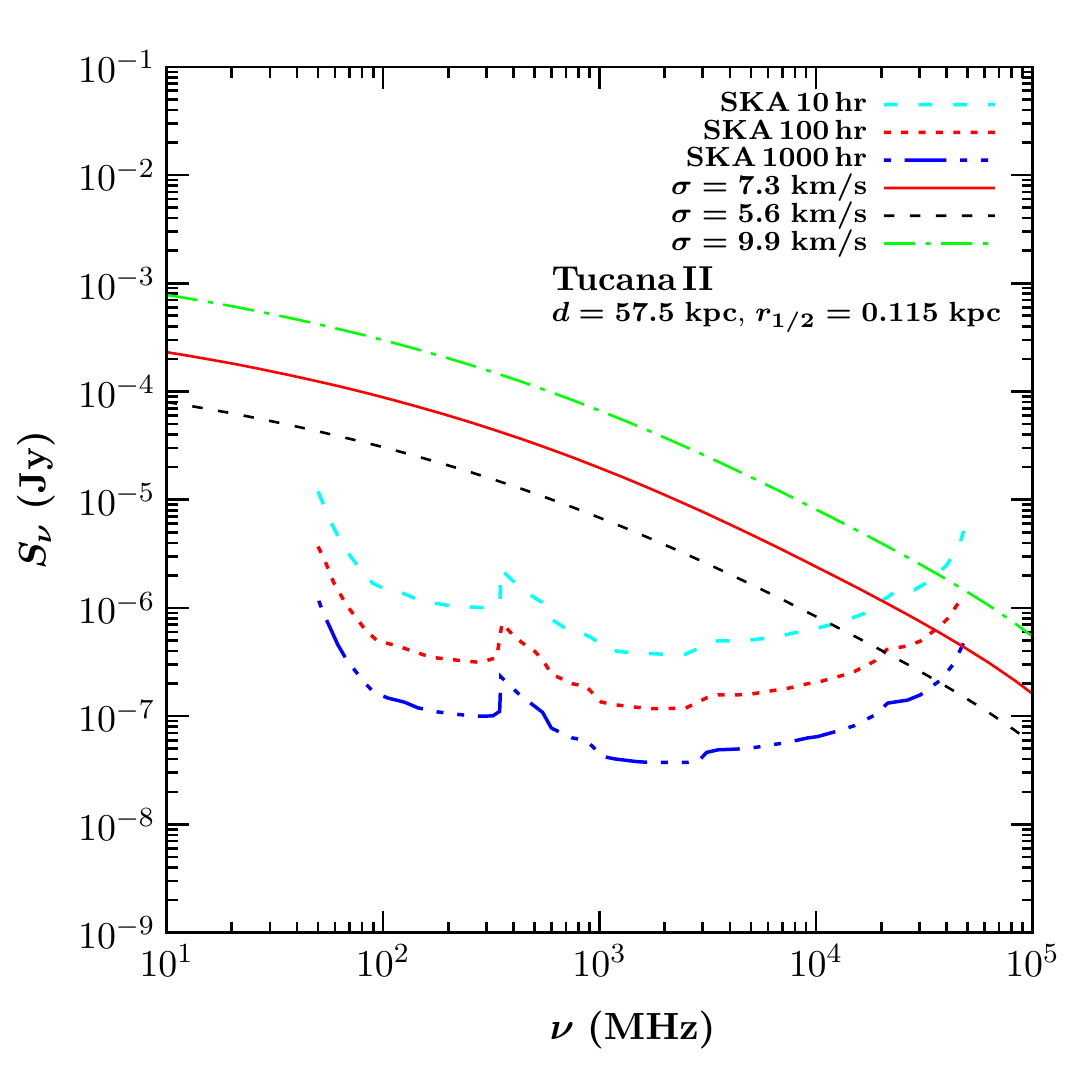}
    \caption{}
    \label{fig:sigma_uncertainty}
  \end{subfigure}
\caption{\em  Individual contributions to the uncertainty in the
  synchrotron flux---resulting from a pair of 200 GeV DM particles in
  Tucana II annihilating into $b\bar b$---due to 1$\sigma$ uncertainties in
  the parameters {\em (a)} $d$, {\em (b)} $r_{\frac{1}{2}}$ and {\em (c)} $\sigma_{l.o.s}$. 
  We use the NFW profile along with $B \, = \, 1\,
  \mu$G, $D_0 = 3 \times 10^{28}$ cm$^2$/s, $\gamma_D = 0.3$ and
  $\langle \sigma v\rangle \, = 10^{-26}$ cm$^3$/s.  }
   \label{figure:uncertainty}
\end{figure}

As with the case for the gamma-ray fluxes, the synchrotron fluxes too
are subject to uncertainties on account of the errors in determining
the astrophysical parameters $d$ (distance to the UFD), $r_{1/2}$
(its half-light radius) and the velocity-dispersion $\sigma_{l.o.s}$. Using
the $1\sigma$ uncertainties in these, as listed in Table
\ref{table:astro_fundamental_param_dwarfs}, we show, in Fig.
\ref{figure:uncertainty}, the consequent uncertainties in the
synchrotron flux for a 200 GeV DM annihilating to $b\bar{b}$ final
state in Tucana II. The choice of the particular UFD to demonstrate
these effects is motivated by it being the one associated with the
highest synchrotron flux. While the actual error would be a
combination of all three contributions, in the absence of adequate
information regarding the correlations between the same, we have not
chosen to attempt this convolution. It should, nonetheless, be obvious
that with the relative error in $d$ already being small, a further
improvement in this measurement is not crucial. With the relative
error in the $r_{1/2}$ measurement being larger, this variable, quite
expectedly, contributes more to the error. Understandably, much more
crucial is the determination of the velocity dispersion, both on
account of the current level of accuracy in its measurement and the
particularly important role that this variable plays in determining
the synchrotron spectrum.

\begin{figure}[h!]
\begin{center}
 \includegraphics[width=.49\linewidth,height=3in]{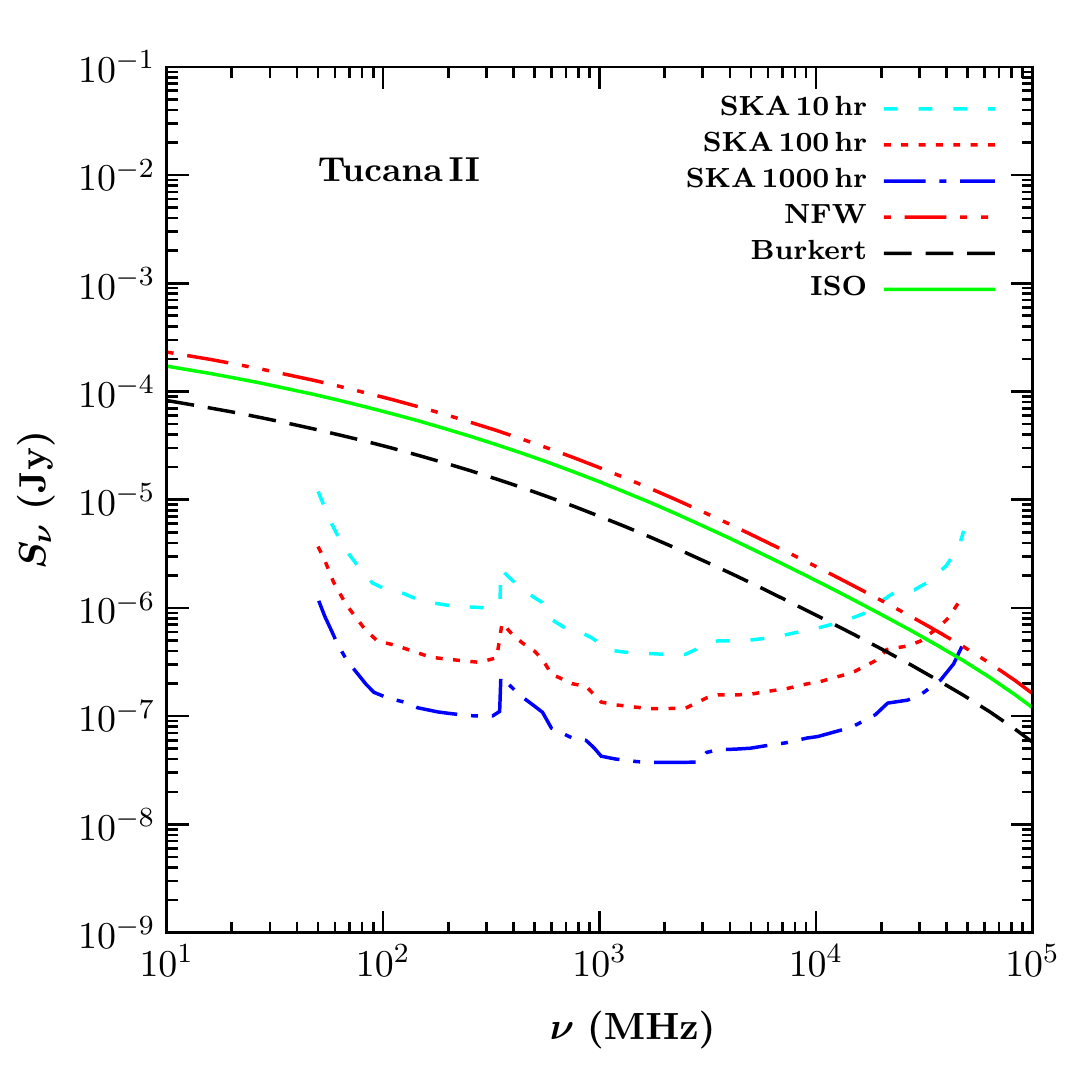}
\end{center}
 \caption{\em Synchrotron flux vs. frequency for Tucana II for different dark matter distribution profiles : NFW, Burkert and ISO.
   We have considered a 200 GeV DM annihilating to $b\bar{b}$ final state,
   as also 
 $B \, = \, 1\, \mu$G, $D_0 = 3 \times 10^{28}$ cm$^2$/s, $\gamma_D = 0.3$ and $\langle \sigma v\rangle \, = 10^{-26}$ cm$^3$/s.}
 \label{figure:sync_profile_dependence}
 \end{figure}

As with the gamma-ray fluxes, the lack of precise knowledge of the DM
density distribution leads to substantial uncertainties.  While we
have, until now, used the NFW profile, in
Fig. \ref{figure:sync_profile_dependence} we display the differences in
predicted flux for Tucana II for NFW, Burkert and ISO profiles. It is
interesting to note that while, for the gamma-ray flux, the NFW
profile closely corresponded to the median prediction, in the present
context it leads to the highest fluxes, while the Burkert profile
leads to the lowest. Fortunately enough, the difference between the
predictions for theoretically favoured NFW profile and the
observationally favoured isothermal profiles is not very significant,
thereby maintaining the robustness of our results.

\begin{figure}[!h]
   \begin{subfigure}[b]{0.45\textwidth}
  \centering
  \includegraphics[width=\textwidth]{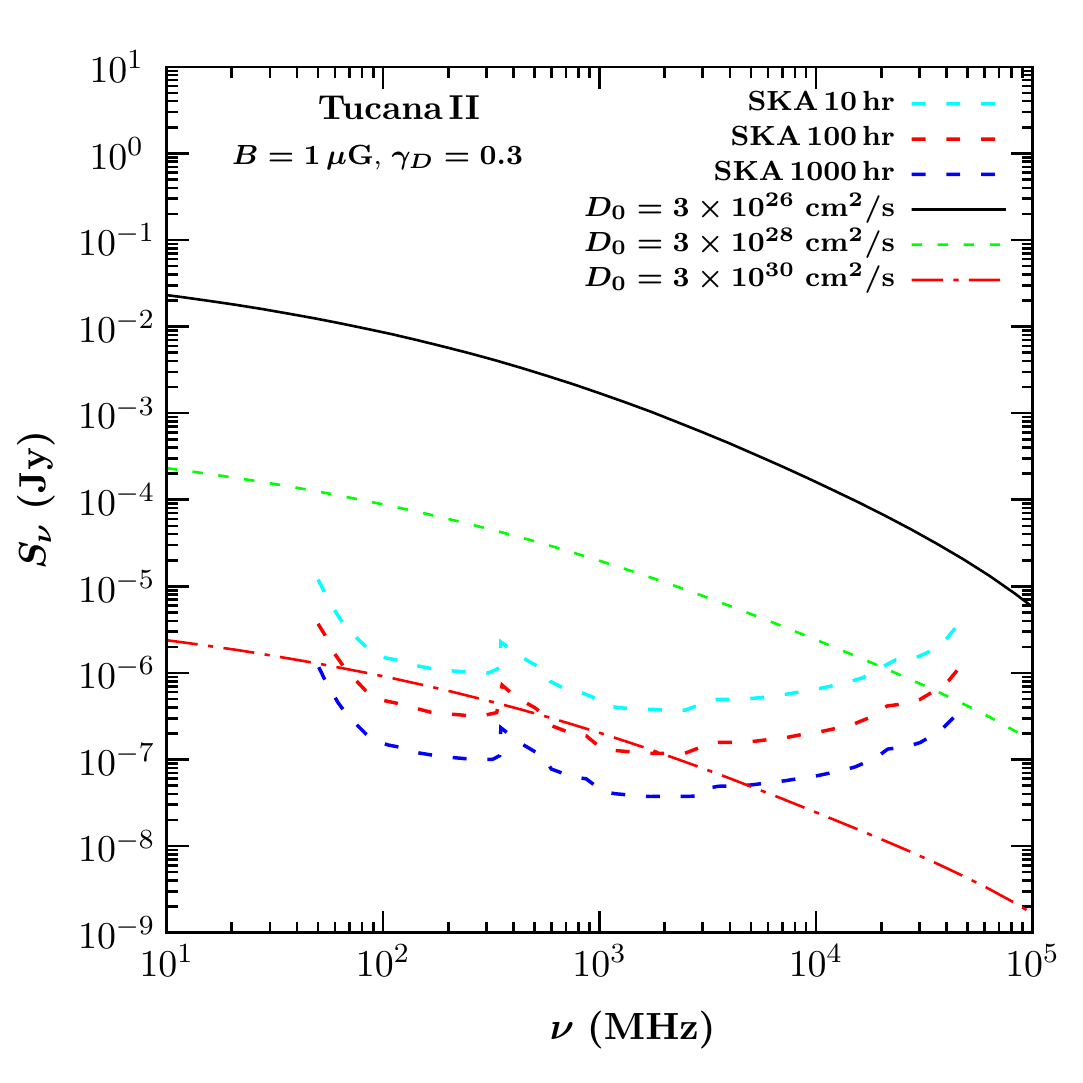}
  \caption{}
   \label{fig:flux_different_D0}
  \end{subfigure}
     \begin{subfigure}[b]{0.45\textwidth}
  \includegraphics[width=\textwidth]{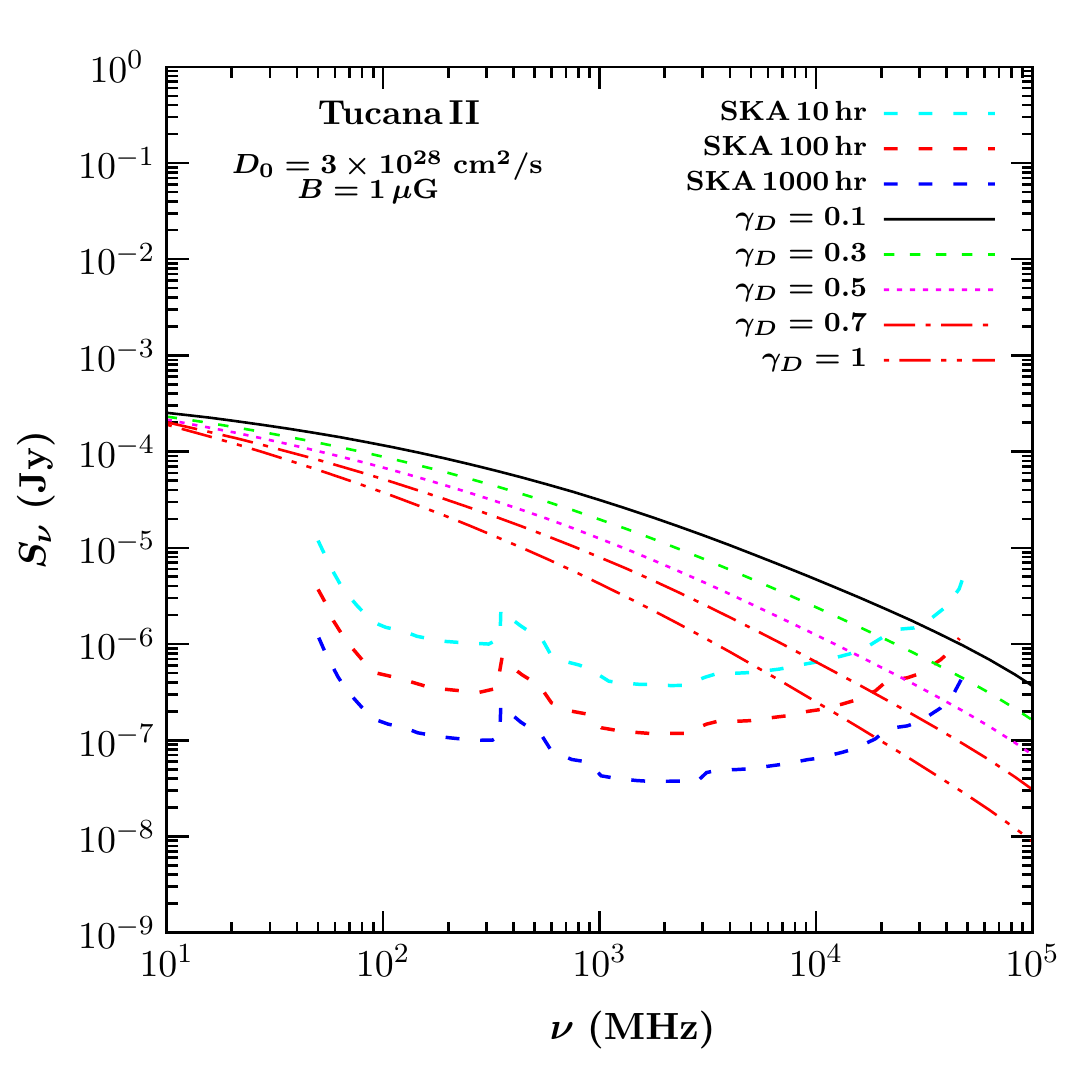}
  \caption{}
  \label{fig:flux_different_gamma}
  \end{subfigure}
     \caption{\em Synchrotron flux versus frequency for different
       values of  {\em (a)} $D_0$ and
       {\em (b)} $\gamma_D$.  The mass of
       DM is 200 GeV and the DM pair annihilates to $b\bar{b}$ final
       state with $\langle \sigma v\rangle \, = 10^{-26}$ cm$^3$/s.
       NFW density profile has been used for DM distribution inside
       the UFD.}
   \label{fig:flux_diff_D0gamma}
\end{figure}

Much more crucial is the dependence on the diffusion parameter $D(E)$,
which too is not known precisely.  In Section \ref{sec:synchr}, we had
indicated the typical ranges of the diffusion constant $D_0$ and
exponent $\gamma_D$, but effected our numerical analysis with
`central' values alone.  In view of the fact even the probability
distributions for these parameters are not well known, ``$1\sigma$''
intervals are not well-defined.  Hence, in
Fig.\ref{fig:flux_diff_D0gamma}, we plot the synchrotron fluxes for
the entire plausible ranges of these parameters, $D_0 \in [3\times
  10^{26}, 10^{30}]$~cm$^2$/s, and $\gamma_D \in [0.1,1]$.  A smaller
value for $D_0$ would allow the $e^\pm$ to see the magnetic field of
the UFD for a longer duration resulting in a larger synchrotron flux.
The overall dependence is close to being a linear one (see
Fig.~\ref{fig:flux_different_D0}). As for $\gamma_D$, the very
definition (eqn. \ref{eqn:diffusion_coefficient}) stipulates that a
larger value enhances (suppresses) $D(E)$ for $E(e^\pm) > 1 \, (< 1)$.
Since high-frequency synchrotron photons would preferentially be
radiated by high-energy $e^\pm$ (see Fig.~\ref{fig:synpower}), a
larger $\gamma_D$ would progressively suppress the synchrotron flux at
higher frequencies; similarly, substantially below $\nu \sim 1$~MHz,
the flux would be enhanced.  The region $\nu \in (1, 5)$~MHz roughly
corresponds to the peak for $E(e^\pm) = 1$~GeV (see
Fig.~\ref{fig:synpower}) and, understandably, the effect is relatively
small in this region (see Fig.~\ref{fig:flux_different_gamma}).

\begin{figure}[h]
 \begin{subfigure}[b]{0.45\textwidth}
  \centering
  \includegraphics[width=\textwidth]{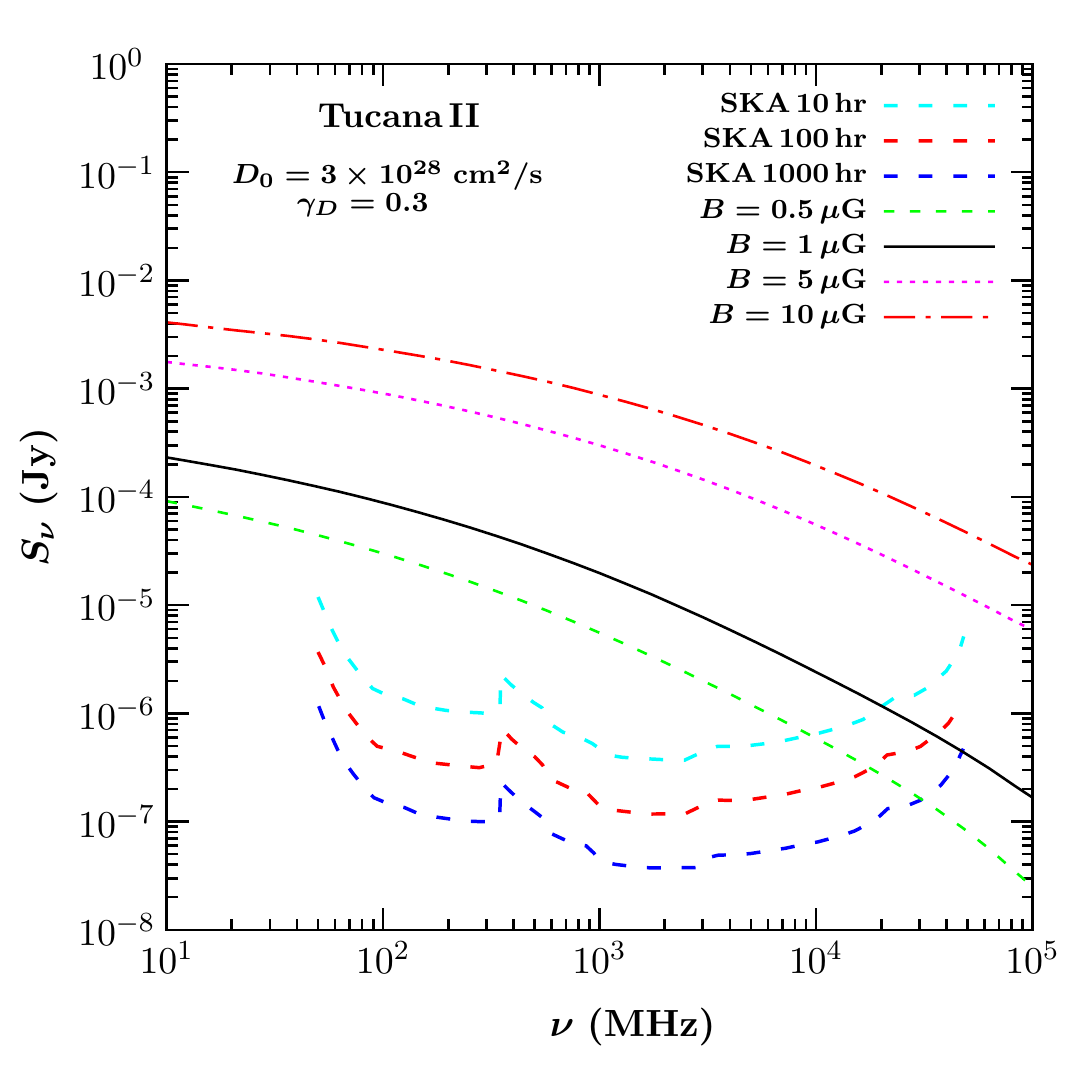}
  \caption{}
  \label{fig:flux_different_B}
  \end{subfigure}
   \begin{subfigure}[b]{0.45\textwidth}
  \centering
  \includegraphics[width=\textwidth]{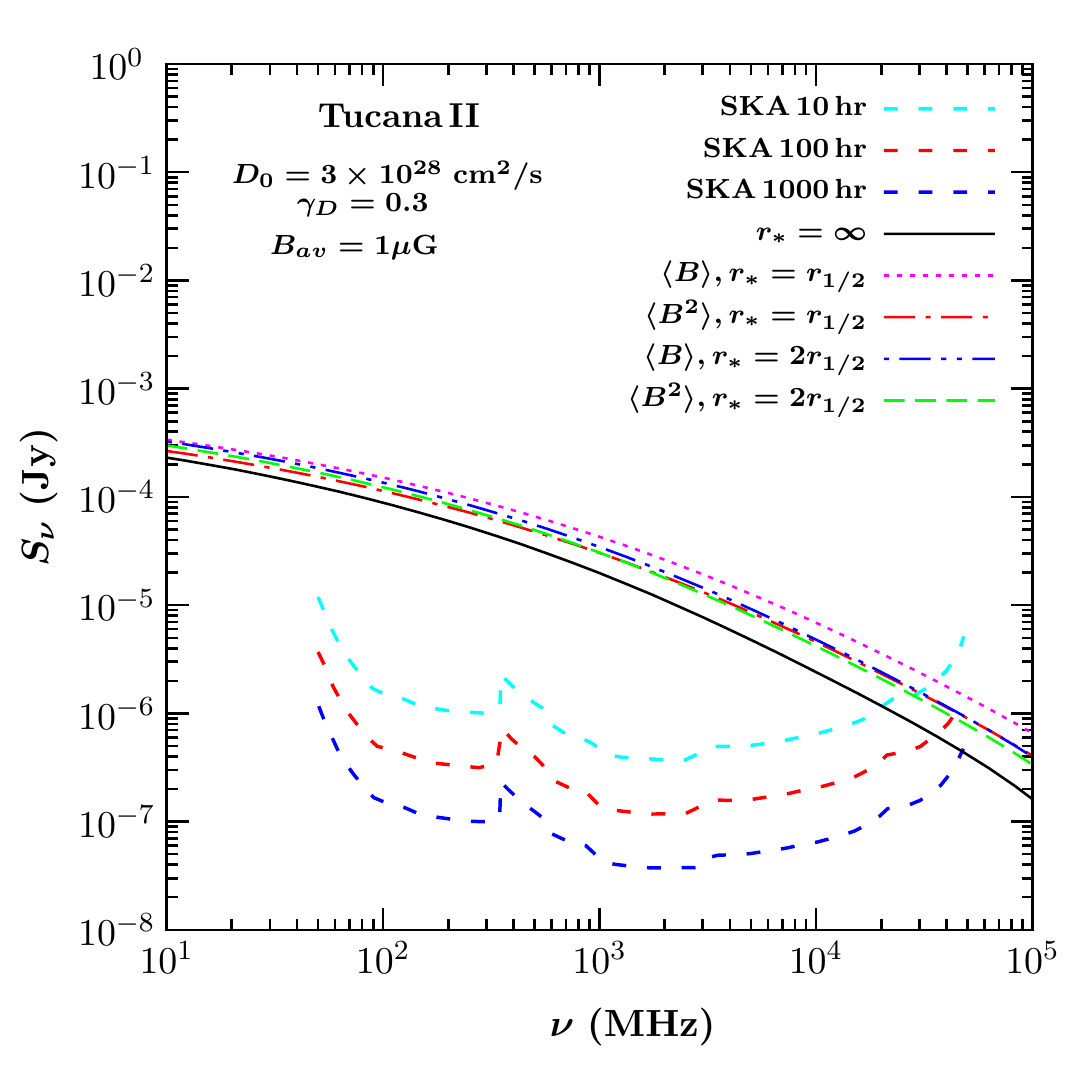}
  \caption{}
   \label{fig:flux_B_profiles}
  \end{subfigure}
   \caption{\em Synchrotron flux as a function of frequency, for $m_\chi = 200\GeV$ with the DM pair annihilating to $b\bar{b}$ final state with
     $\langle \sigma v\rangle \, = 10^{-26}$ cm$^3$/s.  The NFW density
     profile has been used for DM distribution inside the UFD. 
     {\em (a)} For flat magnetic fields of different strengths.
     {\em (b)} Assuming a exponential profile for $B$
     maintaining an average (see text) of $1\mu$G.}
   \label{fig:flux_diff_B}
\end{figure}

Finally, we come to the elephant in the room, namely the size and the
profile of the magnetic field. In Section \ref{sec:synchr}, we had
discussed the uncertainties in these and indicated the typical ranges of
the field $B$. Once again, eschewing ``$1\sigma$'' intervals,
and concentrating first on a flat magnetic field,
we plot, in Fig.\ref{fig:flux_different_B}, 
the synchrotron fluxes for the entire plausible range {\em viz.}, 
$B \in [0.5,10]\,\mu$G. With the magnetic
field driving the synchrotron radiation, the strong dependence of the
flux (see Fig.~\ref{fig:flux_different_B}) is understandable. Indeed,
for small frequencies, the flux is nearly proportional to $B$, with
the dependence growing significantly stronger at larger frequencies,
an effect that is easily understood in term of classical
electrodynamics.

While a majority of DM studies have considered a flat magnetic
  field profile within a dSph, the truth could be very different
  indeed. As discussed in Section \ref{sec:synchr}, a popular choice
  (in the absence of more precise data or simulations) is to consider
  an exponential profile, {\em viz.}  $B = B_0 \, e^{-r/r_*}$ with
  $r_*$ being of the order of the half-light radius of the dSph
  \cite{McDaniel:2017ppt, Regis:2014koa, Cembranos:2019noa}. Assuming
  such a profile, we study next the consequences for the same. To
  compare the results to those for a flat profile $B = B_{\rm av}$,
  one needs to normalize appropriately and we do this by equating the
  average values, with the average taken over the entire diffusion
  zone, {\em viz.}  $0 \leq r \leq r_h$. It should be obvious that, in
  view of the damping of the field, ${\cal O}(1)$ changes in the upper
  limit is of little consequence. Much more important is the value of
  $r_*$ (and, here, we choose two representative values, namely $r_* =
  r_{1/2}$ and $r_* = 2 r_{1/2}$) and the definition of the
  averaging. While equating the magnetic energies would require that $
  \langle B^2\rangle \equiv V^{-1} \int d^3 r B^2(r) = B^2_{\rm av}$
  (with $V$ being the volume of the dSph), some analyses have used
  $\langle |B|\rangle = B_{\rm av}$ instead, and hence we display
  results for both the definitions.  As Fig.\ref{fig:flux_B_profiles}
  shows, for each such profile, the resultant flux would be larger
  than that obtained for the flat profile (namely, $r_* = \infty$)
  with the same average value.  The reason is easy to divine. In each
  case, the central region of the galaxy 
   is associated with a field larger than the average
  value. With the central region also being characterised by a larger
  DM (the progenitors of the charged particles) number density, it
    is conceivable that such
  profiles would lead to larger synchrotron fluxes.    Such arguments, applied naively,
  would indicate that a smaller value for $r_*$ should lead to a
  larger flux. However, it needs to be borne in mind that a very
  fast-falling magnetic field would also imply that only a small
  volume contributes to the flux. With a nontrivial energy dependence
  of $d n_e^\pm/d E$ and an even more complicated form of the
  synchrotron power spectrum (see Sec.\ref{sec:synchr}), the resultant
  convolution should not be expected to follow the aforementioned
  naive argument. Nonetheless, it is well-satisfied in the case of
  averaging $\langle |B|\rangle$. If the averaging algorithm
  involves $\langle B^2 \rangle$, the dependence of $S_\nu$ on the value of $r_*$
  is relatively muted, and there is a crossover at $\nu \sim 1$~GHz.


\section{Discussions and Conclusions}\label{section:conclusion}
Dwarf spheroidal galaxies, apart from being dominated by dark matter,
have almost no millisecond pulsars, and yet possess considerable large
scale magnetic fields. These render them ideal aren\ae\ for indirect
detection of dark matter through photonic signals.  Several attempts
have been made recently to derive strong bounds on annihilation cross
sections of dark matter using both gamma rays and radio
observations \cite{Abdallah:2020sas, Alvarez:2020cmw, HAWC:2019jvm, Halder:2019pro, Oakes:2019ywx, Rinchiuso:2019etv, Rico:2020vlg, Bhattacharjee:2018xem, Hoof:2018hyn, DiMauro:2019frs, Fermi-LAT:2016uux, Beck:2019ukt, Regis:2017oet, Regis:2014tga}. Most of the early
observations concentrated on the dwarf spheroidal galaxies of the
Local group. Our work has been primarily motivated by the new class of
UFDs discovered in the past years using data from Pan-STARRS, Dark
Energy Survey and a few other surveys.  In this paper, we have studied
such signals in two distinct parts of the electromagnetic spectrum,
namely high energy gamma rays on the one hand and radio signals on the
other.  We have performed a search for signatures of particle DM
annihilations in 15 UFDs using data from the Fermi-LAT. Additionally
we have also searched for synchrotron radiation from high-energy electrons
and positrons originating from DM annihilations from these objects
using GMRT and VLA. We have also estimated the spectra of the radio signals
and compared their detectability with the upcoming Square Kilometer
Array (SKA) in the frequency range 50 MHz to 50 GHz.

Eleven years of gamma-ray data collected by the Fermi-LAT
collaboration has failed to show any significant excess from any of
the fifteen ultra faint dwarf spheroidal galaxies. The stringent upper
limits on the $\gamma$-ray flux that these null-results imply are
translated to upper limits on the thermal average of the pair
annihilation cross section $\langle\sigma v\rangle$, for a given value
of DM mass. We have done this for 12 of the UFDs. As for the remaining three
(viz. Triangulum-II, Hydra-II and Tucana-III), since only upper limits
on the corresponding (astrophysical) $J$-factors are available, it is
not possible to use these three to obtain robust upper limits on
$\langle \sigma v \rangle$

The DM could, of course, annihilate into a plethora of channels, with
the relative probabilities having a strong dependence on the
underlying dynamics governing the system. Rather than be slave to a
particular model, we consider instead that the pair annihilation
proceeds {\em exclusively} through one of three channels, {\em viz.},
$b \bar b$, $\tau^+\tau^- $ and $\mu^+ \mu^-$. The ensuing independent
constraints can, then, be combined to constrain the parameter space of
any model where the annihilation is dominated by one of the three
channels.  Indeed, simple scaling arguments can also be used to infer
bounds applicable for annihilations into other light fermions.

As far as the gamma-ray signal is concerned, of all the dwarf
galaxies, Holorologium-I imposed the most stringent limits, and, for
the $b \bar b$ channel, easily outdoes the limits obtained by the
Planck collaboration \cite{Aghanim:2018eyx}.  It needs
to be said here that the limits obtained by us have a significant
dependence on the DM density profile. While most of our results were
derived using the NFW profile, we have considered the Burkert and ISO
profiles as well. In general the Burkert profile led to the strongest
bounds on $\langle \sigma v\rangle$ and the ISO the weakest. Our
limits, thus, represent, typically, the middle-of-the-road
constraints.
We also performed a stacking analysis where we have generated a joint 
likelihood function as a product of individual likelihood functions of UFDs. 
We found that the stacking analysis improves the combined limits modestly 
as compared to the limits obtained from the few best individual UFDs alone.
For brevity's sake, we do not present here a figure for the same.

A further source of photons is the synchrotron radiation, largely from
electrons/positrons resulting from the aforementioned DM annihilations
(produced either directly or as the end product of cascades for
annihilations into heavier Standard Model particles). Emitted as the
$e^\pm$ travel through the magnetic field of the UFDs, the synchrotron
radiation, typically, falls in the radio frequency range.  We have
calculated the upper limit on DM annihilation cross section for
different annihilation final states, i.e., $b\bar{b}$, $\tau^+ \tau^-$
and $\mu^+ \mu^-$ for the UFDs using the available radio flux upper
limits from the GMRT and VLA telescope data.  The rms sensitivity of
the observations range from about a few mJy for GMRT at 147 MHz to few
hundreds of $\mu$Jy for VLA at 1.4 GHz for all the targets of
observations.  Comparing these results with those obtained from the
Fermi-LAT data, we have found that it is the VLA that imposes the
strongest constraints. This holds for a wide range of DM masses, not
merely the representative values of 200 GeV and 2 TeV that we have
used to depict our results.

Turning to the near future, we consider the synchrotron fluxes , over
a wide frequency range (10 MHz to 100 GHz) for a low annihilation
cross section ($\langle \sigma v \rangle = 10^{-26}$ cm$^3/$s) and a
low magnetic field ($B = 1 \mu$G) and a typical diffusion coefficient
($D_0 = 3\times 10^{28}$ cm$^2/$s) and index ($\gamma_D =
0.3$). Comparing these with the (frequency-dependent) sensitivity of
the SKA telescope, we have found that for a DM of mass 200 GeV
pair-annihilating into $b \bar b$, even a 10 hour exposure would be
enough to detect a signal from each of the UFDs. For annihilations
into $\tau^+\tau^-$ or $\mu^+ \mu^-$ pairs, the fluxes are lower, but
100 hours worth of data should suffice. On the other hand, for much
heavier DM (say, a mass of 2 TeV), the number density is smaller, and,
furthermore, the spectrum of the synchrotron radiation is
harder. Consequently, a significantly longer exposure would be
needed. For example, for annihilations into a $b\bar b$ pair, a 10
hour exposure would now suffice only for Triangulum II and Tucana II
(apart from the classic Draco, of course), with 100 hours being
sufficient for the rest. On the other hand, for such heavy DM
annihilating into $\tau^+ \tau^-$ or $\mu^+ \mu^-$, even an exposure
of 1000 hours would be barely enough.

Such quantitative conclusions depend, of course, on the annihilation
rate $\langle \sigma v \rangle$ and it should be appreciated that we
have assumed values significantly lower than that allowed
currently. Also to be remembered is that the fluxes have significant
dependences on many astrophysical parameters that are, yet, not very
well-measured. Investigating the effects of these, we conclude that
the strength of the magnetic field (inside the UFD) and the diffusion
coefficient $D_0$ are the parameters that need to be better measured
with the velocity dispersion of the DM being the other critical
observable.  
Unlike high energy gamma-rays, these synchrotron radiations are
  strongly dependent on the magnetic field and the diffusion
  coefficient via the diffusion process of the electrons through the
  inter-galactic medium.   The current knowledge of
  magnetic field and the diffusion coefficient for the UFDs are not
  very precise.  We have studied the dependence of synchrotron flux on $B$,
  $D_0$ and $\gamma_D$ for a 200 GeV DM annihilating to $b\bar{b}$
  final state in Tucana II. We have found that the flux changes
  significantly with the variation of these parameters.
Moreover, the amount of
synchrotron flux depends on the choice of DM profile for a given UFD.
For Tucana II we compared the flux for NFW, Burkert and ISO profiles,
and found that NFW profile provides largest amount of flux among the
three.
For the newly discovered UFDs, large uncertainties are present in the
measured values of the astrophysical parameters such as the half-light
radius ($r_{1/2}$), heliocentric distance ($d$) and velocity
dispersion ($\sigma_{l.o.s}$).  The radius of the diffusion zone ($r_h$) and
the parameters of the DM profile also depend on these astrophysical
parameters, hence affecting the synchrotron flux from a UFD.  We have
studied the uncertainty in synchrotron flux due to uncertainty in
these parameters.  We have found that uncertainty in flux is largest due to
uncertainty in velocity dispersion compared to uncertainty in
half-light radius and heliocentric distance.

\begin{figure}[h!]
 \includegraphics[width=.3\linewidth]{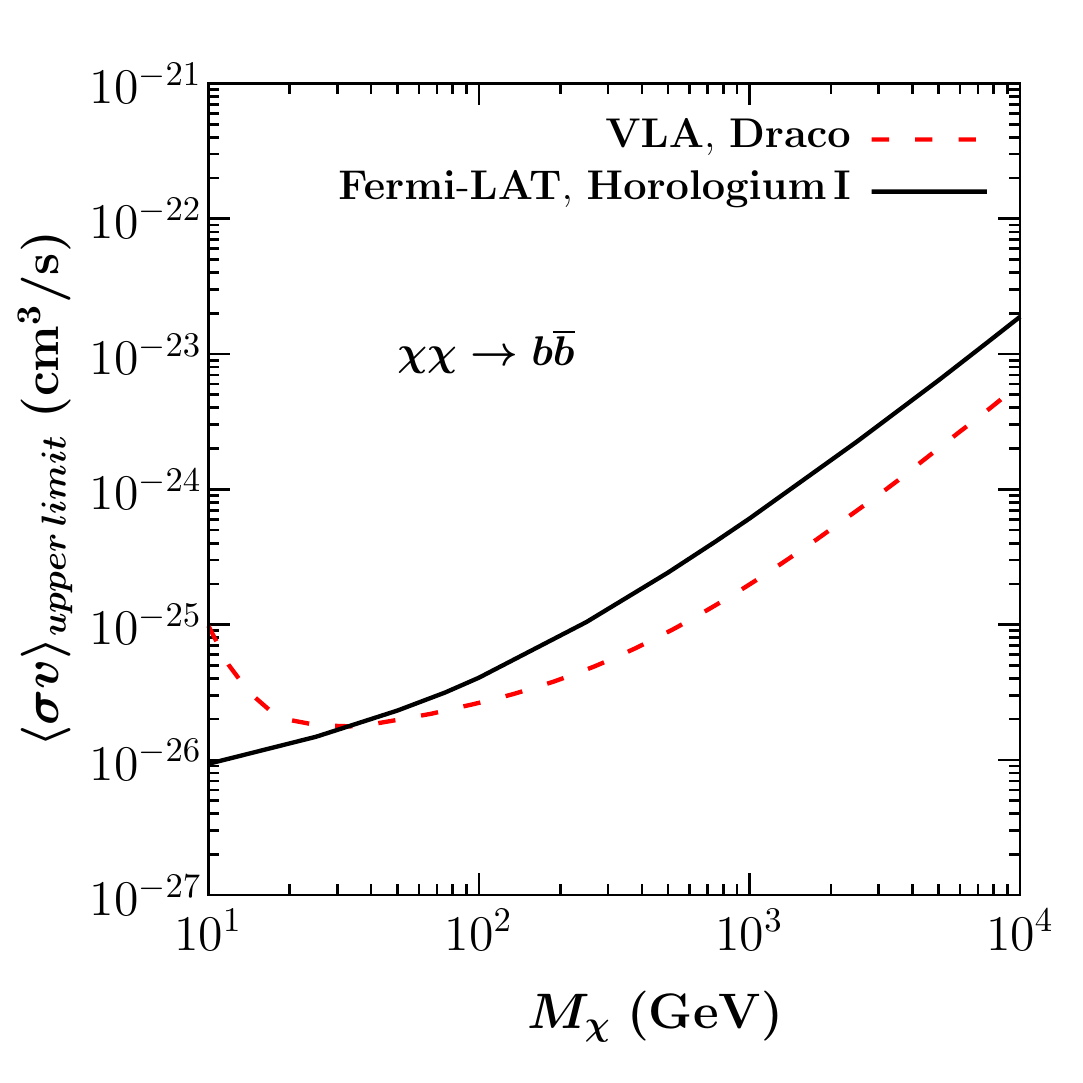}
\hskip 10pt
 \includegraphics[width=.3\linewidth]{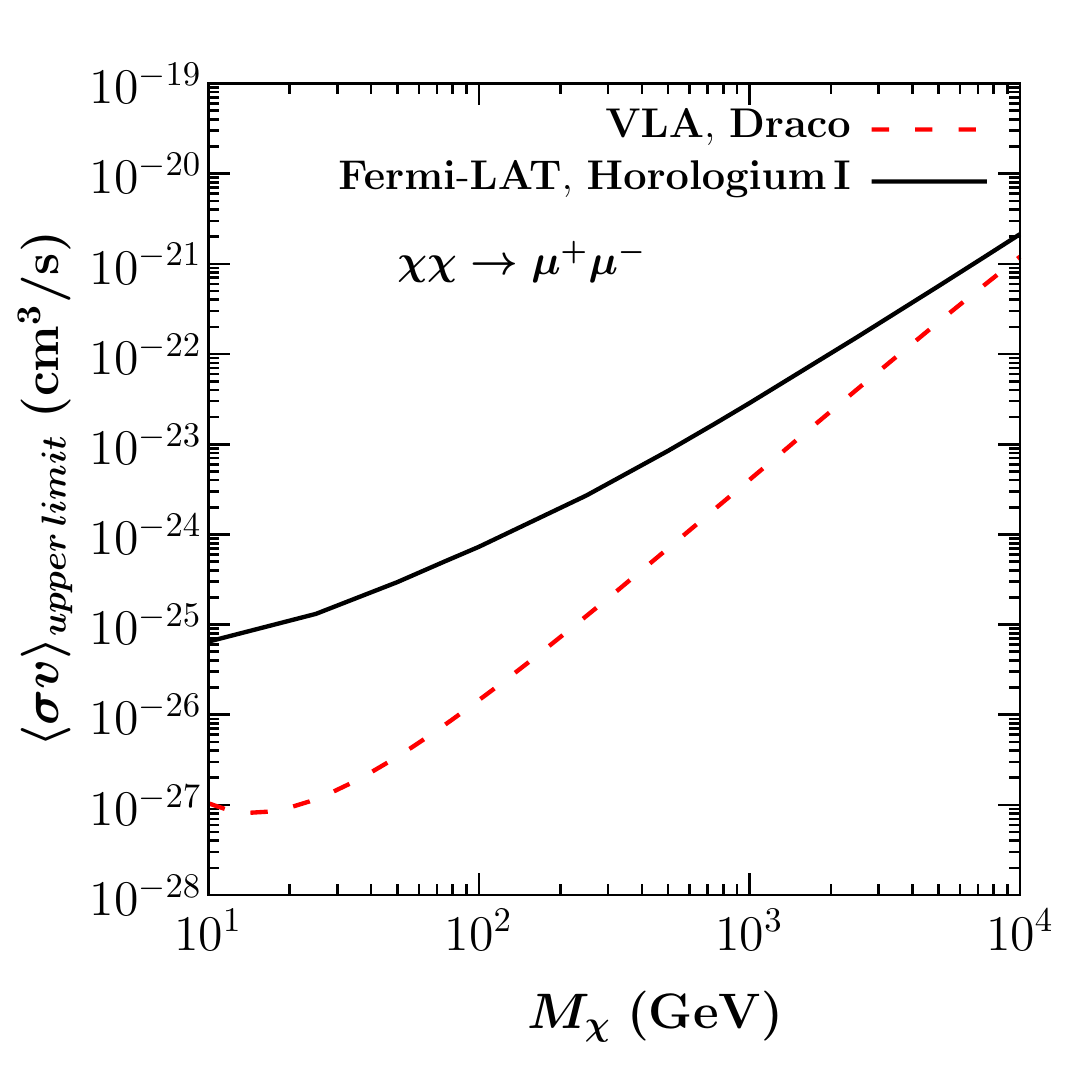}
\hskip 10pt
 \includegraphics[width=.3\linewidth]{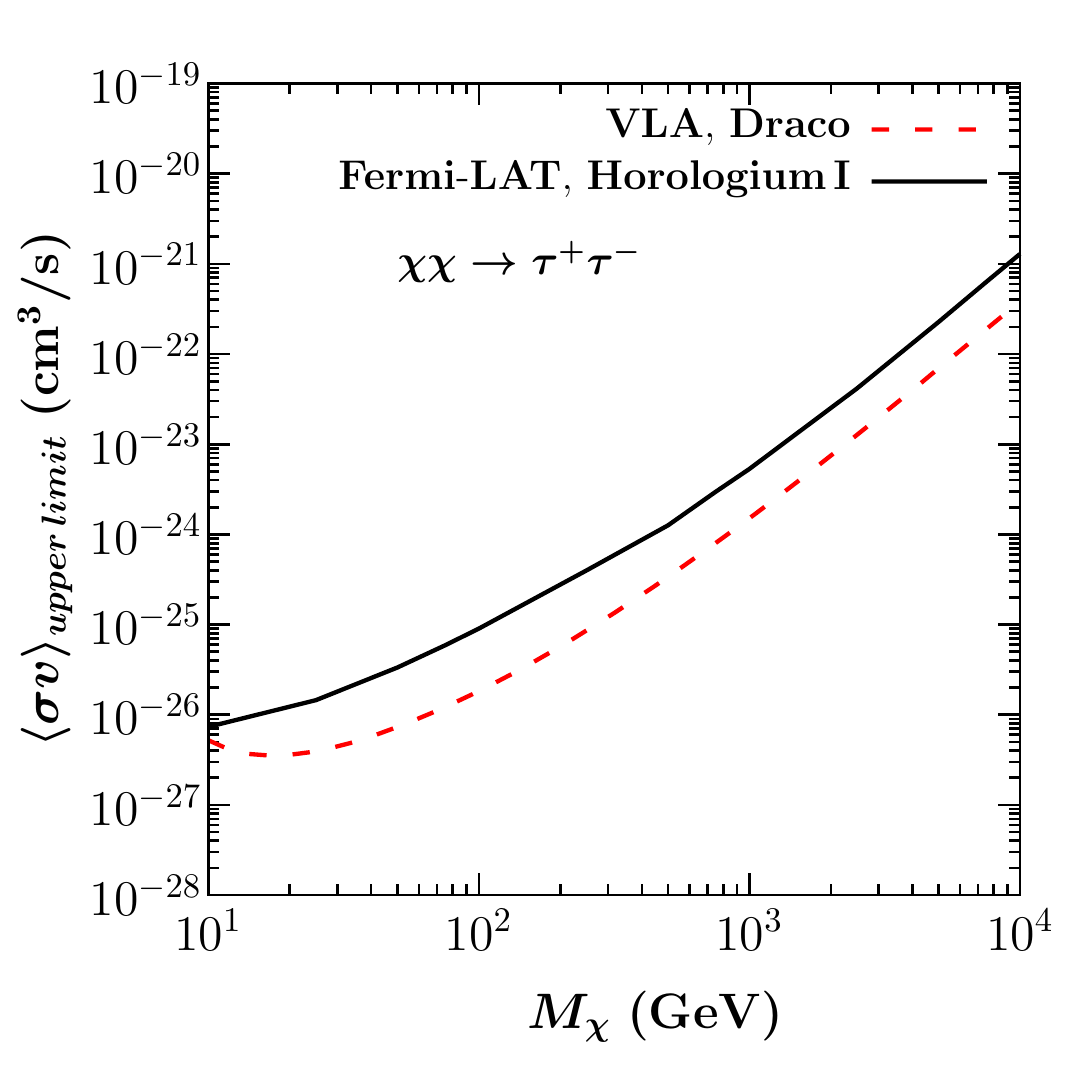}
 
 \caption{\em Comparison of 95 $\%$ C.L. upper limits on $\langle \sigma
 v \rangle$ from VLA and Fermi-LAT. We have compared the best limit
 from VLA (Draco) with the best limit from Fermi-LAT (Horologium
 I). The left, center and right panels are for $\chi \chi \rightarrow
 b \bar{b}\ , \ \tau^+ \tau^-\ \rm{and} \ \mu^+\mu^- $ respectively.
 The VLA limit for Draco has been obtained with the astrophysical
 parameters $B \, = \, 1\, \mu$G, $D_0 = 3 \times 10^{28}$ cm$^2$/s
 and $\gamma_D = 0.3$. NFW density profile has been used for DM
 distribution inside the
 UFD.}  \label{figure:Exclusion_curve_VLA_Fermilat_comparison}
\end{figure}

 Despite these uncertainties, it can, thus,  be safely concluded
that we are about to enter a very interesting era of indirect
detection of DM. A particularly important conclusion is that, even
a photonic signal is best searched for by a combination of multiple
telescopes working at very disparate wavelengths. This is exemplified by
Fig.~\ref{figure:Exclusion_curve_VLA_Fermilat_comparison} which shows the
comparison between the best limits from VLA with the best from
Fermi-LAT for all the three annihilation channels. It is interesting
to note that while, for Fermi-LAT, it is Horologium I that provides 
the best limit, for the VLA, it is Draco that does so. Furthermore,
over the wide mass range $M_\chi \in (10, 10^4)\gev$, if the DM annihilates
preferentially to $\mu^+ \mu^-$ or $\tau^+ \tau^-$, then it is the VLA that
has a better sensitivity, while if the preferred channel is $b\bar{b}$ instead, 
Fermi-LAT is likely to prove a better bet. 

It is heartening to note that apart from collecting a much
increased amount of
data, the Fermi-LAT collaboration is already engaged in efforts
that would allow it to perform improved
event selection with larger collection area \cite{Ackermann:2011wa}.
Furthermore, multi-target multi-instrument searches for
gamma-ray signals from annihilation of dark matter particles in the
mass between 10 GeV and 100 TeV \cite{Ahnen:2016qkx} are in the offing.
Such an approach
can provide rather stringent limits on the cross sections for dark
matter particles and has the ability to probe the widest range of
masses. In the near future, the next generation ground based
atmospheric Cerenkov telescope array (CTA) will be operational
and is expected to have an unprecedented sensitivity over a very wide
range of energy from few tens of GeV to hundreds of
TeV \cite{CTAConsortium:2018tzg} and will have the potential to probe
the thermal relic interaction rate in various astrophysical objects.
At the same time, radio astronomy is entering into a golden era with
the Square Kilometer Array (SKA) project expected to complete the
Phase-I construction of the world's largest radio telescope array soon
and further improve the sensitivity and spectral range in Phase-2 in
future. Along with its pathfinders ASKAP and MeerKAT as also
Low-Frequency Array (LOFAR) which have started operation, multifarious
search strategies can be undertaken.  All these telescope arrays
will be able to reach a sensitivity of a few $\mu$Jy rms noise level
in only a few hours of observation time. Finally,
SKA will have $\sim$ 10
times increase in sensitivity from few tens of MHz to 50 GHz and such
a wide frequency coverage will be important for putting constraints on
dark matter particle mass. Furthermore SKA will also have the
capability to measure the magnetic field in many dSphs via Faraday
rotation measurements which will help one to make more robust
predictions for the expected dark matter induced synchrotron
radiation. 

It is also worthwhile to consider that, were DM
signals to be confirmed by other independent experiments, the signal
(or lack thereof) in future radio-frequency (or gamma-ray)
observatories could, in principle, be used to probe properties (such
as magnetic fields, diffusion parameters and the DM density
distribution) of the dwarf spheroidals.

\vskip 10pt
\noindent
    {\em Note added:} After this paper was completed,
    ref.\cite{Vollmann:2020gtu} appeared, claiming to have corrected
    an error in the Green's function (see eqn.\ref{greens_func}) used
    so far in the literature. We have recomputed $S_\nu$ with the
    Green's function advocated in ref.\cite{Vollmann:2020gtu}, and
    find that while it is now enhanced slightly for small $\nu$, it is
    suppressed by about 40\% at the very largest of
    frequencies. Overall, there would be only a small change in the
    outcome, distinctly smaller than the other uncertainties that we
    have listed.

\begin{acknowledgments}
The authors would like to thank Alex McDaniel for providing us the most recent 
version of the RX-DMFIT code \& various clarifications regarding the 
use of this code, and Stefano Profumo
for discussions on diffusion radii for UFDs.
PB thanks Mousumi Das for helpful discussions on radio analysis.
KD would like to thank Marco Cirelli and Arpan Kar for useful discussions.
The work of PB is supported by the DST INSPIRE Fellowship Scheme, India.
DC acknowledges research grant CRG/2018/004889 of the SERB, India.

\end{acknowledgments}

\setcounter{table}{0}
\renewcommand{\thetable}{A--\Roman{table}}
\appendix
\section{Source details}\label{section:source_details}

We revisit here the dwarf spheroidal galaxies considered in our study,
chosen on account of their very high mass to light ratio and
moderately large velocity dispersion of the stars therein, each of
which points to the conclusion that these UFDs are likely to be very
rich in dark matter content \cite{Baumgardt:2008zt}.  The
corresponding spectroscopic and the photometric studies are mentioned
in Table~\ref{table:astro_fundamental_param_dwarfs}, where the
quantities $M/L$, $d$, $r_{1/2}$ and $\sigma_{l.o.s}$  denote the mass to light
ratio, the heliocentric distance, the half-light radius and the velocity dispersion of each UFD galaxy. 
$\theta_{max}^o$ is the angle made by the outer most star of the UFD.

\begin{table}[!h]
\begin{center}
\begin{tabular}{||p{2.3cm}|p{1.8cm}|p{1.9cm}|p{1.5cm}|p{2.2cm}|p{2cm}||}
\hline 
\hline
Galaxy & M/L $(M_{\odot}/L_{\odot})$ & d (Kpc)  & $r_{1/2}~(pc)$ & $\sigma_{l.o.s}~(km~s^{-1})$ & $\theta_{max}^o$ \\
\hline
Aquarius~II & $1330^{+3242}_{-227}$ & $107.9^{+3.3}_{-3.3}$ & $123^{+22}_{-21}$ & $6.2^{+2.6}_{-1.7}$ & 0.11134 \\
\hline
Carina~II & $369^{+309}_{-161}$ & $37.4^{+0.4}_{-0.4}$ & $77^{+8}_{-8}$ & $3.4^{+1.2}_{-0.8}$ & 0.23\\
\hline
Draco~II & $501^{+1083}_{-421}$ & $20.0^{+3.0}_{-3.0}$ & $12^{+5}_{-5}$ & $3.4^{+2.5}_{-1.9}$ & 0.1\\
\hline
Eridanus~II & $420^{+210}_{-140}$ & $366.0^{+17.0}_{-17.0}$ & $176^{+14}_{-14}$ & $7.1^{+1.2}_{-0.9}$ & 0.062 \\
\hline
Grus~I & $<~2645$ & $120.2^{+11.1}_{-11.0}$ & $52^{+26}_{-26}$ & $4.5^{+5.0}_{-2.8}$ & 0.093\\
\hline
Horologium~I & $570^{+1154}_{-112}$ & $79.0^{+7.0}_{-7.0}$ & $32^{+5}_{-5}$ & $5.9^{+3.3}_{-1.8}$ & 0.0619 \\
\hline
Hydra~II & $<~315$ & $151.0^{+8.0}_{-8.0}$ & $71^{+11}_{-11}$ & $<6.82$ & 0.08509 \\
\hline
Leo~V & $264^{+326}_{-264}$ & $173.0^{+5.0}_{-5.0}$ & $30^{+17}_{-17}$ & $4.9^{+3.0}_{-1.9}$ & 0.077 \\
\hline
Pegasus~III & $1470^{+5660}_{-1240}$ & $215.0^{+12}_{-12}$ & $37^{+14}_{-14}$ & $7.9^{+4.4}_{-3.1}$ & 0.03049\\
\hline
Pisces~II & $370^{+310}_{-240}$ & $183.0^{+15}_{-15}$ & $48^{+10}_{-10}$ & $4.8^{+3.3}_{-2.0}$ & 0.06861\\
\hline
Reticulum~II & $467^{+286}_{-168}$ & $30^{+2}_{-2}$ & $32^{+3}_{-3}$ & $3.4^{+0.7}_{-0.6}$ & 0.24\\
\hline
Tucana~II & $1913^{+2234}_{-950}$ & $57.5^{+5.3}_{-5.3}$ & $115^{+32}_{-32}$ & $7.3^{+2.6}_{-1.7}$ & 0.225\\
\hline
Tucana~III & $<~240$ & $25.0^{+2}_{-2}$ & $43^{+6}_{-6}$ & $<2.18$ & 0.2\\
\hline
Triangulum~II & $<~2510$ & $30^{+2}_{-2}$ & $28^{+8}_{-8}$ & $<6.36$ & 0.15\\
\hline
\hline
\end{tabular}
\end{center}
\caption{\em Properties of our selected UFD galaxies. Parameters $d$, $r_{1/2}$ and $\sigma_{l.o.s}$ has been collected from ~\cite{Pace:2018tin}. 
$M/L$ and $\theta_{max}^o$ has been collected from the references 
\cite{Koposov:2015jla, Kirby:2015ija, Martin:2016, Kim:2016twa, Li:2016utv, Kirby_2017, Walker_2016, Walker:2015mla,
Simon:2016mkr, Torrealba:2016, Li:2018yme, collins:2017, Genina:2016kzg}.}
\label{table:astro_fundamental_param_dwarfs}
\end{table}

\clearpage
\section{Parameters used in Science Tools}
\small
  \begin{table}[!h]
{\small
    \begin{center}
    \begin{tabular}{|ll|}
        \hline \hline
        {\bf Parameter for data extraction\tablefootnote{\url{https://fermi.gsfc.nasa.gov/ssc/data/analysis/scitools/extract_latdata.html}}}  &\\
        \hline\hline
        Parameter & Value \\[-0.3ex]
        \hline \hline
        Radius of interest (ROI) &  $15^{\circ}$ \\[-0.3ex]
        TSTART (MET) & 241976960 (2008-09-01 15:49:19.000 UTC) \\[-0.3ex]
        TSTOP (MET) & 570987500 (2019-02-04 15:38:15.000 UTC) \\[-0.3ex]
        Energy Range & 100 MeV - 300 GeV  \\[-0.3ex]
        \textit{Fermitools} version & \texttt{1.2.1}\tablefootnote{\url{https://fermi.gsfc.nasa.gov/ssc/data/analysis/software/}} \\[-0.3ex]
        \hline \hline
        $~~~~~~~~~~~~~~~~~~~$\texttt{gtselect} for event selection\tablefootnote{\url{https://fermi.gsfc.nasa.gov/ssc/data/analysis/scitools/help/gtselect.txt}\label{gtselect}} &\\[-0.3ex]
        \hline \hline
        Event class & Source type (128)\tablefootnote{\url{https://fermi.gsfc.nasa.gov/ssc/data/analysis/documentation/Cicerone/Cicerone_Data_Exploration/Data_preparation.html}\label{datacut}}  \\[-0.3ex]
        Event type & Front+Back (3)\footref{datacut} \\[-0.3ex]
        Maximum zenith angle cut & $90^{\circ}$\footref{datacut} \\[-0.3ex]
        \hline \hline
        $~~~~~~~~~~~~~~~~~~~$\texttt{gtmktime} for time selection\tablefootnote{\url{https://fermi.gsfc.nasa.gov/ssc/data/analysis/scitools/help/gtmktime.txt}}  & \\[-0.3ex]
        \hline \hline
        Filter applied & $\rm{(DATA\_QUAL>0)\&\&(LAT\_CONFIG==1)}\tablefootnote{\url{https://fermi.gsfc.nasa.gov/ssc/data/analysis/scitools/data_preparation.html}\label{data_preparation}}$ \\[-0.3ex]
        ROI-based zenith angle cut & No \footref{data_preparation}  \\[-0.3ex]
        \hline \hline
        $~~~~~~~~~~~~~~~~~~~$\texttt{gtltcube} for livetime cube\tablefootnote{\url{https://fermi.gsfc.nasa.gov/ssc/data/analysis/scitools/help/gtltcube.txt}}  &\\[-0.3ex]
        \hline \hline
        Maximum zenith angle cut ($z_{cut}$) & $90^{\circ}\tablefootnote{\url{https://fermi.gsfc.nasa.gov/ssc/data/analysis/documentation/Cicerone/Cicerone_Likelihood/Exposure.html}}$ \\[-0.3ex]
        Step size in $cos(\theta)$ & 0.025  \\[-0.3ex]
        Pixel size (degrees) & 1 \\[-0.3ex]
        \hline \hline
        $~~~~~~~~~~~~~~~~~~~$\texttt{gtbin} for 3-D (binned) counts map\tablefootnote{\url{https://fermi.gsfc.nasa.gov/ssc/data/analysis/scitools/help/gtbin.txt}}  & \\[-0.3ex]
        \hline \hline
        Size of the X $\&$ Y axis (pixels) & 140 \\[-0.3ex]
        Image scale (degrees/pixel) & 0.1 \\[-0.3ex]
        Coordinate system & Celestial (CEL) \\[-0.3ex]
        Projection method & AIT \\[-0.3ex]
        Number of logarithmically uniform energy bins & 24 \\[-0.3ex] 
        \hline \hline
        $~~~~~~~~~~~~~~~~~~~$\texttt{gtexpcube2} for exposure map\tablefootnote{\url{https://fermi.gsfc.nasa.gov/ssc/data/analysis/scitools/help/gtexpcube2.txt}}  &\\[-0.3ex]
        \hline \hline
        Instrument Response Function (IRF) & $\rm{P8R3\_SOURCE\_V2}\tablefootnote{\url{https://fermi.gsfc.nasa.gov/ssc/data/analysis/documentation/Pass8_usage.html}\label{pass8}}$  \\[-0.3ex]
        Size of the X $\&$ Y axis (pixels) & 400 \\[-0.3ex]
        Image scale (degrees/pixel) & 0.1 \\[-0.3ex]
        Coordinate system & Celestial (CEL) \\[-0.3ex]
        Projection method & AIT \\[-0.3ex]
        Number of logarithmically uniform energy bins & 24 \\[-0.3ex] 
        \hline \hline
        $~~~~~~~~~~~~~~~~~~~$diffuse models and Source model XML file\tablefootnote{\url{https://fermi.gsfc.nasa.gov/ssc/data/analysis/user/make4FGLxml.py}}  &\\[-0.3ex]
        \hline \hline
        Galactic diffuse emission model & $\rm{gll\_iem\_v07.fits}\tablefootnote{\url{https://fermi.gsfc.nasa.gov/ssc/data/access/lat/BackgroundModels.html}\label{background}}$  \\[-0.3ex]
        Extragalactic isotropic diffuse emission model & $\rm{iso\_P8R3\_SOURCE\_V2\_v1.txt}\footref{background}$ \\[-0.3ex]
        Source catalog & 4FGL  \\[-0.3ex]
        Extra radius of interest &  $5^{\circ}$ \\[-0.3ex]
        Spectral model &  DMFit Function\cite{Jeltema:2008hf} \tablefootnote{\url{https://fermi.gsfc.nasa.gov/ssc/data/analysis/scitools/source_models.html}} \\[-0.3ex]
        \hline \hline
    \end{tabular}
    \end{center}
}
\vskip -15pt
\caption{\em Parameters used in \texttt{Science Tools} for \textit{Fermi}-LAT data analysis}
    \label{table:fermi_lat_parameters}
\end{table} 

\normalsize
\clearpage
\bibliographystyle{JHEP}
\bibliography{ref.bib}
\end{document}